\documentclass[preprint,11pt]{elsarticle}
\usepackage{listings}
\usepackage{longtable}
\usepackage{lscape}
\usepackage{geometry}
\usepackage{float}
\usepackage{adjustbox}
\usepackage[justification=centering]{caption}
\usepackage{rotating}
\usepackage{soul}
\usepackage{amsmath,amsfonts}

\usepackage[utf8]{inputenc} % for french accents
\usepackage{lmodern}            % Latin Modern (improves Computer Modern)
\usepackage{microtype}          % Beautiful microtypography
\usepackage{newtxtext,newtxmath}  % Times-like, professional
\usepackage{ifthen}
\usepackage{graphicx}
\usepackage{hyperref}
\usepackage{multirow}
\usepackage{array} % for aligning table columns
\usepackage{arydshln} % for dashed lines in tables
\usepackage[many]{tcolorbox} % for bordered theorem boxes

\usepackage{xspace}
\usepackage{xcolor}
\usepackage[normalem]{ulem} 			% \sout macro
\usepackage{amsmath,amsfonts}

\usepackage{caption} % for subfigures
\usepackage{subcaption} % for subfigures

\newcolumntype{L}[1]{>{\raggedright\let\newline\\\arraybackslash\hspace{0pt}}p{#1}}
\newcolumntype{C}[1]{>{\centering\let\newline\\\arraybackslash\hspace{0pt}}p{#1}}
\newcolumntype{R}[1]{>{\raggedleft\let\newline\\\arraybackslash\hspace{0pt}}p{#1==}}

\newboolean{showcomments}
\setboolean{showcomments}{true}

\ifthenelse{\boolean{showcomments}}
{
	\newcommand{\nb}[3]{
		{\colorbox{#2}{\bfseries\sffamily\scriptsize\textcolor{white}{#1}}}
		{\textcolor{#2}{\sf\small \textit{#3}}}}
	 
	\newcommand{\bnote}[2]{\fbox{\color{blue}\bfseries\sffamily\scriptsize#1}
    	{\color{blue}\sf\small \textit{#2}}}
	\newcommand{\old}[1]{{\color{gray}\sout{#1}}} % old to be removed
	\newcommand{\del}[1]{\old{#1}} % please remove
	\newcommand{\ins}[1]{{\textcolor{blue}{\uline{#1}}}} % please insert	
	\newcommand{\ugh}[1]{{\textcolor{red}{\uwave{#1}}}} % please rephrase	
	\newcommand{\chg}[2]{{\textcolor{red}{\sout{#1}}}{\ra}\textcolor{blue}{\uline{#2}}} % please change
	 
	\newcommand{\fix}[1]{\bnote{FIX}{#1}}
}{
	\newcommand{\bnote}[2]{}
	\newcommand{\nb}[3]{}
	\newcommand{\old}[1]{}
	\newcommand{\del}[1]{}
	\newcommand{\ins}[1]{}
	\newcommand{\ugh}[1]{}
	\newcommand{\chg}[2]{}
	
	\newcommand{\fix}[1]{}
} 

\newcommand{\hide}[1]{}

\usepackage[many]{tcolorbox}  % For colored boxes
\newenvironment{boxB}{
  \begin{tcolorbox}[colback=white, colframe=black, rounded corners]
}{
  \end{tcolorbox}
}

\graphicspath{{figures/}{newFigures/}}
\newcommand{\commented}[1]{}

\newcommand{\eg}{\emph{e.g.,}\xspace}
\newcommand{\ie}{\emph{i.e.,}\xspace}
\newcommand{\etal}{\emph{et al.,}\xspace}
\newcommand{\ct}[1]{{\textsf{#1}}\xspace}

\newcommand{\compl}{\textsc{Complishon}\xspace}
\newcommand{\baseline}{\textit{Baseline}\xspace}
\newcommand{\semantics}{\textit{Semantics-Based}\xspace}
\newcommand{\lateralpackage}{\textit{Lateral Packages}\xspace}
\newcommand{\directdependencies}{\textit{Direct Package Dependencies}\xspace}
\newcommand{\transitivedependencies}{\textit{Two-Level Package Dependencies}\xspace}

\newcommand{\locallevel}{\textit{Local Level}\xspace}
\newcommand{\unitlevel}{\textit{Unit Level}\xspace}
\newcommand{\completelevel}{\textit{Complete Level}\xspace}
\newcommand{\class}{\textit{class names}\xspace}
\newcommand{\method}{\textit{selectors}\xspace}
\newcommand{\etc}{\textit{etc.}\xspace}

\usepackage{url}            
\makeatletter
\def\url@leostyle{%
  \@ifundefined{selectfont}{\def\UrlFont{\sf}}{\def\UrlFont{\small\sffamily}}}
\makeatother
\definecolor{main}{HTML}{828282}    % setting main color to be used
\definecolor{sub}{HTML}{E0E0E0}     % setting sub color to be used

\tcbset{
    sharp corners,
    colback = white,
    before skip = 0.2cm,    % add extra space before the box
    after skip = 0.5cm      % add extra space after the box
}                           % setting global options for tcolorbox

\newtcolorbox{cbox}{
    enhanced, % for a fancier setting,
    boxrule = 0pt, % clearing the default rule
    borderline = {0.75pt}{0pt}{main}, % outer line
    borderline = {0.75pt}{2pt}{sub} % inner line
}

\journal{Journal of Computer Languages}
\graphicspath{{figures/}{../fig/}}

\usepackage{tikz}
\newcommand*\circled[1]{\tikz[baseline=(char.base)]{
            \node[shape=circle,draw,inner sep=2pt] (char) {\scriptsize #1};}}

\usepackage[many]{tcolorbox}
\usepackage{mdframed}
\usepackage{paralist}

\begin{document}
\begin{frontmatter}

\title{Evaluating Package-Level Scoping Strategies \\ for Repository-Level Code Completion in Pharo}

\author{Omar AbedelKader\fnref{label1}}
\author{Stéphane Ducasse\fnref{label1}}
\author{Guillermo Polito\fnref{label1}}
\author{Oleksandr Zaitsev\fnref{label2}}
\author{Romain Robbes\fnref{label3}}

\fntext[label1]{Univ. Lille, Inria, CNRS, Centrale Lille, UMR 9189 CRIStAL, Park Plaza, Parc scientifique de la Haute-Borne, 40 Av. Halley Bât A, 59650 Villeneuve-d'Ascq, France }
\fntext[label2]{CIRAD, UMR SENS, 34000 Montpellier, France}
\fntext[label3]{CNRS, University of Bordeaux, Bordeaux INP, LaBRI, UMR5800, 33400, Talence, France}

%%%%%%%%%%%%
%%%%%%%%%%%%
\begin{abstract}
Source code does not exist in isolation.  Code is structured in project repositories composed of packages, and large applications are often composed of multiple projects. This paper investigates whether token-level completion engines can take advantage of such information.

In this context, Pharo offers a sophisticated completion engine based on language semantics using \emph{heuristics}, \ie ordered sequences of lazy fetchers that retrieve and rank completion candidates. These heuristics can be recomposed or extended to support various activities (\eg live programming or history usage navigation). The default heuristics rely on the language's semantics and scoping rules. While this system is powerful, it does not account for package concerns of any sort (neither repository structure nor package dependencies). As a result, it does not prioritize classes within the same package or project, nor the classes that they may use, treating all global names equally. 

In this paper, we introduce and systematically evaluate several new heuristics that address this limitation: \lateralpackage, \directdependencies, and \transitivedependencies heuristics. Our benchmark runs over the method call sites of \ct{219} packages, comprising \ct{4,535} classes and a total of \ct{35,972} methods. This codebase covers major projects such as Iceberg, Moose, Roassal, Seaside, and Spec, ensuring a diverse and realistic benchmark for assessing the effectiveness of our proposed heuristics. We evaluate two key concerns: the completion of class and method names. Rather than proposing a new learning model, this work evaluates whether explicit package dependencies, already available in modular software systems, can serve as a lightweight structural signal for improving completion ranking.

Results show that the \directdependencies heuristic improves Mean Reciprocal Rank (MRR): for 3-character prefixes, MRR increases from \ct{0.20} to \ct{0.44} for class-name completion and from \ct{0.10} to \ct{0.41} for method-name completion. These results indicate that package-aware completion strategies provide more accurate and relevant suggestions than the default Semantics-Based strategy in the evaluated projects. Because \directdependencies outperformed the default strategy for class-name completion, the Pharo Industrial Consortium decided to introduce it in Pharo 13.

\end{abstract}

\end{frontmatter}

%%%%%%%%%%%%
%%%%%%%%%%%%
\section{Introduction}\label{introduction}

Applications are often composed of multiple projects structured around Git repositories. In such environments, code does not exist in isolation: projects are organized into repositories composed of multiple packages with explicit dependencies. At the same time, modern code completion systems are increasingly expected to be intelligent, predicting not only syntactically valid tokens but also semantically relevant ones, thereby reducing the effort required to navigate large codebases \cite{Biba22a, Jin25a}. 

This paper investigates \emph{whether token-level completion engines can take advantage of package structures and their explicit dependencies}.
Repository-level code completion \cite{Izad22a, Zhan23a} denotes techniques that exploit the full project structure, packages, cross-file relationships, and dependency information, rather than treating the global namespace as a flat set of entities. 
This work extends our previous package-aware completion approach \cite{Abed25a}. That earlier work introduced the intuition that packages belonging to the same project can provide useful completion candidates. In contrast, the present paper provides a broader and more systematic evaluation: it distinguishes lateral package relationships, direct package dependencies, and two-level package dependencies. Moreover, it evaluates them at unit, local, and complete completion levels; it compares their behavior for both \class and \method completion; and finally, it studies their responsiveness in the production Pharo completion engine.

This article takes the Pharo programming environment as a testbed for a thorough evaluation \cite{Blac09a}.  Pharo is an open-source object-oriented reflective programming language. It is a descendant of Smalltalk. Pharo also includes a development environment that follows the Smalltalk tradition of a live, image-based system in which programs are developed, inspected, and modified while the system is running. This setting makes code completion particularly important: developers interact continuously with a large live object space, where classes, methods, packages, and dependencies evolve together during development. The Pharo programming environment comes with \compl: a sophisticated code completion engine. Developed by G. Polito, \compl employs a modular and lazy architecture. Its default configuration is based on semantic heuristics that consider both the type of entity (\eg \class, \method \footnote{A selector is the term used to refer to the name of a method in Pharo/Smalltalk. (See \ref{vocab})}) and its location in the program structure (\eg prioritizing classes in the current class hierarchy before superclasses). Unlike traditional alphabetical or purely syntactic completion systems, such as those criticized by Bruch \etal \cite{Bruc09a} and Robbes \etal \cite{Robb10c}, \compl is grounded in the idea that completion quality can be significantly improved by taking into account a context.

\compl was designed to address two main challenges faced by many code completion engines \cite{Biba22a}: \emph{accuracy of the answer} and \emph{time} to find it. 
\compl supports code completion in the context of large industrial code bases. For instance, companies using Pharo in production, such as Lifeware, maintain systems exceeding 30 million lines of code with several tens of thousands of classes. In these contexts, completion suggestions must not only be semantically relevant but also efficiently scoped to avoid information overload. In addition, developers expect answers within one hundred milliseconds \cite{Biba22a}. 
 
Even using advanced default heuristics, \compl, like many completion engines, treats the global namespace as flat, walking through global entities (\class, \method ) in a non-deterministic order other than prefix match. One significant omission in the default configuration of \compl is awareness of project package structure and package dependencies. 

\paragraph{\textbf{Vocabulary}} \label{vocab}
To ensure consistency throughout this paper, we adopt a unified vocabulary and define two key terms as follows:
\begin{itemize}
	\item \textbf{Class Names.} In Pharo, the names of classes, global variables, shared variables, and shared pools start with an uppercase. 	Class names and global variables are all stored within the global namespace. In the rest of the article when we use the term \class we refer 	to class and global variable names. 
	\item \textbf{Method Names.} We use the term \method to refer to method names, following the idiomatic usage in Pharo and Smalltalk literature.
\end{itemize}

\paragraph{\textbf{Contributions}} This article evaluates package-level and repository-level scoping as a lightweight source of context for code completion in a live, dynamically typed programming environment. The contribution is empirical and system-oriented rather than algorithmic in the sense of proposing a new statistical or neural model: we formulate, implement, and systematically evaluate dependency-aware completion strategies within the \compl architecture.

\begin{itemize}
	\item We formulate package-aware and repository-level dependency-based completion heuristics that exploit the current package, lateral packages, direct package dependencies, and two-level package dependencies.
	\item A systematic comparison between several completion scoping strategies (\lateralpackage, \directdependencies, 	\transitivedependencies) assessing how each heuristic prioritizes and ranks candidates in large modular codebases.
	\item An evaluation of these strategies across multiple levels of granularity, specifically: (1) in isolation (\unitlevel), (2) when combined with the package where the completion is requested (\locallevel), and (3) when integrated with the existing semantic-based completion engine 		(\completelevel).
	\item An evaluation of the difference between the completion of \class and \method.
\end{itemize}

\paragraph{\textbf{Nature of the contribution}}
The contribution of this paper is empirical and system-oriented. We study whether package dependency information, which is already present in modular software systems, can be used as an effective scoping signal for code completion. The novelty lies in formulating several package-level scoping strategies, integrating them into the \compl architecture, and evaluating them systematically across completion targets, completion levels, projects, ranking metrics, and responsiveness measurements. This positions dependency-based scoping as a lightweight complement to existing statistical, neural, type-based, and history-based completion approaches.

\paragraph{\textbf{Research Questions}} 
Building upon the motivation and contributions outlined above, this work aims to systematically evaluate the role of package-aware and dependency-based strategies in improving \compl. We formulate the following research questions:

\begin{itemize}
    \item \textbf{RQ1:} Can leveraging package structure and dependency information improve the relevance and accuracy of code completion in Pharo compared to the default semantics-based approach?
    \item \textbf{RQ2:} Which dependency strategy (\lateralpackage, \directdependencies, \transitivedependencies) achieves the best performance in terms of completion accuracy and ranking quality (\eg Mean Reciprocal Rank)?
    \item \textbf{RQ3:} Does package awareness affect completion performance differently for \class and for \method ?
    \item \textbf{RQ4:} Can dependency-based strategies enhance completion relevance without compromising responsiveness and performance in large modular Pharo systems?    
\end{itemize}

\paragraph{\textbf{Outline}} This article is structured as follows: Section~\ref{background} gives an overview of the \compl engine and its modular heuristic-based architecture. Section~\ref{approach} presents our approach and outlines our hypothesis for package awareness suggestions, specifically describing the different heuristics that we will assess. Section~\ref{methodology} presents the evaluation methodology. Section~\ref{classevaluation} presents the evaluation results and a discussion for the strategies applied to \class, while Section~\ref{methodevaluation} provides the corresponding results and discussion for \method. Section~\ref{limitations} discusses the findings and their implications for completion systems in dynamic, large-scale environments. Section~\ref{responsiveness} evaluates the responsiveness of the approach. Section~\ref{relatedwork} situates our work within the broader landscape of static code completion.
	
%%%%%%%%%%%%
%%%%%%%%%%%%
\section{Background: The Complishon Engine and its Limits}\label{background}

\compl, the Pharo completion engine, is based on a modular architecture based on heuristics, which are a sequence of lazy fetchers. It builds a dedicated context for code completion, considering elements such as source text and caret position, where the caret is the current insertion point in the editor. The generation and presentation of suggestions are managed by the IDE, which also integrates variations such as case sensitivity filtering and adaptive configuration based on the structure and semantics of the current code environment (workspace, live objects, \etc). \compl leverages AST-based analysis and a double-dispatch mechanism to adapt dynamically to the surrounding context. This adaptability employs fetcher-based heuristics triggered by a Visitor \cite{Gamm95a} matching AST nodes, informed by parsing and typing processes, to refine output and dynamically eliminate redundant or irrelevant suggestions. 

\subsection{Architecture}

\begin{figure}[htbp]
\begin{center}
   \includegraphics[width=0.7\linewidth]{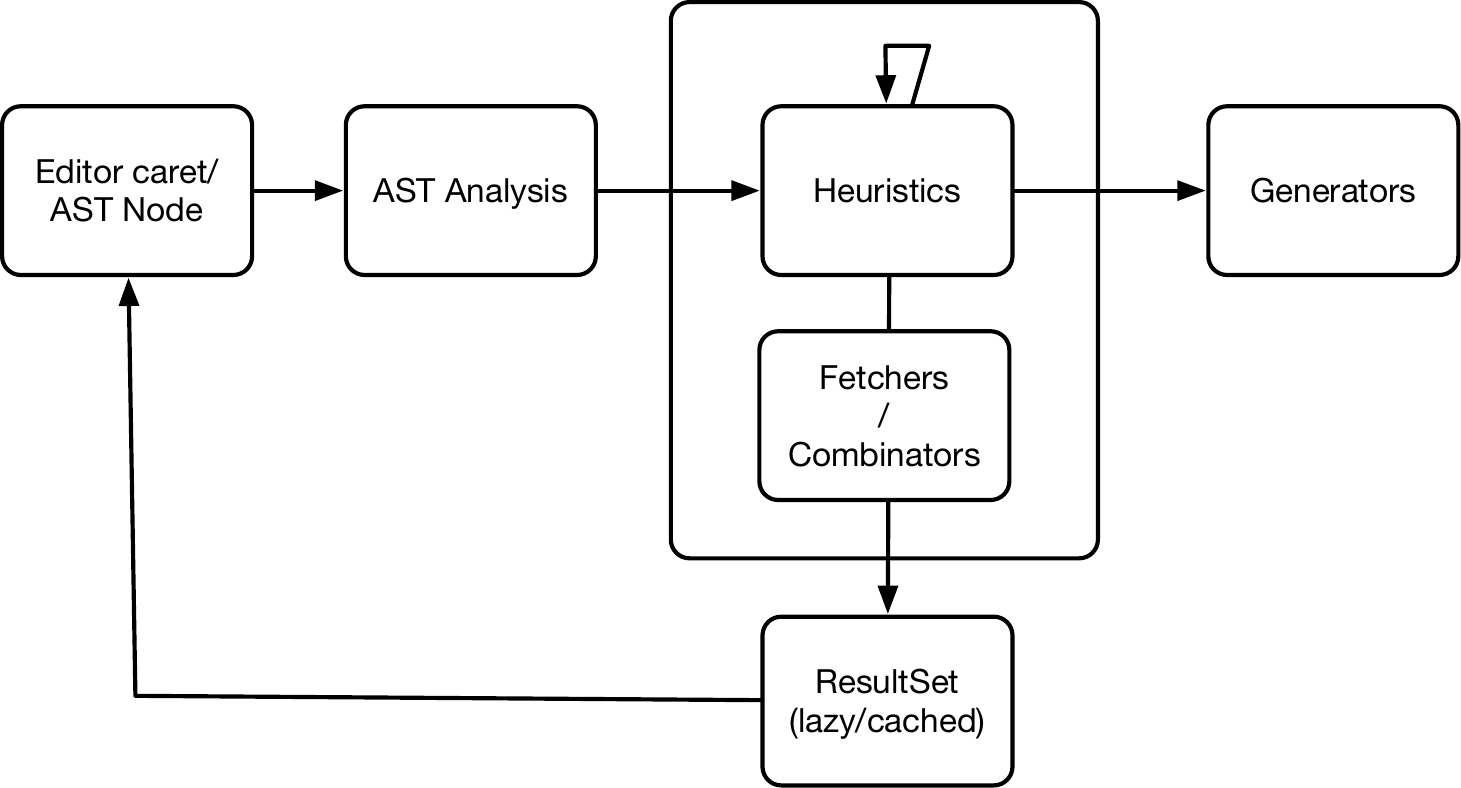}
\end{center}
\caption{\textbf{\compl's architecture: Heuristics are sequences of fetchers. Fetchers are elementary query units that lazily request information and can pass control to other fetchers in the sequence of a heuristic.}}
\label{fig:complishon}
\end{figure}

The heuristics are modular, specialized for different code elements such as \class and \method, and systematically connected in sequences forming a robust and comprehensive filtering framework. This chaining process includes semantics-based strategies such as prioritizing instance variables before superclass variables, \ct{self}/\ct{super} message suggestions, inherited methods, and inferred initialization constructs. It supports live programming within debuggers and inspectors. Up to version 12, Pharo used this default configuration. The Pharo completion engine consists primarily of three key components: Heuristics, Lazy Fetchers, and a lazily cached Result Set (see Figure~\ref{fig:complishon}). 
Note that the strategies we evaluate are expressed as \compl heuristics (See Section~\ref{approach}). 

\paragraph{\textbf{Heuristics}} 
A heuristic is a sequence of lazy fetchers that provide semantic guidance for the completion process by analyzing the Abstract Syntax Tree (AST) node located at the cursor (editor caret) and selecting the appropriate fetchers for completion suggestions. These heuristics are structured in a Chain of Responsibility Design Pattern \cite{Gamm95a}, allowing a heuristic to pass handling responsibilities down the chain if it cannot process the current AST node itself. By default, Pharo is configured to use a heuristic, called \semantics, that follows the language scope and semantics. When such a heuristic fails, the default is to look in the global namespace, just filtering names based on a pattern without any deterministic order.

\paragraph{\textbf{Lazy Fetchers}} 
Fetchers, implemented using Generators \footnote{A generator is a routine that can be used to control the iteration behavior of a loop --- https://en.wikipedia.org/wiki/Generator\_(computer\_programming)}, lazily retrieve and filter potential code completion candidates based on the context and user input, significantly optimizing performance and memory usage. A decorator pattern further enhances fetchers by preventing duplicate suggestions, particularly crucial in scenarios involving method inheritance, ensuring results remain relevant and unique. Fetchers use specialized filters (\eg \ct{CoBeginsWithFilter}) to match completion suggestions with the user's partially typed input. For \class, the \ct{CoGlobalVariableFetcher} looks for names in the global namespace. It requests the namespace for names matching a prefix. Note that such a namespace is unsorted, therefore, it will give the first name matching a given pattern without any ordering. 

\paragraph{\textbf{Result Set}}
The Result Set component serves as a lazy, cached store that accumulates the suggestions provided by fetchers only as required, further enhancing efficiency. 

\subsection{Limitations of the existing approach}

Although \compl is effective in identifying global entities such as \class and \method, it currently does not leverage the package structure of a project effectively. As a result, it treats all global names uniformly, offering no preference to entities located in the same or related packages. It is then a good case study to assess the impact of project structure on the completion quality.

When invoking code completion within the \ct{SpPresenter} class, located in the \ct{Spec2-Core} package, \compl proposes candidates from the global namespace without taking package-local context into account. As a result, locally relevant classes such as \ct{SpPresenterBuilder}, \ct{SpTextPresenter}, or \ct{SpApplication}, which are defined in the same package, may be ranked below unrelated globally visible classes. This behavior degrades the developer experience, especially in large codebases where many globally accessible entities are available. Recent work on code completion \cite{Prok15a, Biba22a, Chis15b} shows that narrowing the search space can lead to more relevant suggestions. Approaches such as Bayesian completion models \cite{Prok15a} and log-based ranking strategies used in commercial IDEs \cite{Biba22a} follow this direction by using contextual or structural information to guide ranking. Our work is aligned with this idea but focuses on the package structure of Pharo systems. Following the principles of moldable development \cite{Chis15b}, which promote tools that adapt to the structure of the system at hand, we extend \compl's architecture with new heuristics that take package boundaries and dependencies into account.

%%%%%%%%%%%%
%%%%%%%%%%%%
\section{Package-Level Code Completion}\label{approach}

We present now all the strategies that we designed and evaluated for \class (Section~\ref{classevaluation}) and \method 
 (Section~\ref{methodevaluation}). We define several new strategies that leverage package structure to improve completion prioritization. 
 
A strategy is implemented as a sequence of \compl heuristics with extension points, enabling fine-grained control over completion fetchers and filtering logic. All the strategies default to the \baseline strategy. Each of the strategies represents a possible setup developers can use. In particular, the second one, the \semantics strategy, is the default one in Pharo 12.

We present here the completion heuristic strategies for the \completelevel (\ie the setup that developers use).

\begin{figure}[htbp]
\begin{center}
   \includegraphics[width=\linewidth]{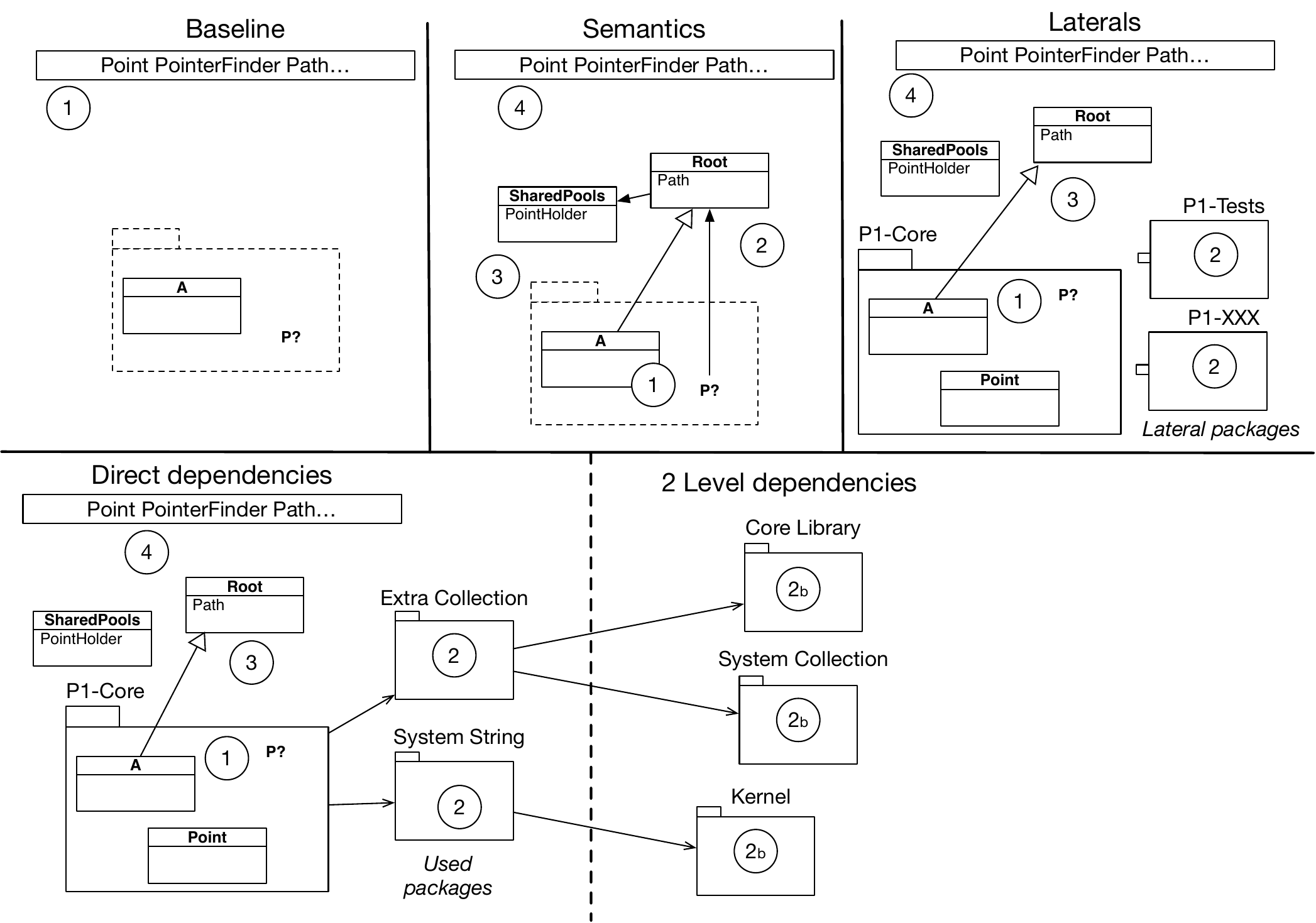}
\caption{\textbf{Different Heuristic Strategies: In the \baseline, all global variables are visible without a package context. In \semantics, the language scopes are followed. In \lateralpackage, variables within the same package are considered. In the \lateralpackage level, name resolution follows a structured order: current package (P1-Core), then related packages (\ct{P1-Test}, \ct{P1-Extension}), and finally \semantics. In \directdependencies provider, used packages are looked one level.}}
\label{fig:flatvspackag}
\end{center}
\end{figure}

 \paragraph{\textbf{\baseline Strategy}}
The \baseline strategy is a straightforward heuristic (see Figure~\ref{fig:flatvspackag} ). It establishes a fundamental point of reference for evaluating advanced completion strategies. It uses a global sorter, providing suggestions based solely on global namespace availability without considering semantic nuances, package boundaries, or dependency relationships (in other words, all names matching a prefix are sorted alphabetically).  While this approach yields broader suggestion coverage, it often returns irrelevant matches, resulting in low accuracy. Nonetheless, this baseline is crucial for quantifying and comparing the effectiveness of other heuristics. For \class, it is composed of a single fetcher that looks in the global namespace. When all the preceding fetchers fail, the baseline's fetcher is the one used. It is the default of the \semantics heuristic as well as of all the \completelevel strategy.

\paragraph{\textbf{\semantics Strategy}}
The \semantics strategy enhances code completion by taking into account the language scoping semantics. For example, global names are looked up in shared variables (accessible at the level of the class and the subinstances), shared pools (groups of shared variables accessible in unrelated class hierarchies). Similarly, method completion follows inheritance and supports the semantics of \ct{this}/\ct{super} pseudo variables. It supports possible methods and variable shadowing. Finally, it supports local variables, workspace-defined identifiers\footnote{A workspace is a kind of REPL where the developer can interactively define and execute code snippets.}, or debugger-specific information during live programming sessions. Using contextual and lexical analysis, this sequence of heuristics prioritizes semantically relevant identifiers, which improves suggestion precision compared to the baseline. 

This is the completion setup of Pharo 10, 11, and 12\footnote{\compl was first introduced in Pharo 10}. It is the second strategy depicted in Figure~\ref{fig:flatvspackag}: it shows that the completion package is ignored but that the name lookup follows the semantics of the language. For non-local variables, the completion is looked up in the class \circled{1}, its superclass \circled{2}, its shared variables/pool variables \circled{3}, defaulting to the global namespaces as a last resort \circled{4}.

 \paragraph{\textbf{\lateralpackage Strategy}} \label{sec:laterals}
The \lateralpackage strategy adds awareness of the package structure to the completion. It prioritizes identifiers defined in packages that are laterally related to the current package. In this paper, we operationalize a lateral, or ``friend'', package as a package that belongs to the same repository and shares the same project-level naming prefix as the current package. For example, packages such as \ct{Spec2-Core}, \ct{Spec2-Adapters}, and \ct{Spec2-Tests} are considered lateral packages because they belong to the same project family. This definition follows the package-aware intuition introduced in our previous work \cite{Abed25a}, but here we make the selection rule explicit and evaluate its impact separately from dependency-based strategies.

This choice has consequences for the results. Lateral packages may include useful sibling components, but they may also include tests, examples, or optional extensions that are not directly used by the current package. Therefore, \lateralpackage can improve recall but may also introduce noise compared to \directdependencies, which relies on explicit package usage.

In Figure~\ref{fig:flatvspackag}, the completion first looks in the package where the completion is requested \circled{1}, then in lateral packages within the same project family \circled{2}. If no suitable candidate is found, the strategy falls back to the semantic-based model: it searches the superclass hierarchy \circled{3a}, then shared variables and pool variables \circled{3b}, and finally the global namespace as a last resort \circled{4}.

\paragraph{\textbf{\directdependencies Strategy}}
Packages often require behavior from other packages. This creates a dependency relation between one package and the packages it uses. When coding within a package to define new functionalities, developers are likely to refer to entities of the packages they depend on. This is the behavior that is captured by the \directdependencies heuristics. It includes identifiers from both the current package and directly dependent ones. This method aligns closely with modular design principles and respects encapsulation boundaries, providing developers with precise, contextually relevant suggestions \cite{Mart00b}. Note that the heuristic implementation dynamically computes the package dependency graph based on the state of the program. It does not use static import statements. This approach fits better with the dynamic nature of the development, where several packages can be edited at the same time, even if they are not within the same project but are dependencies. For example, the Pillar project \cite{Arlo16a} depends on the Microdown language \cite{Duca20a}, and tooling may require some API changes, editing all packages at the same time. 

In Figure~\ref{fig:flatvspackag}: the completion looks in the package where the completion is requested \circled{1}, then the packages it depends on (here \ct{Extra Collection} and \ct{System String}) \circled{2}, then follows the semantic-based model \ie superclass \circled{3}, its shared variables/pool variables \circled{3}, defaulting to the global namespace as a last resort \circled{4}. 

 \paragraph{\textbf{\transitivedependencies Strategy}}
The \transitivedependencies (or transitive dependencies) strategy is an extension of the \directdependencies approach, which collects identifiers from the current package and all dependent packages recursively at two levels. While this expensive scope improves identifier discoverability, it also introduces higher complexity and an increased number of suggestions. From a developer's perspective, it should be noted that the transitive dependencies are not the same as the direct dependencies because the transitive ones do not contain \class that are directly used. Therefore, the likelihood that an identifier of the transitive dependency is requested is clearly lower than a direct one.

In Figure~\ref{fig:flatvspackag}: the completion looks in the package where the completion is requested \circled{1}, then the packages it depends on (here \ct{Extra Collection} and \ct{System String}) \circled{2}, then follows the second level dependent packages (\ct{Core Library}, \ct{System Collection}, and \ct{Kernel})\circled{2b} and then follows the semantic-based model \ie superclass \circled{3}, its shared variables/pool variables \circled{3}, defaulting to the global namespace as a last resort \circled{4}.

%%%%%%%%%%%%
%%%%%%%%%%%%
\section{Methodology}\label{methodology}

When a fetcher fails to provide a match, the lookup passes to its next fetcher. While such a logic is good for the completion engine's modularity and overall performance, it can hamper an adequate assessment of the performance of an individual fetcher or a given sequence. This is why our evaluation methodology is structured around different levels of characterization. 

\subsection{\textbf{Different characterization levels}} \label{charact}
To assess the real impact of a fetcher, we measure it individually from the rest of the completion strategy and in the context of a heuristic. This is why we introduced three levels of characterization for our evaluation. The first two levels measure fetchers while the third one evaluates a complete heuristic chain of fetchers.

\begin{figure}[htbp]
\begin{center}
   \includegraphics[width=0.5\linewidth]{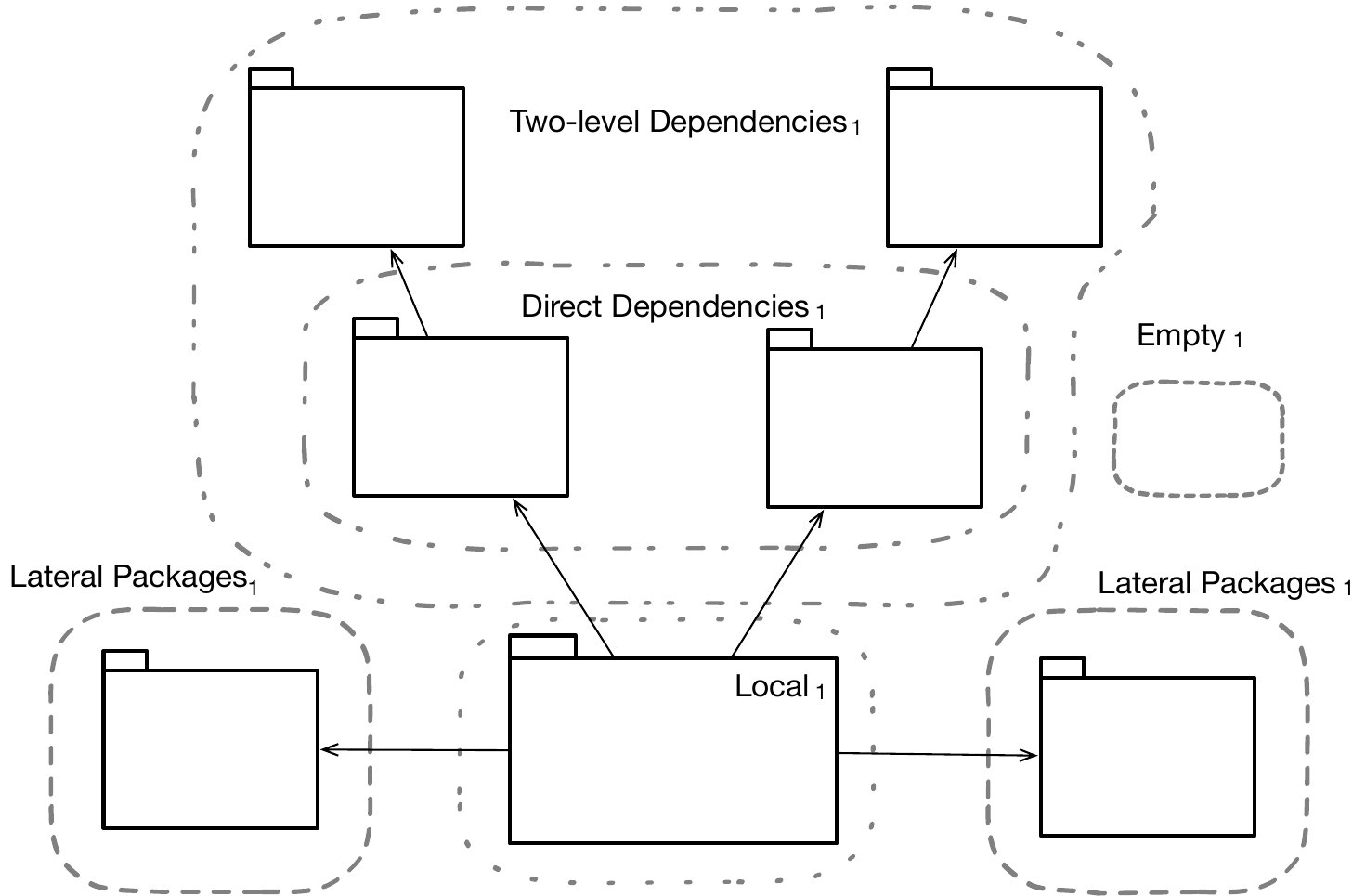}
\end{center}
\caption{\textbf{Elementary levels.} }
\label{fig:unit}
\end{figure}

\begin{itemize}
\item \textbf{Level 1: \unitlevel.} This level only looks in the identified packages. Figure~\ref{fig:unit} shows them: $Local_{1}$ looks exclusively in the package where the completion is requested. With $\lateralpackage_{1}$, the completion looks only in the packages within the same code repository (see Section~\ref{sec:laterals}) - local package is excluded. With $\directdependencies_{1}$, the completion  only looks in the packages that the current package depends on - note that the $Local_{1}$ is excluded. $\transitivedependencies_{1}$ is the same as the previous one, but repeated once. Finally, we also consider the ${Empty}$ unit: it never proposes any completion and serves as the ultimate baseline.

\item \textbf{Level 2: \locallevel.} This level is the combination of $Local$ and one of the preceding unit element - except $Empty$. We have $Local_{2}=Local_{1}$, $Laterals_{2}=Local_{1} + Laterals_{1}$, $\directdependencies_{2}=Local_{1} + \directdependencies_{1}$, and $\transitivedependencies_{2}=Local_{1} + \transitivedependencies_{1}$. 
 
\item \textbf{Level 3: \completelevel.} This level represents a full configuration of the completion engine. It includes the local level as well as the fetchers of the described strategies explained in Section~\ref{approach}  (See Figure~\ref{fig:flatvspackag}).  Note that when a fetcher is failing the next one is requested to propose candidates. 
\end{itemize}

\paragraph{\textbf{About the $Empty$ fetcher of the unit level}}
Having the $Empty$ unit is important. It returns no completion and represents the worst-case execution. Having it in our evaluation ensures that we are evaluating a fetcher in isolation and not the fallback to the following default behavior, as this is the case on the \completelevel. 

\paragraph{\textbf{About the \locallevel}}
The idea behind the \locallevel is driven by design considerations stating that packages should be highly cohesive \cite{Mart00b} \cite{Riel96a}. We want to understand whether locality is worth it in the different heuristics, but without the defaulting strategy that the \completelevel follows. Therefore, the evaluation of this level takes into account the local package in addition to each of the packages identified by the configuration.

\subsection{\textbf{Benchmark Logic}}
To evaluate the effectiveness of the strategies, we based our benchmark approach on the one introduced by Robbes \etal \cite{Robb08a, Robb10c}. Although their original benchmark relied on a change-based repository of program history, our approach adapts the core idea to a static analysis, taking both \class and \method benchmarking, making it applicable in contexts where historical data is unavailable. 

The essence of the benchmark is to test whether a completion engine correctly suggests the \emph{original} names of \class/\method when only partial prefixes are provided. The fundamental insight is that by systematically rewriting every \class access site with increasingly longer prefixes (ranging from 2 to 8 characters), we can invoke the completion engine and evaluate whether it ranks the correct name high in its suggestions. This method simulates realistic completion scenarios and allows us to measure not just correctness but also ranking performance and user effort reduction. Our benchmarking approach, implemented via the \ct{StaticBenchmarks} classes, carries out the following steps for each method in a given package:

\begin{enumerate}
\item \textbf{AST Extraction}: The method's AST is computed.
\item \textbf{For Class Filtering}: For each name we check whether it is a class.
\item \textbf{For Method filtering}: Each call site (method invocation) is identified. 
\item \textbf{Name Masking}: The name (of class or method) is programmatically shortened to generate several prefixes of increasing lengths, from 2 up to 8 characters (or the full name length, if shorter). For example, the class \ct{OrderedCollection} yields prefixes such as \ct{Or}, \ct{Ord}, \ct{Orde}, etc. The selector \ct{at:put:} yields prefixes such as \ct{at}, \ct{at:}, \ct{at:p}, etc.
\item \textbf{Completion Invocation}: For each prefix (and then position in the source code), the completion engine is triggered as if the user was requesting suggestions after typing that fragment.
\item \textbf{Result Logging}: The engine's output is analyzed to determine if the original name appears among the top 10 suggestions. If found, we record the rank at which it was suggested. If not, it is considered a failure for that prefix length. We limit the number of suggestions to 10    because the completion objective is to propose the correct candidate within the top choices \cite{Biba22a,Robb08a}. 
\end{enumerate}

This process is repeated for all the classes in all the methods of the targeted package. 
We provide a modular API allowing execution across individual classes or entire packages.

\subsection{\textbf{Metrics}}
Our benchmark focuses on prefixes up to length eight: In practice, developers rarely type more than five to eight characters before expecting a completion suggestion. Typing beyond this point provides little added value, as the prefix is already highly specific and the potential completion set has effectively collapsed. Prior research on word prediction and typing behavior (\eg \cite{Shain24a,Wobb06a}) shows that the informational gain from additional characters decreases rapidly, while the cognitive effort of continued typing increases. For prefixes longer than eight characters, the benchmarks no longer return any suggestions. In these cases, the benchmark result is always zero.

We collect the following metrics:

\begin{enumerate}
	\item \textbf{Mean Reciprocal Rank (MRR)}: Measures the average reciprocal rank position of the first correct prediction. Formally, given a set of queries $Q$, the MRR is calculated as:

$$
\text{MRR} = \frac{1}{|Q|} \sum_{i=1}^{|Q|} \frac{1}{\text{rank}_i}
$$

where $\text{rank}_i$ is the rank position of the first correct prediction for the $i$-th query. MRR emphasizes the importance of placing the first correct result as high as possible and is particularly suitable when only one relevant result is sufficient per query \cite{Mitr18a}.

\item \textbf{Timing Metrics}: Includes total and average completion times and memory usage, evaluated per prefix length.

\end{enumerate}

\subsection{\textbf{Principal Component Analysis}}

To complement the quantitative metrics presented above, we applied Principal Component Analysis (PCA) \cite{Joll02a} to the multidimensional benchmark data to explore dominant sources of variance and structural relationships among the evaluated completion heuristics. PCA was used here as an unsupervised exploratory technique to identify clusters of heuristics that behave similarly across evaluation metrics and projects, following common practice in software analytics and recommender system evaluation \cite{Mann99a}. For each project and level (\unitlevel, \locallevel, and \completelevel), we constructed a matrix where rows correspond to completion strategy (\baseline, \semantics, \lateralpackage, \directdependencies, and \transitivedependencies) and columns correspond to prefix-length specific performance indicators (MRR and the numerical columns 2–8 extracted from the benchmark tables). All metrics were standardized using z-score normalization (via \ct{StandardScaler}) to ensure comparability across projects with differing absolute performance scales. PCA was then applied using the implementation from \ct{scikit-learn} \footnote{Python library for machine learning tasks.}, operating on the standardized metric matrix. The first two principal components (PC1 and PC2) were retained, capturing the majority of the total variance across heuristics. We present the two-dimensional PCA projections as scatterplots, where each point corresponds to a heuristic.

\subsection{\textbf{Project Selection}}

Based on the statistical analysis of Pharo code by Zaitsev \etal \cite{Zait20a} we selected projects that ensure a broad domain diversity, including web development, visualization, software analysis, and user-interface frameworks, while also focusing on projects that demonstrated active maintenance, extensive test suites, and representative structural attributes, such as method length distributions and language feature usage patterns. We selected Pharo projects reflecting significant diversity in domain, size, and development activity. The chosen projects span essential areas such as visualization (Roassal), software analysis (Moose), web application development (Seaside), user interface framework (Spec), and version control systems (Iceberg). 

\begin{table}[htbp]
\centering
\footnotesize
\begin{tabular}{llllllll}
\hline
\textbf{Framework/Bib}&\#\textbf{Packages}&\#\textbf{Classes}&\#\textbf{Defined Classes}&\# \textbf{Methods}&$\rho_{\text{int}}$&$R_{\text{int}}$&$R_{\text{ext}}$ \\
\hline
Iceberg & 15 & 641 & 553 & 5173 & 0.41 & 245 & 333 \\ \hline
Iceberg-tests& 5 & 130 & 129 & 936 & 0.05 & 18 & 155 \\ \hline
Moose & 31 & 632 & 497 & 5450 & 0.33 & 195 & 582 \\ \hline
Moose-tests & 41 & 283 & 258 & 2115 & 0.13 & 28 & 458 \\ \hline
Roassal & 39 & 691 & 527 & 6413 & 0.20 & 158 & 987 \\ \hline
Roassal-tests & 13 & 104 & 102 & 820 & 0.00 & 0 & 315 \\ \hline
Seaside & 30 & 653 & 537 & 5082 & 0.25 & 274 & 285 \\ \hline
Seaside-tests & 12 & 219 & 211 & 1487 & 0.16 & 31 & 262 \\ \hline
Spec & 22 & 858 & 703 & 6892 & 0.30 & 258 & 306 \\ \hline
Spec-tests & 11 & 324 & 312 & 1604 & 0.10 & 57 & 287 \\ \hline
\textbf{TOTAL} & \textbf{219} & \textbf{4535} & \textbf{3829} & \textbf{35972} & - & - & - \\
\hline
\end{tabular}
\caption{\textbf{Overview of Selected Pharo Projects for Benchmark}}
\label{bench}
\end{table}

Table~\ref{bench} provides a comprehensive overview of the selected Pharo projects used in our benchmarking evaluation. It includes the total number of packages, classes, and defined classes (in Pharo, a class can be extended in another package). By definition, we make a distinction between the classes defined in a package vs. the classes that are defined in another package but extended by the package, and the methods analyzed in each project. Additionally, to help understanding the behavior of the completion, the table presents key metrics such as the number of internal references ($R_{\text{int}}$), the number of external references ($R_{\text{ext}}$), and the ratio of internal references ($\rho_{\text{int}}$). A higher $\rho_{\text{int}}$ indicates stronger intra-package cohesion, reflecting more frequent references to internal entities rather than external ones. For example, Iceberg shows a relatively high $\rho_{\text{int}}$ (0.41), suggesting significant internal cohesion, whereas Roassal-Tests exhibits the lowest $\rho_{\text{int}}$ (0), implying a greater reliance on external references, as expected for a test package.This is the case for most test packages. Typically, they reference things from other packages (that are being tested). Test packages show a lower ratio because test classes rarely refer to other test classes. These metrics provide context for understanding how package structure and usage patterns may influence the effectiveness of our package-aware heuristic. In the following, we provide brief descriptions of the selected projects to better situate their roles within the benchmark.

\paragraph{\textbf{Spec}}
Spec is Pharo's user-interface framework. It is organized around presenters, widgets, and adapters, making it representative of modular UI-oriented systems \cite{DeHon24a}.

\paragraph{\textbf{Roassal}}
Roassal is a visualization engine for Pharo. It provides graphical primitives and layouts for building interactive visualizations, and depends on several external packages \cite{Berg16c}.

\paragraph{\textbf{Seaside}}
Seaside is a component-based web application framework for Smalltalk and Pharo. Its architecture combines reusable components, sessions, and rendering packages \cite{Duca07a, Duca10a}.

\paragraph{\textbf{Iceberg}}
Iceberg is Pharo's Git-based version control tool. It is composed of packages for repository management, UI integration, and Git backend interaction \cite{Poli20y}.

\paragraph{\textbf{Moose}}
Moose is a platform for software and data analysis in Pharo. It combines meta-modeling, importing, querying, and visualization facilities across several packages \cite{Anqu20a}.

%%%%%%%%%%%%
%%%%%%%%%%%%
\vspace{-0.2cm}
\section{Class Name Completion Evaluation}\label{classevaluation}	

\subsection{\textbf{Overall Three-Level Results}}
Across all five projects, similar patterns emerge. For the sake of space, we analyze in detail the results obtained for the Roassal project, then we discuss the \completelevel for all the other projects. In addition, once the configuration behavior in isolation is understood, the \completelevel is more interesting since it represents a programmer's perspective and real use case.

We selected Roassal because, unlike Spec or Seaside, it is not a pure framework but a library that depends on several external frameworks, notably Spec. This makes Roassal a representative case of a typical project that combines internal structure with external dependencies.

\subsubsection{\textbf{Roassal}}
\begin{figure}[htbp]
    \centering
    \begin{subfigure}[b]{0.45\textwidth}
        \includegraphics[width=\linewidth]{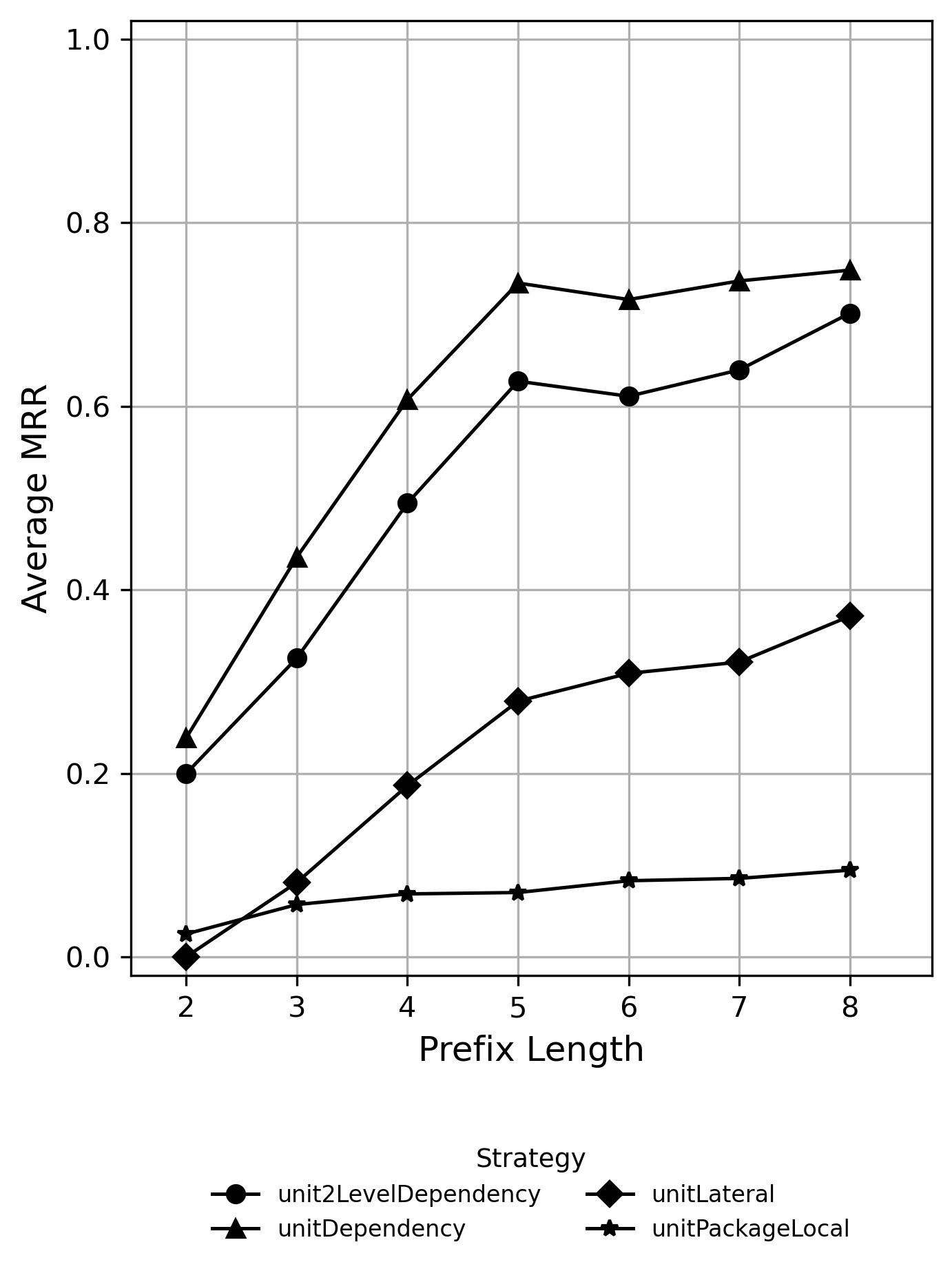}
        \caption{\textbf{\unitlevel for \class in Roassal.}}
        \label{fig:classesroassalUnit}
    \end{subfigure}
	\hfill
	   \begin{subfigure}[b]{0.45\textwidth}
        \includegraphics[width=\linewidth]{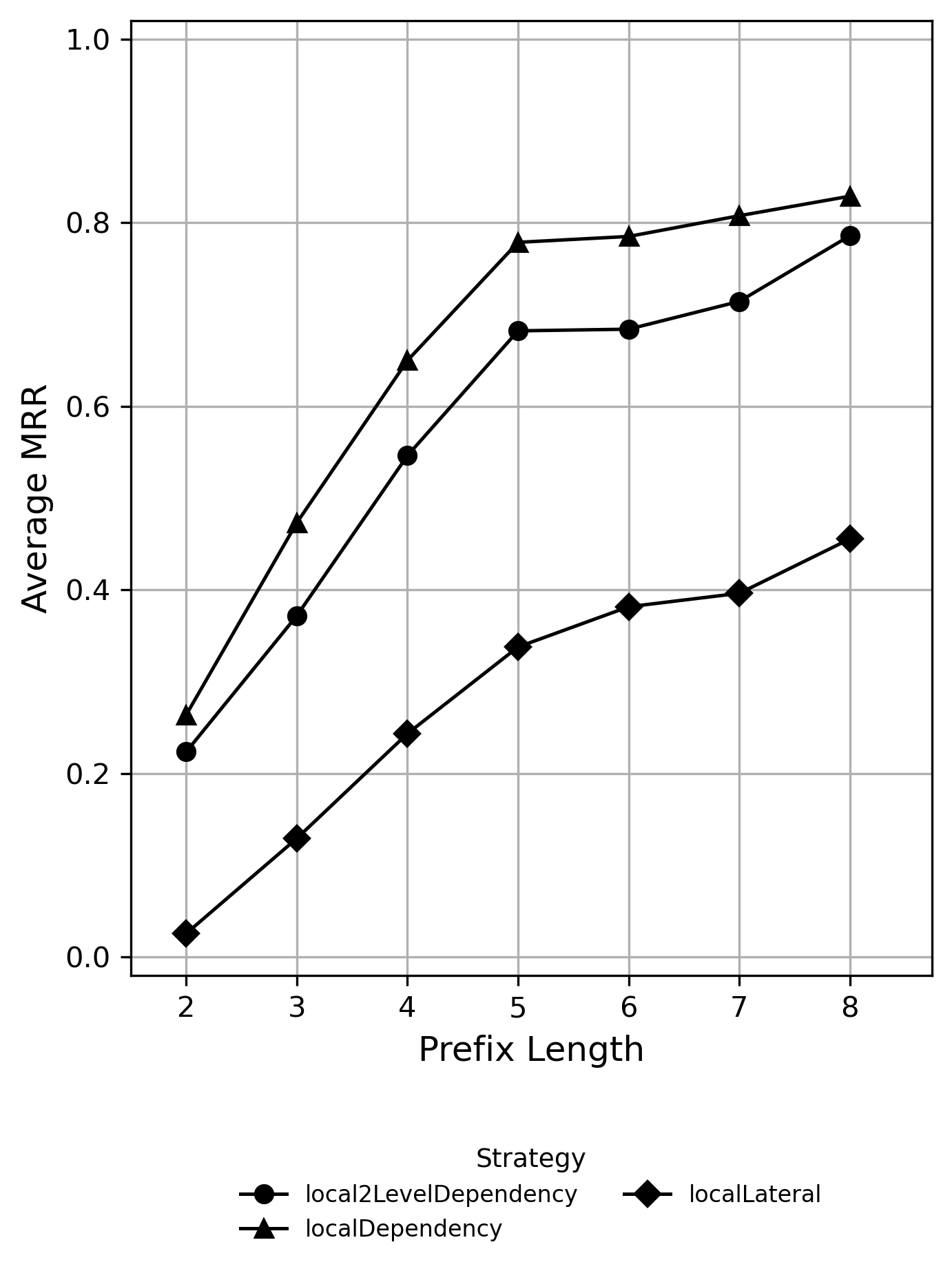}
        \caption{\textbf{\locallevel for \class in Roassal. }}
        \label{fig:classesroassalLocal}
    \end{subfigure}
    \hfill
    \caption{\textbf{\unitlevel VS \locallevel for \class in Roassal.}}
 \end{figure}

\paragraph{\textbf{\unitlevel Analysis}}
Figure~\ref{fig:classesroassalUnit} shows that the \directdependencies and \transitivedependencies configurations achieve the best results across most prefix lengths, particularly in the 2–5 character range, where their performance peaks ($\sim$0.72) before starting to stabilize. This indicates that dependency-aware heuristics are highly effective even when limited contextual information is available. Given that most Roassal \class begin with the prefix \ct{RS} (described later), this performance demonstrates that dependency-based filtering successfully reduces noise within the large candidate space. The \lateralpackage configuration shows a gradual but steady improvement across prefix lengths, reaching an average performance of approximately 0.36 at eight prefixes, a notably long and demanding input length. This suggests that lateral relationships, connections between sibling units within the same structural level, do contribute to completion quality, although less effectively than the \directdependencies or \transitivedependencies.

\paragraph{\textbf{\locallevel Analysis}}
At the \locallevel, the completion starts to look in the local package before following packages identified by each of the heuristics. This level accentuates the impact of package-level cohesion and reveals how often developers reference classes within the same package. Figure~\ref{fig:classesroassalLocal} demonstrates that the heuristics rank similarly to those at the \unitlevel.  This confirms that explicit dependency information remains the strongest predictor of completion relevance, even when local context is expanded.

\begin{figure}[!h]
    \centering
    \begin{subfigure}[b]{0.45\textwidth}
        \includegraphics[width=\linewidth]{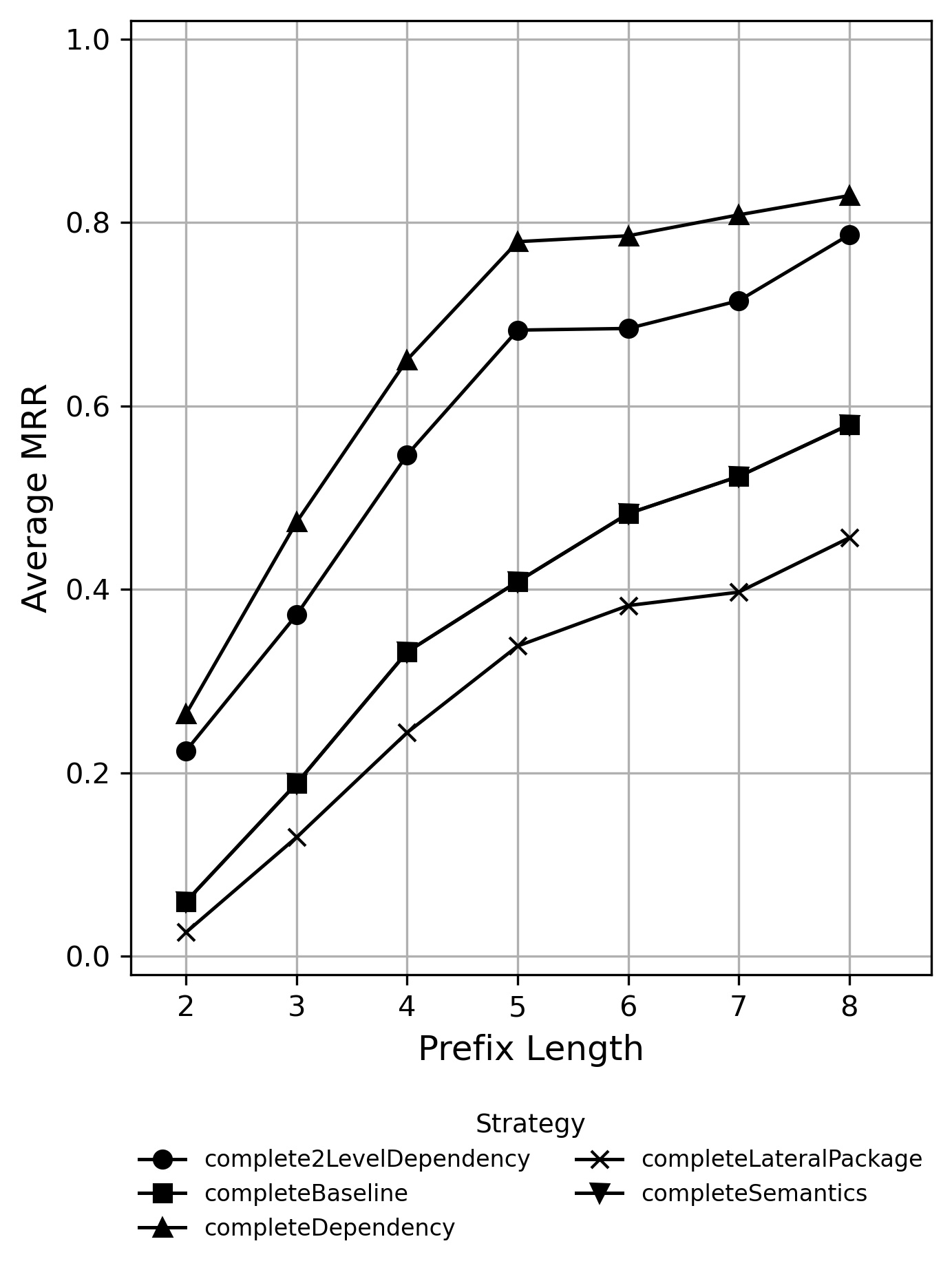}
        \caption{\textbf{\completelevel for \class in Roassal.}}
        \label{fig:classesroassalComplete}
    \end{subfigure}
	\hfill
	  \begin{subfigure}[b]{0.45\textwidth}
        \includegraphics[width=\linewidth]{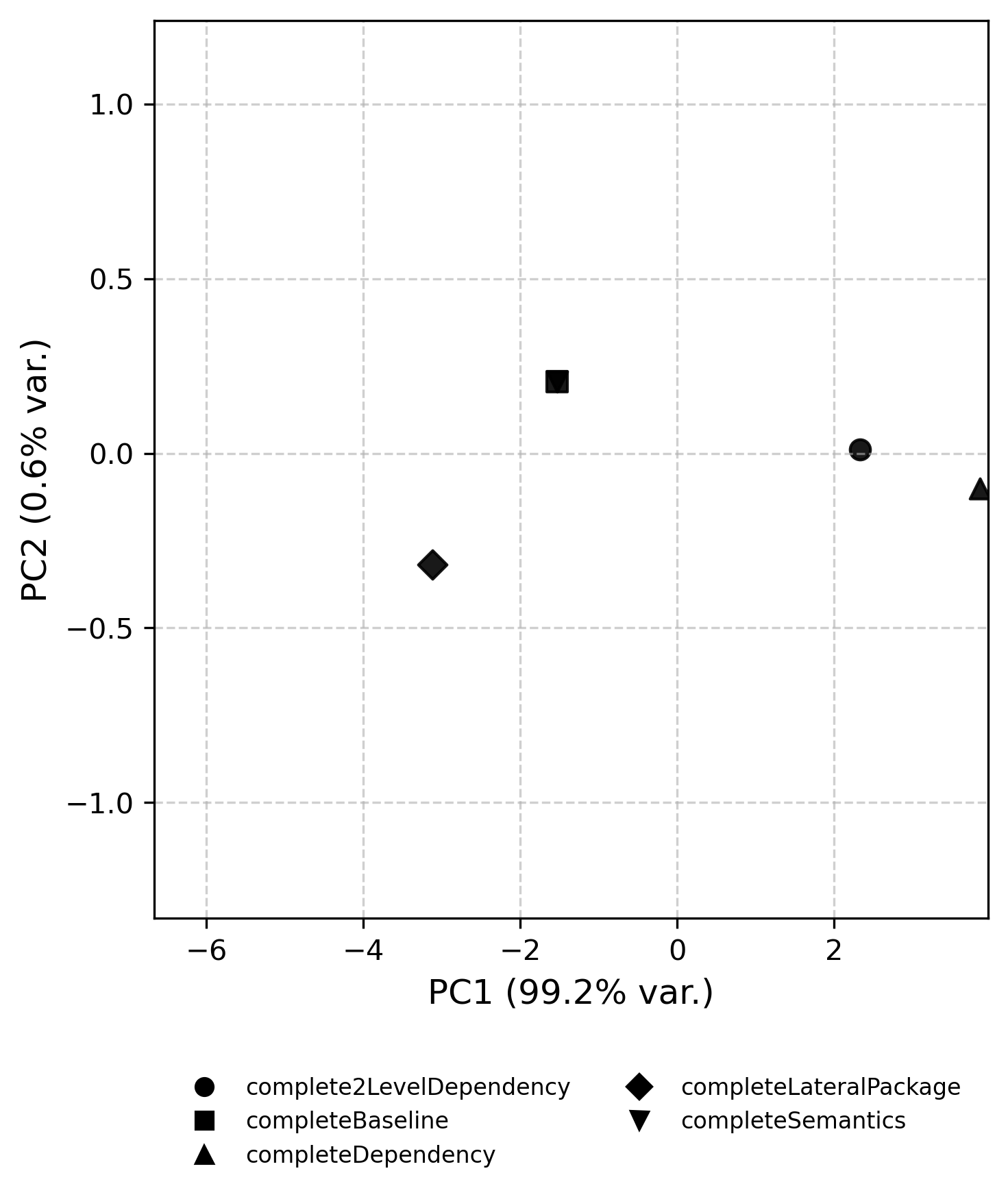}
        \caption{\textbf{PCA of Roassal \class at the \completelevel}.}
        \label{fig:classespcaroassalcomplete}
    \end{subfigure}
	      \hfill
    \caption{\textbf{\completelevel VS PCA for \class in Roassal}}
 \end{figure}

\paragraph{\textbf{\completelevel Analysis}}
The \completelevel corresponds to the configuration closest to real-world usage and is therefore the most representative scenario for practical completion. In this setup, the engine leverages all available heuristics in an ordered manner. Its goal is to reflect how developers typically develop in multi-package systems. As shown in Figure~\ref{fig:classesroassalComplete}, the same pattern persists, \directdependencies yields the highest MRR, closely followed by \transitivedependencies. This confirms the robustness of dependency-aware approaches in realistic settings. The difference between \directdependencies and \transitivedependencies is probably due to the way dependent packages are walked through and the noise induced by the specific code context. 

A similar ranking persists for \lateralpackage and \semantics strategies. The \lateralpackage heuristic benefits from the global scope, reaching nearly 0.5 at higher prefix lengths,demonstrating that lateral relations between packages can sometimes hint at complementary or co-used components. However, when a package refers to a class defined in its lateral packages, such packages have a high chance to be in the list of its dependent packages too. 

\paragraph{About \semantics not displayed in \completelevel charts}
Note that in the figures, the \semantics strategy is not visible because it has the same results as the \baseline: its plot is hidden by that of the \baseline. This is due to the fact that for \class, the \semantics model takes into account rare language features such as pool and shared variables that are rarely or not at all used in the selected projects. Therefore, it is nearly equivalent to the \baseline.

\paragraph{\textbf{Roassal PCA Analysis}}
Figure~\ref{fig:classespcaroassalcomplete} summarizes the Complete Level results for class-name completion in Roassal. The \directdependencies and \transitivedependencies heuristics are positioned close to each other, while the \semantics and \baseline strategies overlap in the PCA projection, appearing on top of each other. The \lateralpackage heuristic is separated from these groups in the PCA projection, indicating a distinct performance profile. Because the directions of the PCA axes do not themselves represent better or worse performance, we interpret only the relative distances among the strategies. We therefore use the PCA plot as a visual summary of the MRR trends rather than as independent evidence of performance superiority.

\begin{figure}[H]
    \centering
    \begin{subfigure}[b]{0.45\textwidth}
        \includegraphics[width=\linewidth]{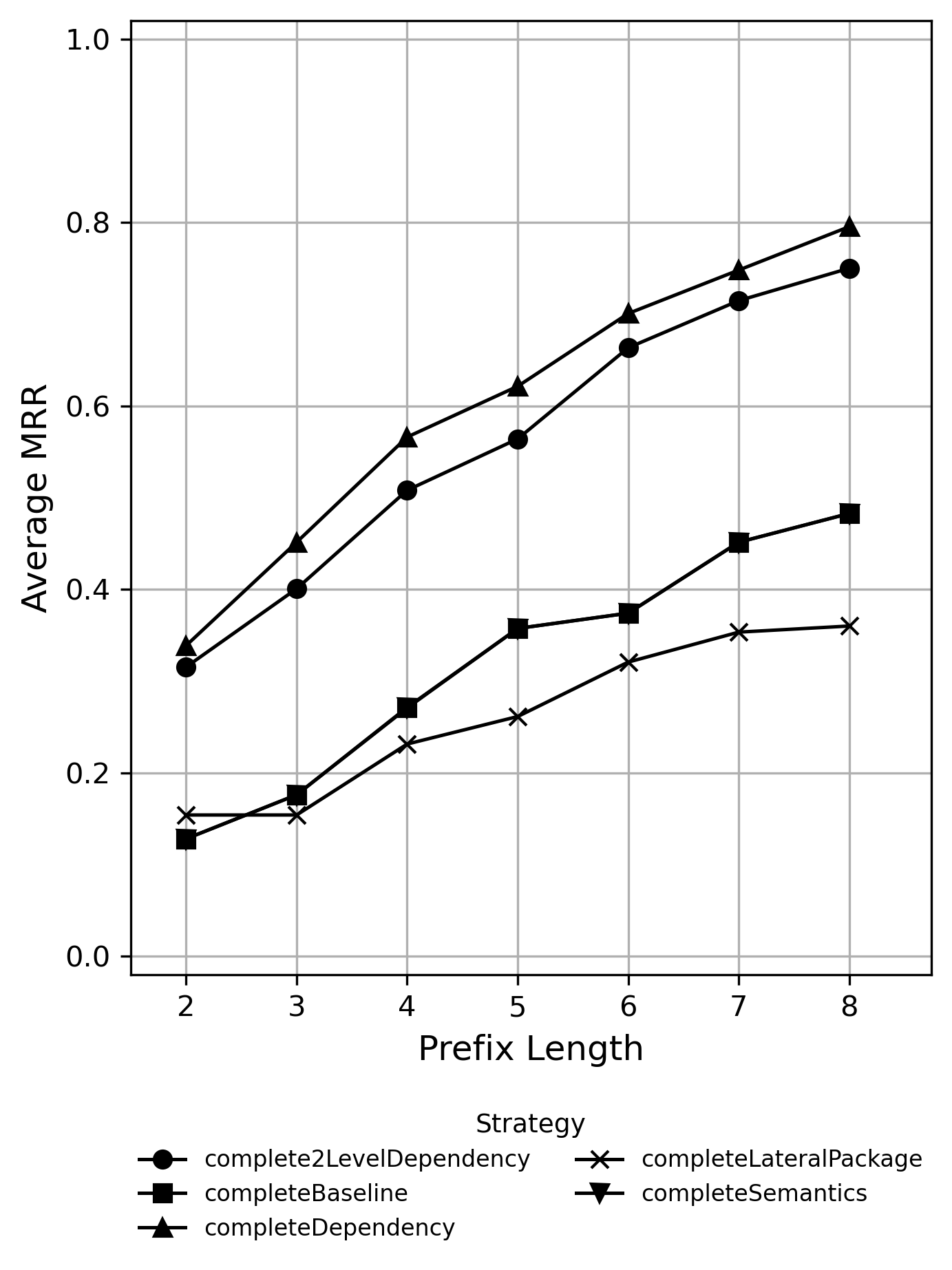}
        \caption{\textbf{\completelevel for \class in Iceberg.}}
        \label{fig:classesicebergComplete}
    \end{subfigure}
    \hfill
    \begin{subfigure}[b]{0.45\textwidth}
        \includegraphics[width=\linewidth]{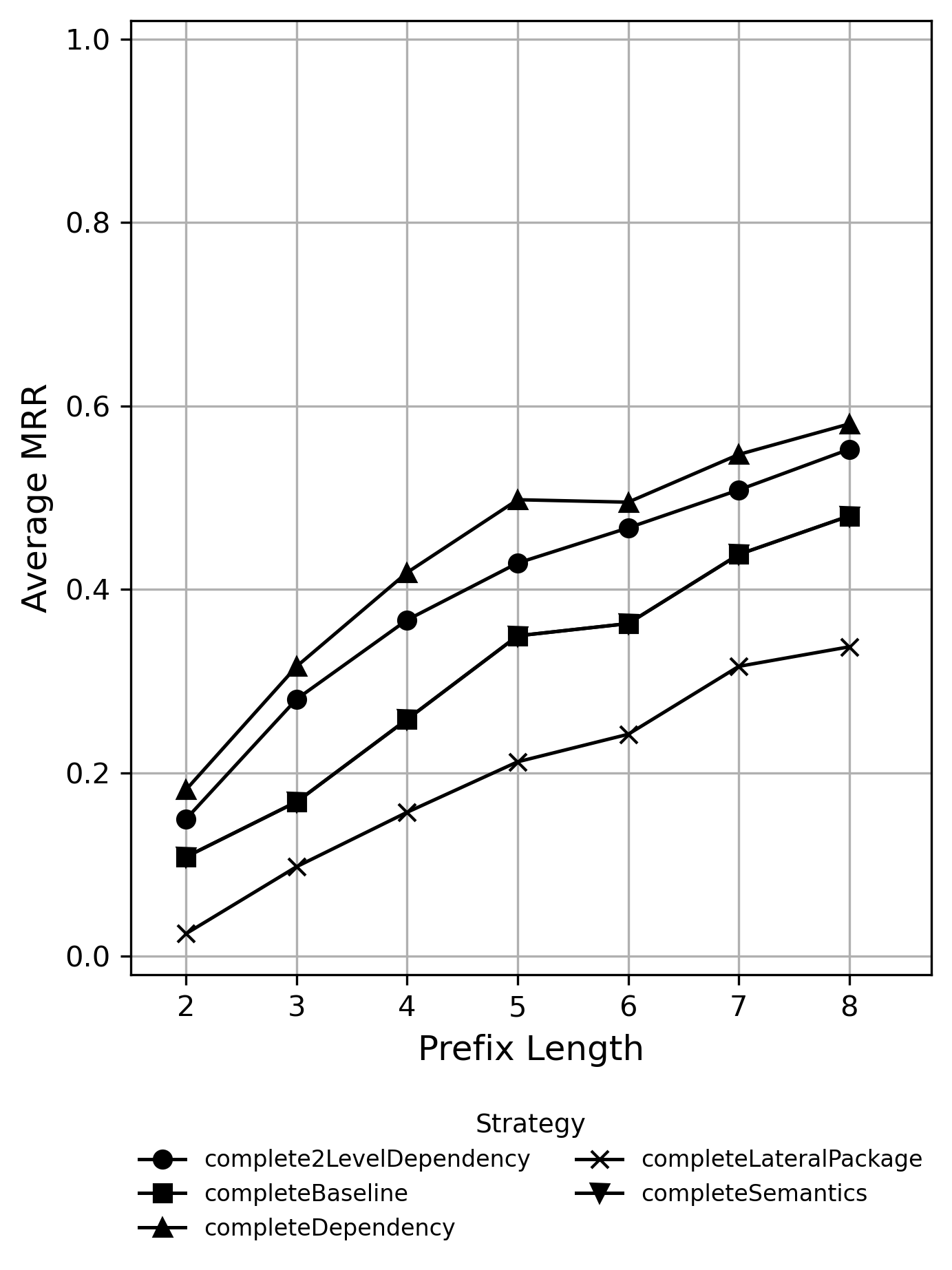}
        \caption{\textbf{\completelevel for \class in Seaside.}}
        \label{fig:classesseasideComplete}
    \end{subfigure}
    \caption{\textbf{\completelevel for \class in Iceberg and Seaside.}}
\end{figure}

\subsubsection{\textbf{Iceberg}}

In Figure~\ref{fig:classesicebergComplete}, both \directdependencies and \transitivedependencies outperform all others, reaching over 0.75 in average performance by Top-8, demonstrating that leveraging structural dependencies between tightly coupled packages (\eg \ct{Iceberg-Core}, \ct{Iceberg-Git}, and \ct{Iceberg-UI}) yields the most accurate completions. In contrast, \baseline and \semantics show limited gains, indicating that lexical or semantic similarity alone cannot capture Iceberg's internal relationships. The \lateralpackage achieves only moderate improvements, suggesting that information flow in Iceberg is primarily unidirectional and hierarchical. Overall, the results highlight that taking into account package dependency is the dominant factor driving completion accuracy in this project.

\subsubsection{\textbf{Seaside}}
In Seaside, dependency-driven heuristics again lead, but the relative advantage between \directdependencies and \transitivedependencies narrows compared to Iceberg. Seaside's architecture, built around reusable web components, exhibits a web of both direct and indirect dependencies, meaning that classes in test, session, and component packages reference each other extensively.  \lateralpackage remains weak, confirming that package naming conventions lookups are not predictive of completion relevance in web-framework architectures.

\subsubsection{\textbf{Spec}}
The Spec project demonstrates similar trends to Iceberg, but with subtler MRR differences. \directdependencies achieves the best results overall, driven by the high cohesion of the core \ct{Spec2-Core}, \ct{Spec2-Adapters}, and \ct{Spec2-Widgets} packages. \transitivedependencies produces marginally lower results, likely due to the framework's shallow dependency hierarchy. The global trend remains: the closer the heuristic adheres to explicit dependency relations, the higher its predictive precision.

\begin{figure}[H]
    \centering
    \begin{subfigure}[b]{0.45\textwidth}
        \includegraphics[width=\linewidth]{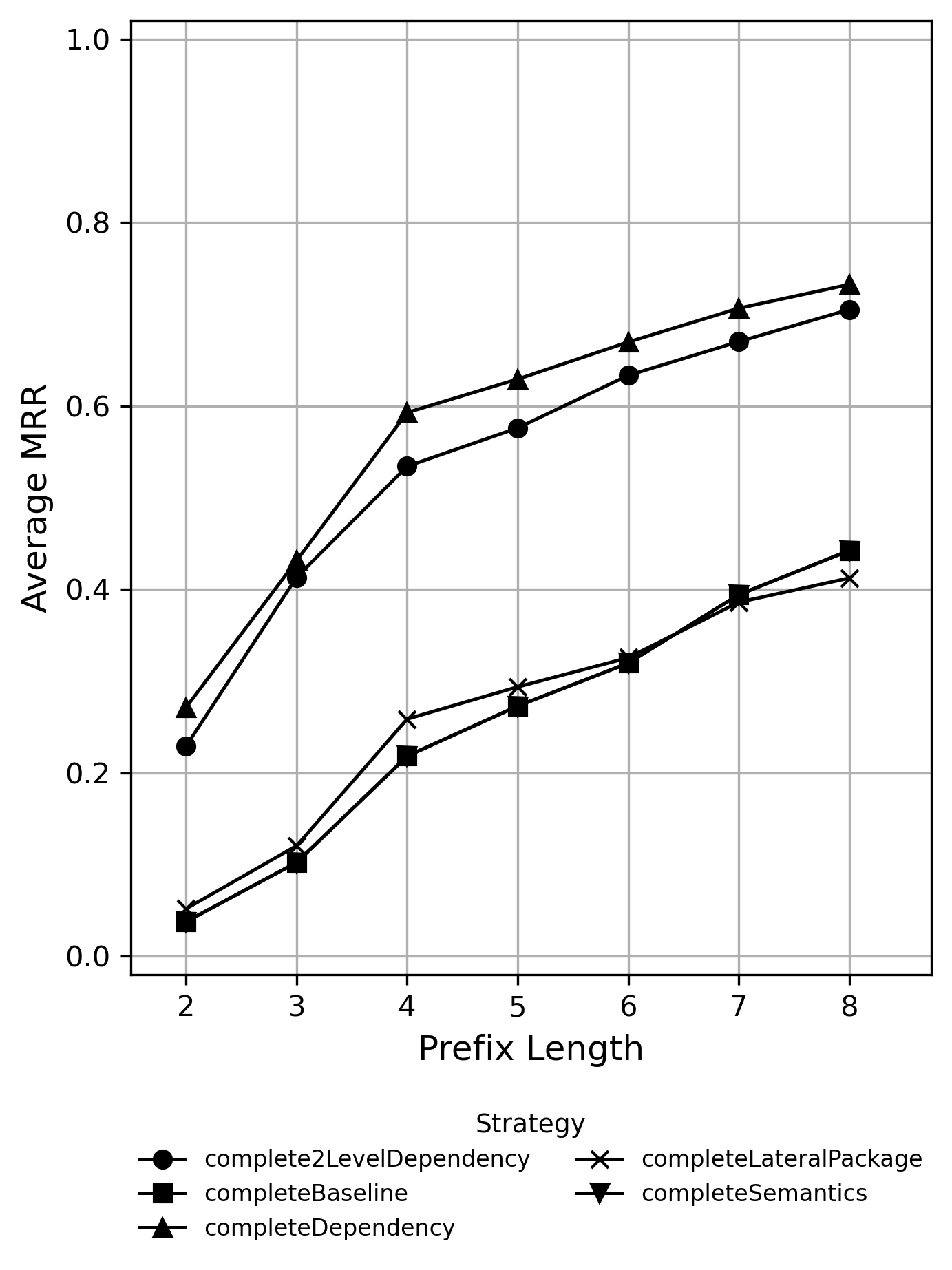}
        \caption{\textbf{\completelevel for \class in Spec.}}
        \label{fig:classesspecComplete}
    \end{subfigure}
    \hfill
    \begin{subfigure}[b]{0.45\textwidth}
        \includegraphics[width=\linewidth]{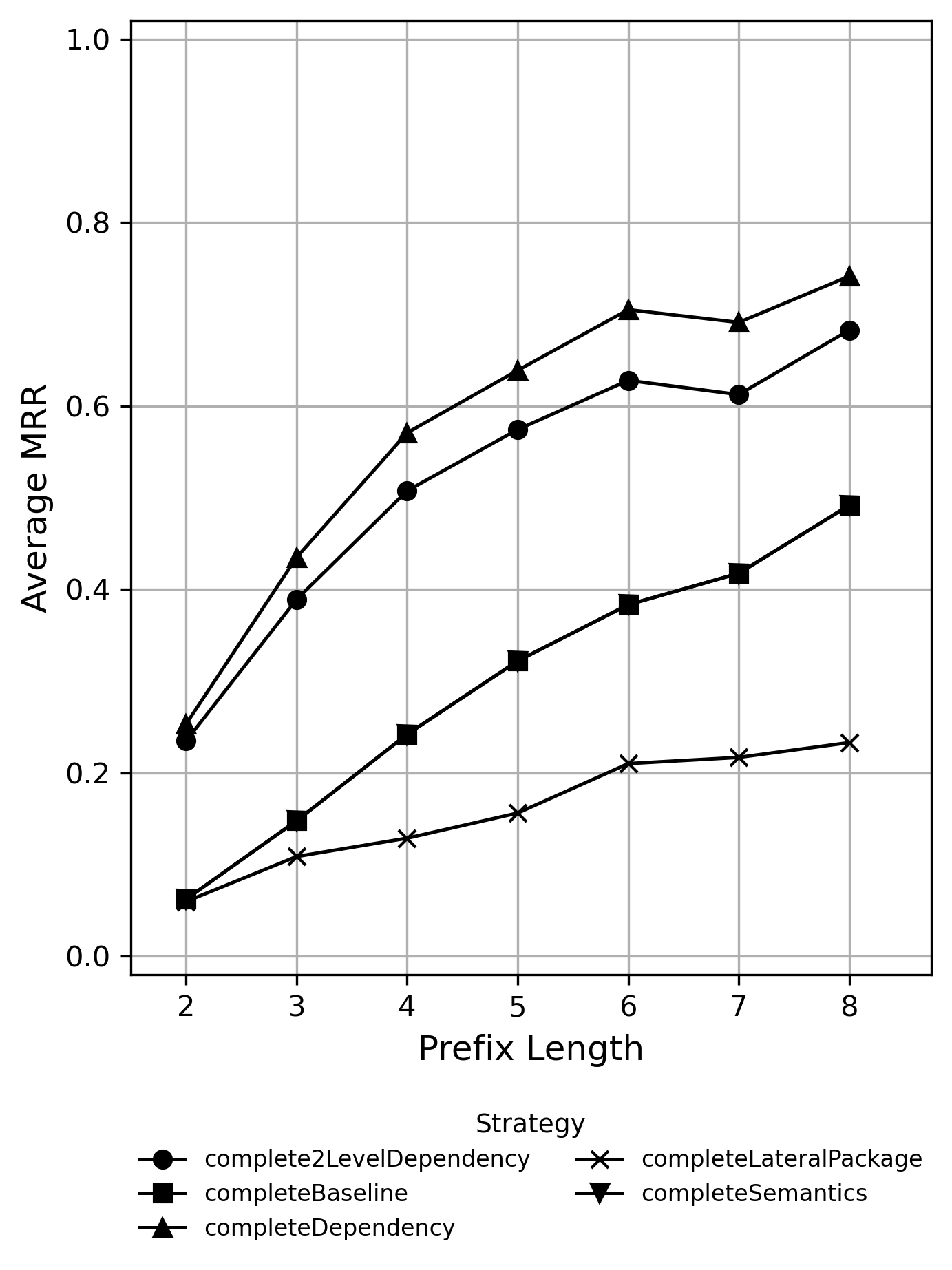}
        \caption{\textbf{\completelevel for \class in Moose.}}
        \label{fig:classesmooseComplete}
    \end{subfigure}
    \caption{\textbf{\completelevel for \class in Spec and Moose.}}
\end{figure}

\subsubsection{\textbf{Moose}}
The results in Figure \ref{fig:classesspecComplete} and \ref{fig:classesmooseComplete} illustrate that, although dependency-based strategies remain the strongest performers, their advantage in Moose is slightly less pronounced than in more cohesive systems such as Iceberg and Spec. Both \directdependencies and \transitivedependencies steadily outperform all other strategies, reaching around 0.7 in average performance at Top-8, but the overall improvement over the \baseline is smaller.  This reflects Moose's highly modular architecture, where functionality is distributed across cross-cutting frameworks (\eg Fame, Roassal) that dilute direct dependency signals. 
The \baseline achieves moderate gains. In contrast, \lateralpackage again lags behind, showing that lateral relationships between packages offer little predictive value.

\begin{boxB}
\textbf{Answer to RQ1: Impact of Package Awareness}

\emph{Can leveraging package structure and dependency information improve the relevance and accuracy of code completion in Pharo compared to the default semantics-based heuristic?}

The results demonstrate that leveraging package structure and dependency information significantly improves the relevance and accuracy of code completion compared to the default semantics-based heuristic. Across all evaluated projects and prefix lengths, dependency-aware strategies consistently outperform both the \baseline and \semantics configurations in terms of Mean Reciprocal Rank (MRR) and Top-K accuracy.

In particular, the semantics-based heuristic behaves similarly to the baseline for \class completion, as global namespace lookup dominates once local scopes are exhausted. Package-aware heuristics, by contrast, effectively constrain the candidate space to structurally relevant entities, resulting in earlier and more accurate rankings.
\end{boxB}

\begin{boxB}
\textbf{Answer to RQ2: Comparison of Dependency Heuristics}

\emph{ Which dependency heuristic (\lateralpackage, \directdependencies, \transitivedependencies ) achieves the best performance in terms of completion accuracy and ranking quality (\eg Mean Reciprocal Rank)?}

Among the evaluated dependency-aware strategies, the \directdependencies heuristic achieves the best overall performance. It consistently delivers the highest MRR across projects, particularly for short prefixes where completion ambiguity is highest.

While \transitivedependencies occasionally improves recall by expanding the search scope, it often introduces additional noise, slightly degrading ranking precision in cohesive systems. The \lateralpackage heuristic shows weaker and less stable performance, indicating that package proximity alone is not a reliable predictor of developer intent.
\end{boxB}

\subsection{\textbf{Discussion}} \label{classdiscussion}

The \class completion results across all projects exhibit consistent and interpretable trends that align with both the Principal Components Analysis (PCA) and Mean Reciprocal Rank (MRR) analysis. For space reasons, the main text discusses the Roassal PCA plot in detail, while Appendix~\ref{appendix} presents the corresponding \completelevel PCA plots for all evaluated projects. These results suggest that dependency-based heuristics are reliable predictors of completion accuracy in the Pharo projects that we evaluated. The PCA groups the \directdependencies and \transitivedependencies configurations apart from all others, highlighting their shared behavioral coherence. In contrast, \lateralpackage, and \baseline configurations form more dispersed or orthogonal groups. 

\paragraph{\textbf{MRR}}
Across all systems, \directdependencies achieves the best trade-off between precision and stability, maintaining high performance from the earliest prefixes (2–5 characters) where completion is most critical \cite{Bruc09a, Robb08a}. The MRR curves confirm this advantage; in each project, the \directdependencies and \transitivedependencies heuristics consistently dominate the Top-K rankings, especially in cohesive frameworks such as Iceberg and Spec. 

\begin{figure}[H]
    \centering
       \begin{subfigure}[b]{0.49\textwidth}
        \includegraphics[width=1\textwidth]{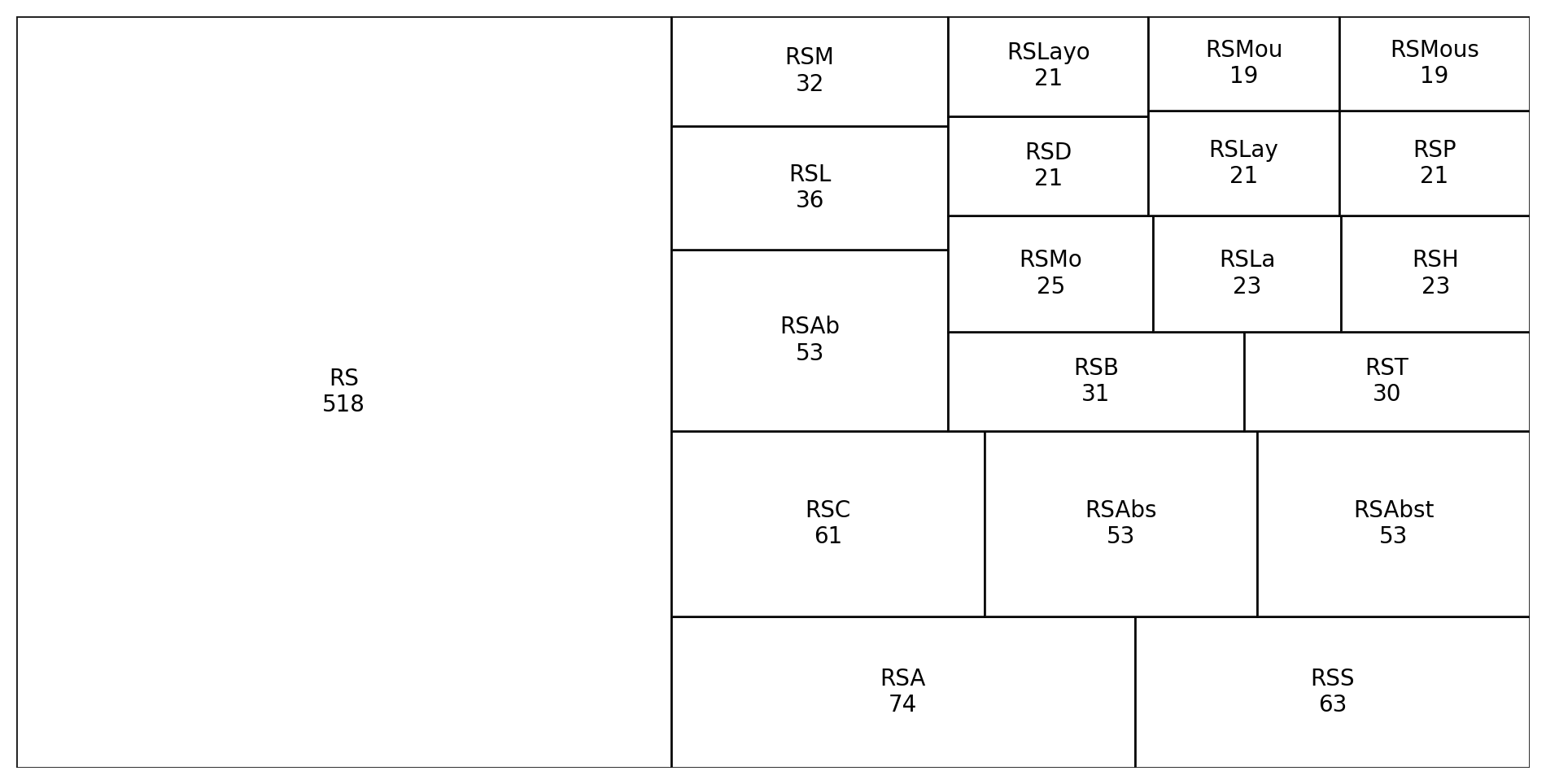}
        \caption{\textbf{Tree Maps of the \class in Roassal.}}
        \label{fig:classesprefixroassal}
    \end{subfigure}
    	\hfill
	   \begin{subfigure}[b]{0.49\textwidth}
        \includegraphics[width=1\textwidth,]{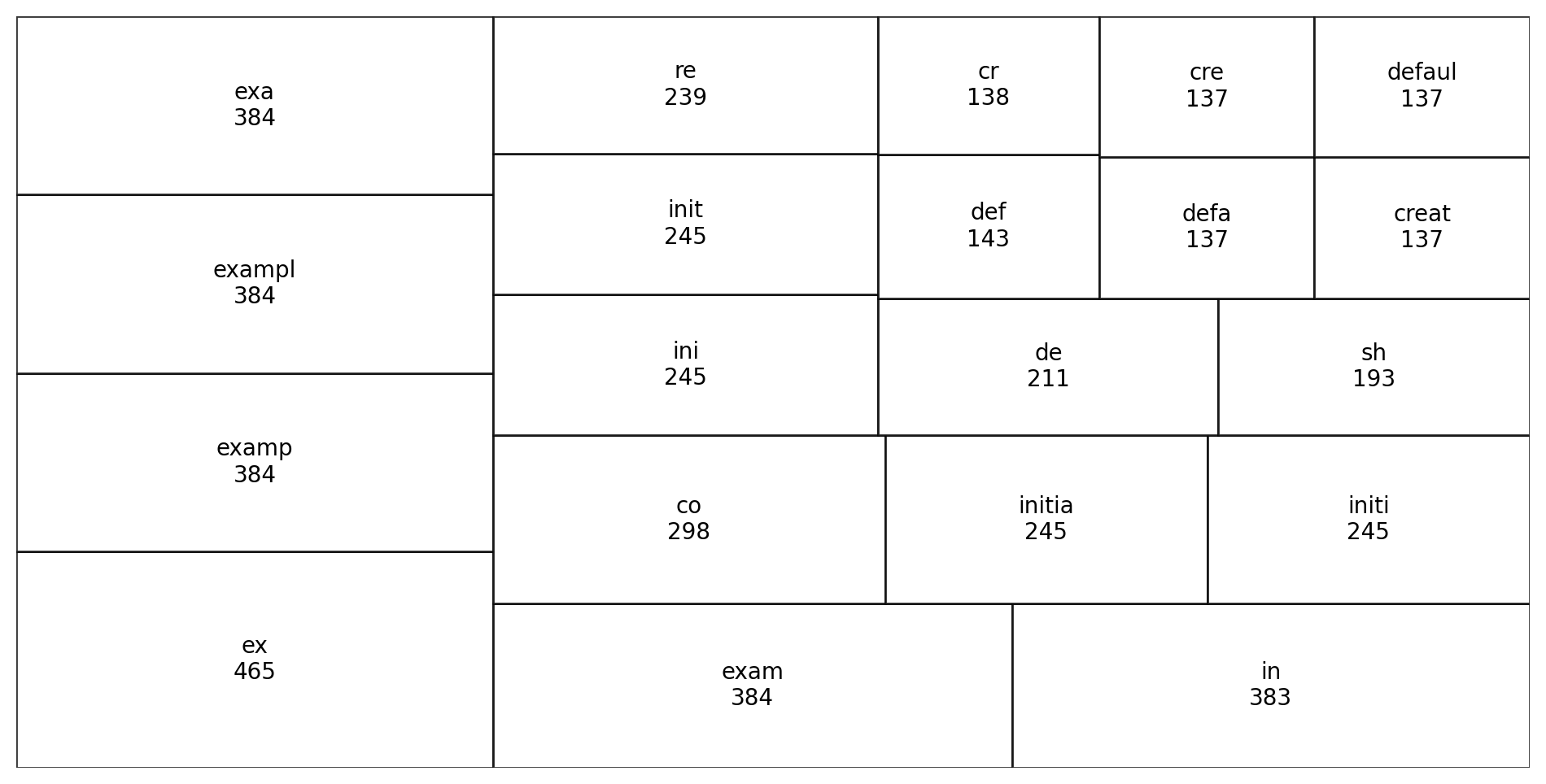}
        \caption{\textbf{Tree Maps of the \method in Roassal.}}
        \label{fig:methodprefixroassal}
       \end{subfigure}
    \hfill
    \caption{\textbf{Tree Maps for \class and \method in Roassal}. Figure~\ref{fig:classesprefixroassal} and Figure~\ref{fig:methodprefixroassal} explain why prefix length affects completion quality. In Roassal, many class names share project-specific prefixes such as \ct{RS}. For very short prefixes, many candidates remain indistinguishable, so structural information such as package dependencies has a stronger effect on ranking. As prefixes become longer, lexical filtering already removes many irrelevant candidates, and the difference between heuristics naturally decreases. The same phenomenon appears for selectors, although selector names are more widely reused across packages and therefore remain harder to disambiguate.}
 \end{figure}

\paragraph{\textbf{Dependency-Aware Behavior Across Projects}}
 The PCA result first observed in Roassal (as shown in Figure~\ref{fig:classespcaroassalcomplete}) generalizes across all five projects (see Appendix \ref{appendix}). Dependency-based heuristics dominate in high-cohesion frameworks such as Iceberg ($\rho_{\text{int}}=0.41$) and Spec ($\rho_{\text{int}}=0.30$), where the majority of relevant classes are located in the same packages (see Table~\ref{bench}). 
 
 In these systems, the first principal component captures most of the variance, showing that explicit dependency information is the principal explanatory factor for completion accuracy. In contrast, lower-cohesion projects such as Roassal ($\rho_{\text{int}}=0.20$) and Moose-tests ($\rho_{\text{int}}=0.13$) show a broader PCA group dispersion, reflecting that completion relevance is diluted by cross-framework dependencies and weaker internal coupling. Seaside lies between these extremes: its component-based design allows both \directdependencies and \transitivedependencies to perform similarly well, demonstrating that shallow but interconnected dependency chains remain beneficial for frameworks with intertwined hierarchies.

\paragraph{\textbf{Impact of Structural Cohesion}}
The correlation between completion performance and structural cohesion ($\rho_{\text{int}}$) is consistent across the dataset (see Table~\ref{bench}). Projects with higher internal cohesion (Iceberg, Spec) form compact PCA clusters with high explained variance, demonstrating that dependency signals are both strong and structurally meaningful. Loosely coupled systems (Roassal, Moose) show broader, less defined clusters, suggesting that cross-package dependencies introduce noise that weakens the predictive role of explicit relationships. This confirms that dependency-based heuristics are most effective when modular boundaries align with architectural semantics, whereas diffuse inter-package coupling reduces their discriminative strength.

\paragraph{\textbf{Interpretation of Prefix Sensitivity}}
Completing a given input is impacted by the use of prefix of variables: for example, when many classes start with \ct{FAMIX}, typing only \ct{FA} does not provide enough information to complete relevant \class candidates. Therefore, we describe the prefix situation of projects to assess their impact on our evaluation. 
 Prefix sensitivity (as shown in Figure~\ref{fig:classesprefixroassal} and Figure~\ref{fig:methodprefixroassal}  ) provides an additional dimension to interpret the observed dynamics. Evaluation as reported in Figure~\ref{fig:classesroassalComplete} shows that for short prefixes, where the search space is largest and ambiguity highest, dependency-aware heuristics exhibit the greatest separation and deliver the strongest gains. This effect diminishes as prefixes grow longer and the candidate pool narrows, leading to a natural convergence among all strategies.  In practical terms, this means that dependency-based strategies offer the most benefit precisely when the completion task is hardest, \eg when fewer characters have been typed. Investigating how to take into account prefix lengths is future work.

%%%%%%%%%%%%
%%%%%%%%%%%%
\section{Selector Completion Evaluation}\label{methodevaluation}

To answer \textbf{RQ3:} \emph{Does package awareness affect completion performance differently for \class and for \method}, we now present the result of the strategies applied to call site level \emph{selector} completion. 

\subsection{\textbf{Overall Three-Level Results}}
We evaluate \method completion on the same five projects (Roassal, Spec, Iceberg, Seaside, and Moose) using the same benchmark protocol from Section~\ref{methodology}. In contrast to \class completion, the selector-completion results show a much smaller separation between the three package-aware strategies. The \lateralpackage, \directdependencies, and \transitivedependencies heuristics often obtain close MRR values, and the best-performing strategy can vary depending on the project and prefix length.

This suggests that package awareness is useful for selector completion, but that selector completion is less directly explained by package dependencies than class-name completion. Selectors are shared across classes, inherited, overridden, and reused through polymorphic protocols. As a result, expanding the package scope can improve recall, but it does not always produce a clear ranking advantage.

\subsubsection{\textbf{Roassal}}
\begin{figure}[H]
    \centering
    \begin{subfigure}[b]{0.45\textwidth}
	\includegraphics[width=\linewidth]{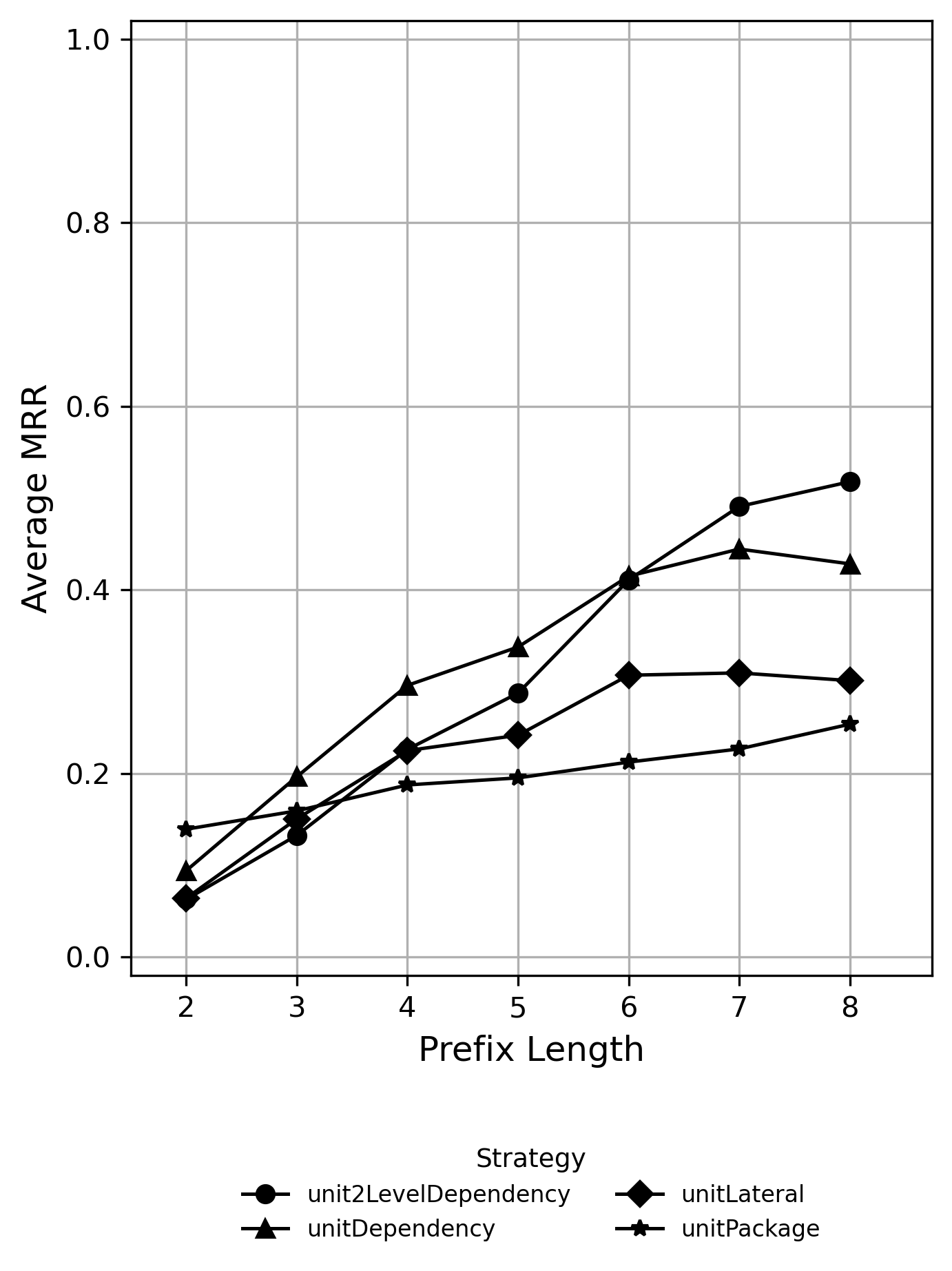}
	\caption{\textbf{\unitlevel for \method in Roassal.}}
	\label{fig:methodroassalUnit}
    \end{subfigure}
	\hfill
	   \begin{subfigure}[b]{0.45\textwidth}
		\includegraphics[width=\linewidth]{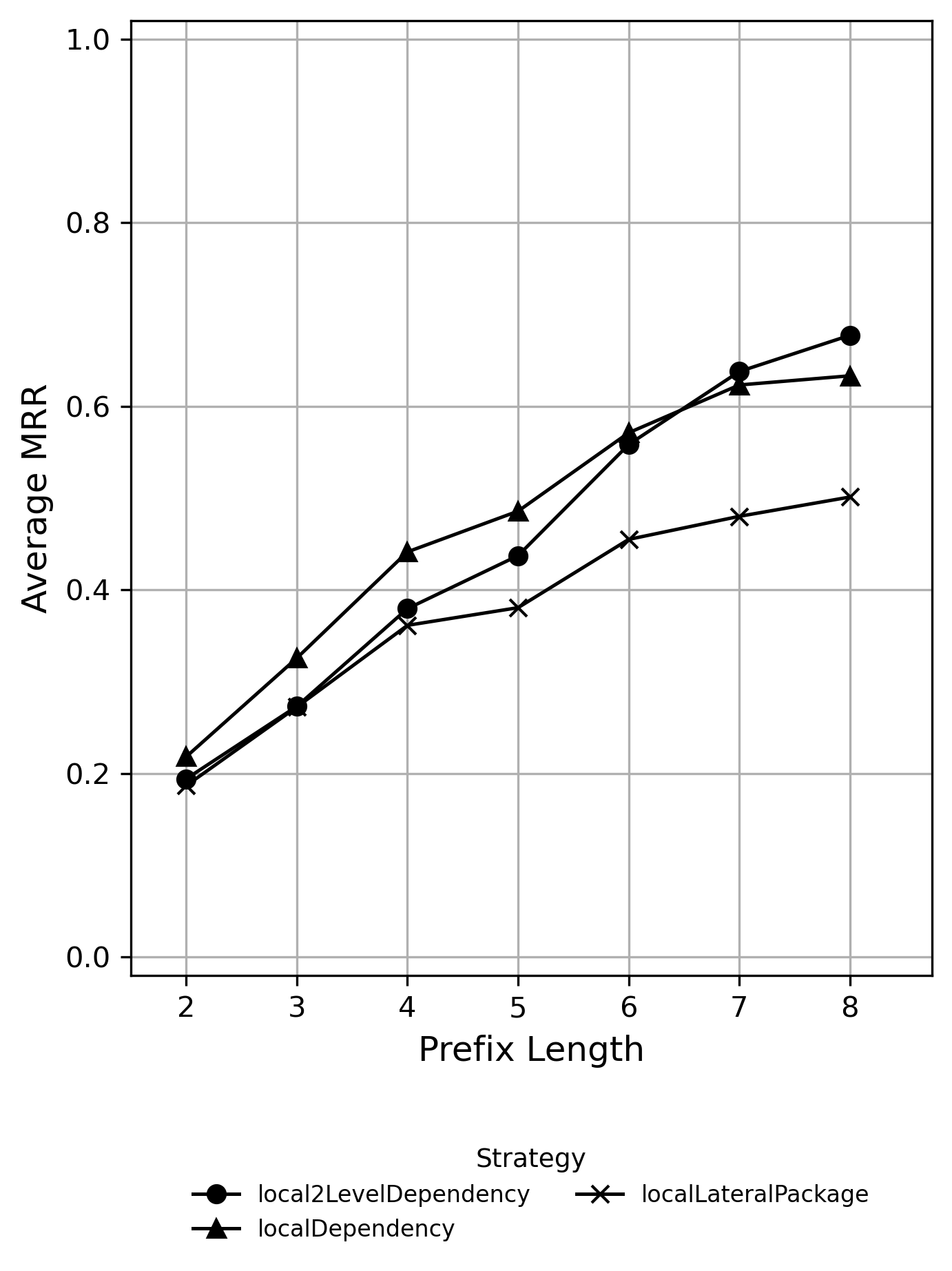}
		\caption{\textbf{\locallevel for \method in Roassal.}}
		\label{fig:methodroassalLocal}
    \end{subfigure}
    \hfill
    \caption{\textbf{\unitlevel VS \locallevel for \method in Roassal}}
 \end{figure}

\paragraph{\textbf{\unitlevel Analysis}}
Figure~\ref{fig:methodroassalUnit} shows that at the \unitlevel, both \directdependencies and \transitivedependencies achieve the highest performance, with the latter slightly outperforming the former for longer prefix lengths (above Top-5). Their steady and nearly parallel growth curves up to $\sim$0.5 indicate that short-range lexical and structural dependencies within a single class context remain moderately effective for \method prediction.  The \lateralpackage and \emph{unitPackage} heuristics improve marginally but stay under 0.25 on average, confirming that inter-class lateral relations are weak predictors of \method within a single unit. This pattern highlights that, at the class level, method invocation patterns depend primarily on direct package proximity.

\paragraph{\textbf{\locallevel Analysis}}
At the \locallevel, the inclusion of all \method within the same package substantially enhances average performance (Figure~\ref{fig:methodroassalLocal}). Both \directdependencies and \transitivedependencies dominate, achieving values around 0.65 at Top-8, with nearly identical progression curves. The improvement over the \unitlevel confirms that package-level context introduces relevant co-usage signals, especially between related visualization components in Roassal, such as \ct{RSShape}, \ct{RSCanvas}, and \ct{RSGroup}. This reinforces that forward-scoped lexical and structural relationships, rather than reciprocal or sibling connections, drive method completion in cohesive graphical packages. The smooth growth of both dependency-based strategies also suggests that local reuse within Roassal is tightly structured around dependency hierarchies rather than unstructured references.

\begin{figure}[H]
    \centering
    \begin{subfigure}[b]{0.45\textwidth}
	\includegraphics[width=\linewidth]{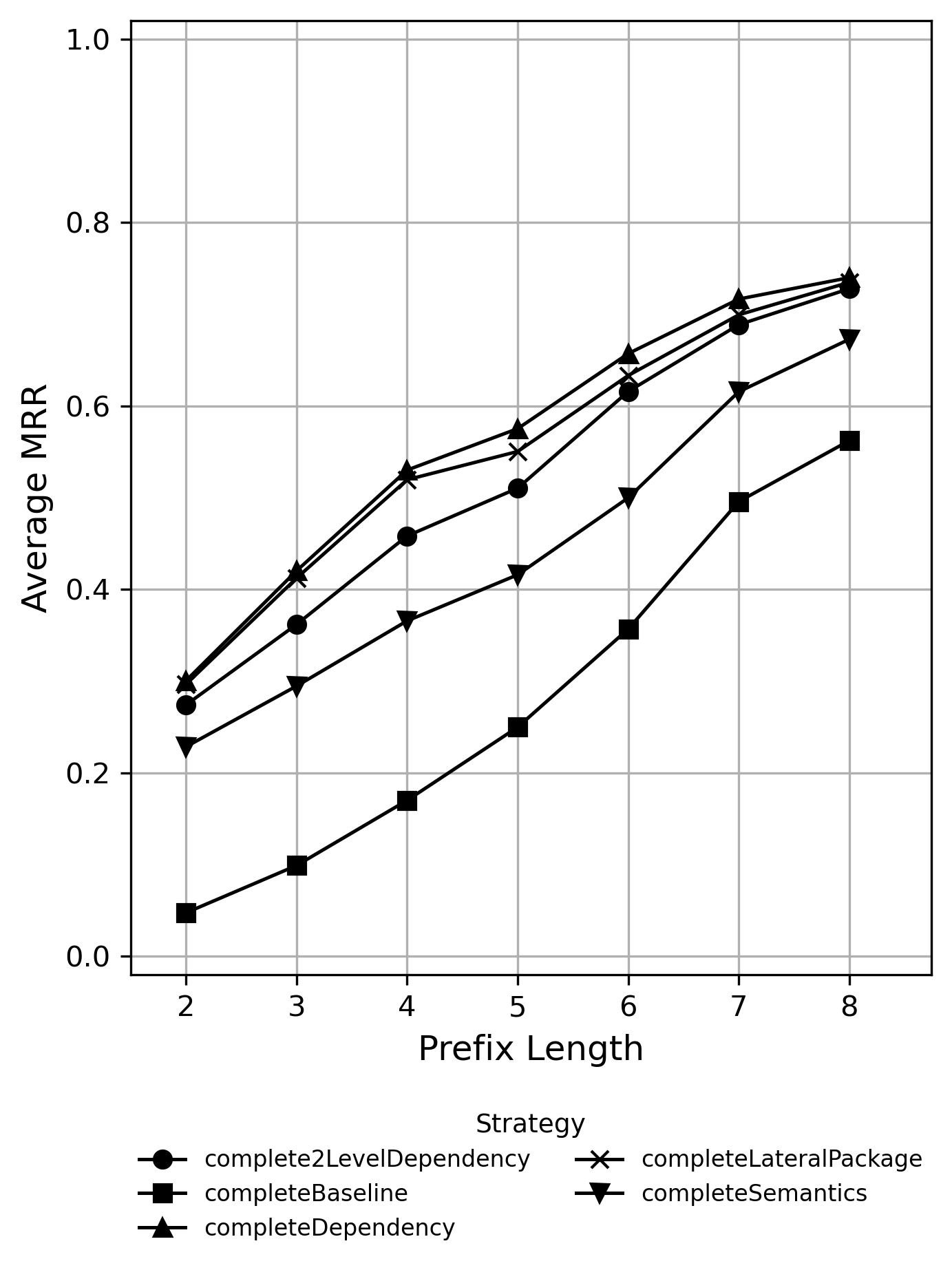}
	\caption{\textbf{\completelevel for \method in Roassal.}}
	\label{fig:methodroassalComplete}
    \end{subfigure}
	\hfill
	   \begin{subfigure}[b]{0.45\textwidth}
        \includegraphics[width=\linewidth]{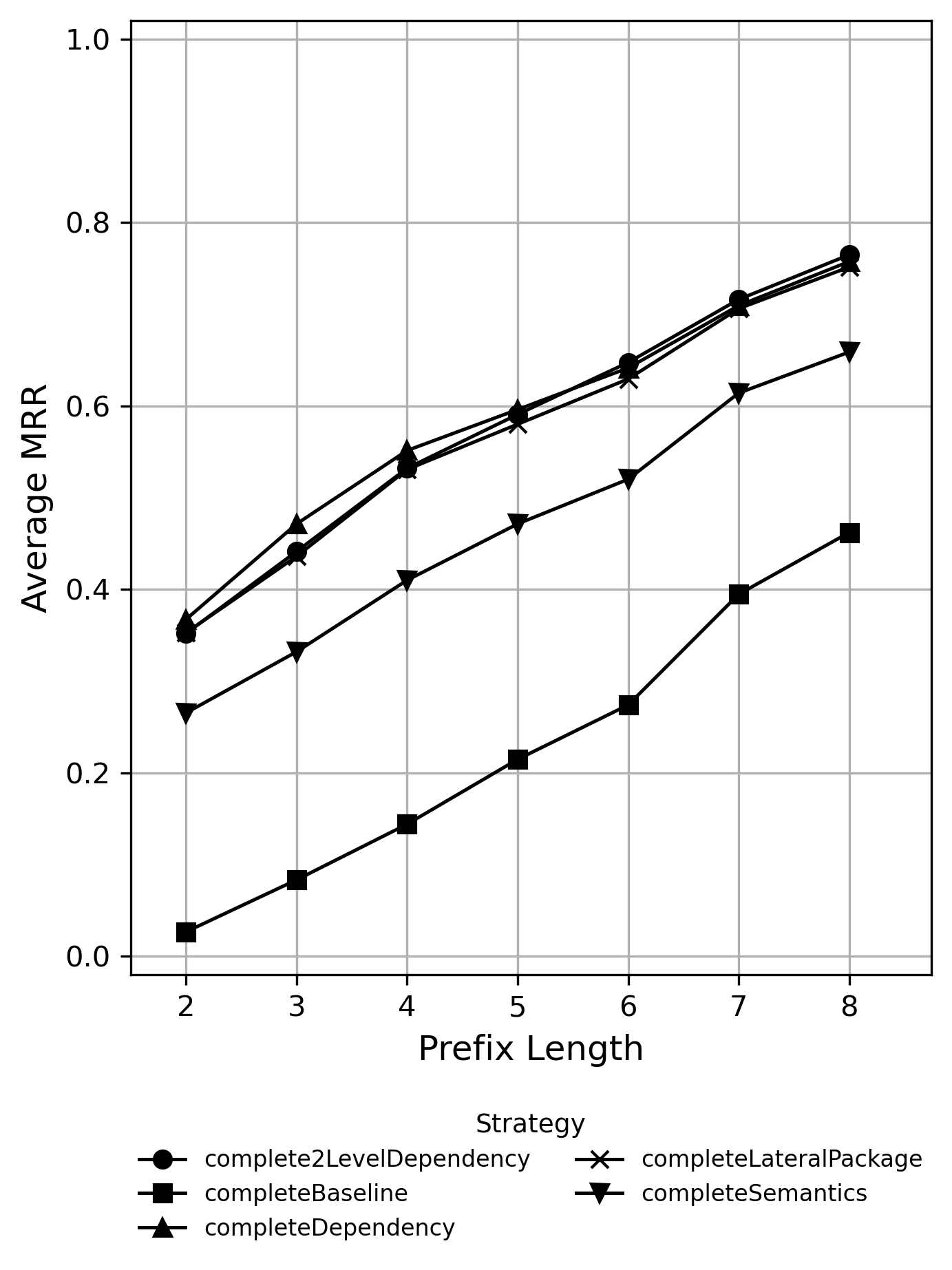}
        \caption{\textbf{\completelevel for \method in Iceberg.}}
        \label{fig:methodicebergComplete}
    \end{subfigure}
    \hfill
        \caption{\textbf{\completelevel for \method in Roassal and Iceberg}.}
 \end{figure}

\paragraph{\textbf{\completelevel Analysis}}
At the \completelevel (Figure~\ref{fig:methodroassalComplete}), rather than a single dominant heuristic, the results show that the top performing strategies are close to each other. The \lateralpackage and \directdependencies configurations achieve the highest overall performance, both exceeding 0.7 at longer prefix lengths, with nearly parallel trajectories throughout. These curves suggest that Roassal's method completion benefits from cross-package associations and direct dependency information, reflecting the compositional and reusable nature of its visualization components. The \transitivedependencies follow closely, indicating that \transitivedependencies awareness still contributes meaningful predictive power when the full project context is available. In contrast, \semantics display moderate but stable growth, reaching around 0.6 on average, suggesting that lexical similarity and inverse relations provide supportive but not leading signals. Finally, the \baseline remains the weakest performer across all ranks, highlighting that unspecialized lexical matching alone cannot capture the structural or behavioral regularities of Roassal's method usage patterns.

\subsubsection{\textbf{Iceberg}}
In Iceberg, the three package-aware strategies obtain very close results (Figure~\ref{fig:methodicebergComplete}). Although \transitivedependencies is competitive, it does not dominate uniformly across all Top-K ranks: for some prefix lengths, \directdependencies performs as well as or slightly better than the two-level dependency strategy. This indicates that extending the dependency scope can help, but the additional candidates introduced by the second dependency level do not always improve ranking precision.

The proximity between \directdependencies and \lateralpackage suggests that several relevant selector candidates are already captured by nearby project packages. In contrast, \semantics remains below the best package-aware strategies, showing that language-level lookup alone does not fully capture the project-level regularities present in Iceberg.

\subsubsection{\textbf{Seaside}}
\begin{figure}[H]
    \centering
    \begin{subfigure}[b]{0.45\textwidth}
        \includegraphics[width=\linewidth]{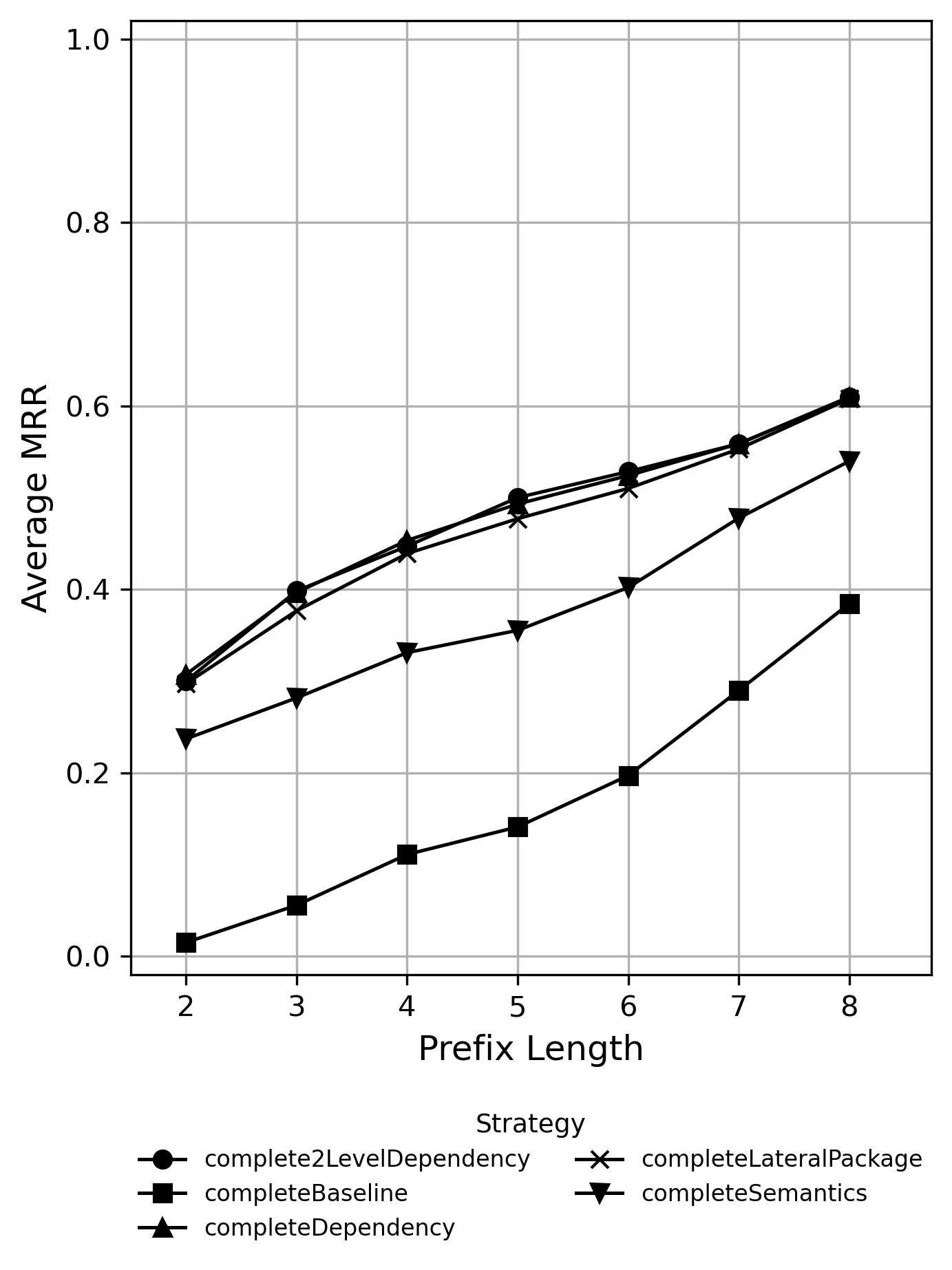}
        \caption{\textbf{\completelevel for \method in Seaside.}}
        \label{fig:methodseasideComplete}
    \end{subfigure}
	\hfill
	   \begin{subfigure}[b]{0.45\textwidth}
        \includegraphics[width=\linewidth]{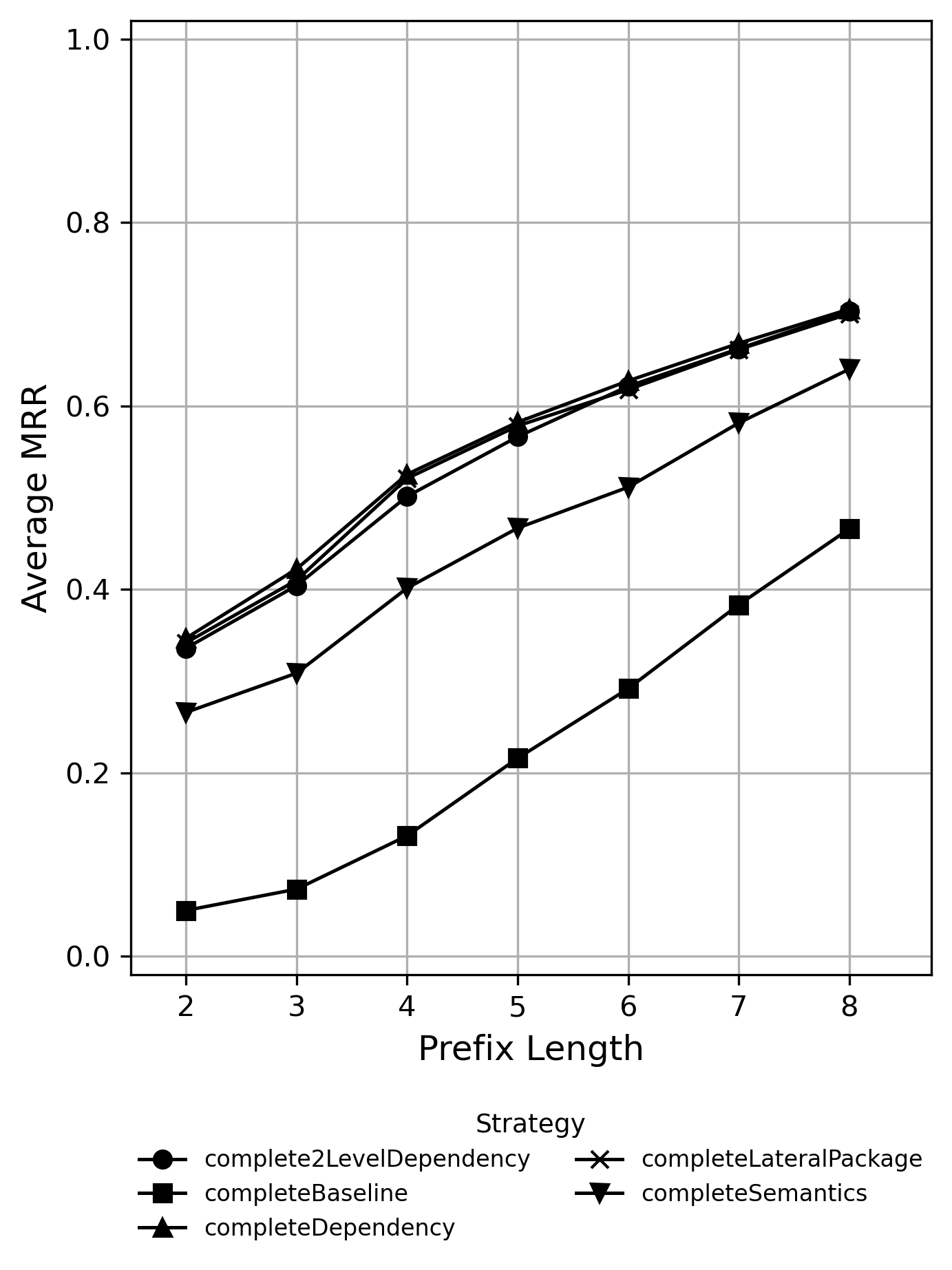}
        \caption{\textbf{\completelevel for \method in Spec.}}
        \label{fig:methodspecComplete}
    \end{subfigure}
    \hfill
        \caption{\textbf{\completelevel for \method in Seaside and Spec}.}
 \end{figure}

In Seaside, all three package-aware strategies show nearly identical performance (Figure~\ref{fig:methodseasideComplete}). This makes Seaside an important case because it has the weakest average selector-completion behavior among the evaluated projects. One likely explanation is that Seaside defines and uses many generic selectors across component, session, rendering, and callback abstractions. Such selectors are reused polymorphically across packages, which reduces the discriminative power of package-level scoping.

Therefore, the Seaside results should not be interpreted as evidence that one dependency strategy clearly dominates. Instead, they show a limitation of package-aware selector completion: when a framework relies heavily on shared protocols and generic method names, dependency information helps constrain the search space but does not necessarily provide enough information to rank the correct selector substantially higher.

\subsubsection{\textbf{Spec}}
In Spec, the \directdependencies strategy achieves the best performance across all Top-K ranks, with \lateralpackage close behind and \transitivedependencies trailing these two but clearly ahead of the remaining methods (Figure~\ref{fig:methodspecComplete}). This ordering reflects Spec's UI composition style, presenters, widgets, and adapters collaborate primarily through short call chains within or between adjacent packages such as \ct{Spec-Core} and \ct{Spec-Adapters}, so immediate dependency signals and intra-layer relations capture most valid completions. The \semantics remains weaker than the structural methods, suggesting that type resolution alone cannot model the dynamic binding and delegate patterns prevalent in Spec. As expected, the \baseline is consistently lowest, underscoring the advantage of leveraging even shallow dependency structure for method completion in this framework.

\subsubsection{\textbf{Moose}}
\begin{figure}[H]
    \centering
    \begin{subfigure}[b]{0.45\textwidth}
        \includegraphics[width=\linewidth]{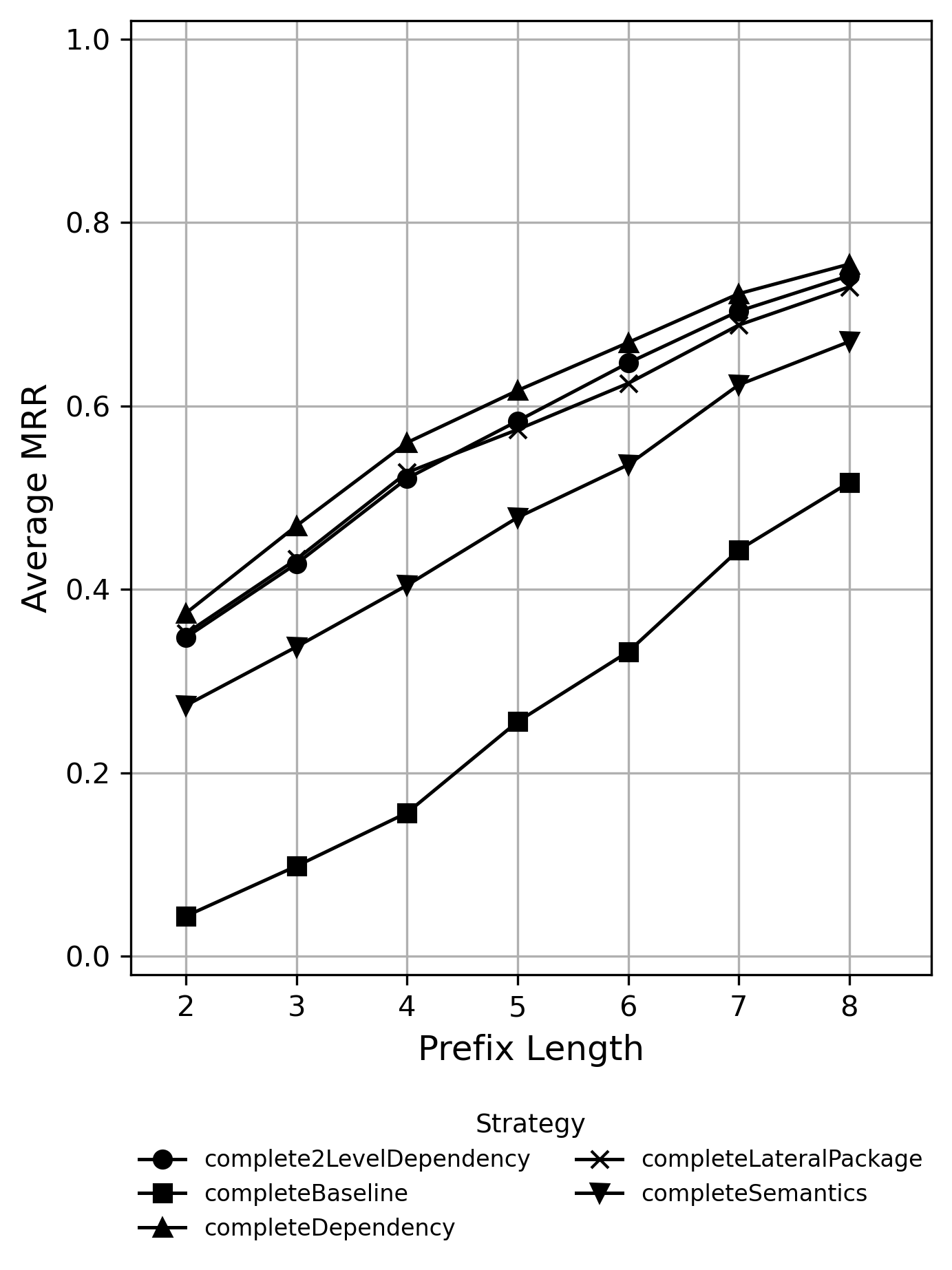}
        \caption{\textbf{\completelevel for \method in Moose.}}
        \label{fig:methodmooseComplete}
    \end{subfigure}
	\hfill
	   \begin{subfigure}[b]{0.45\textwidth}
        \includegraphics[width=\linewidth]{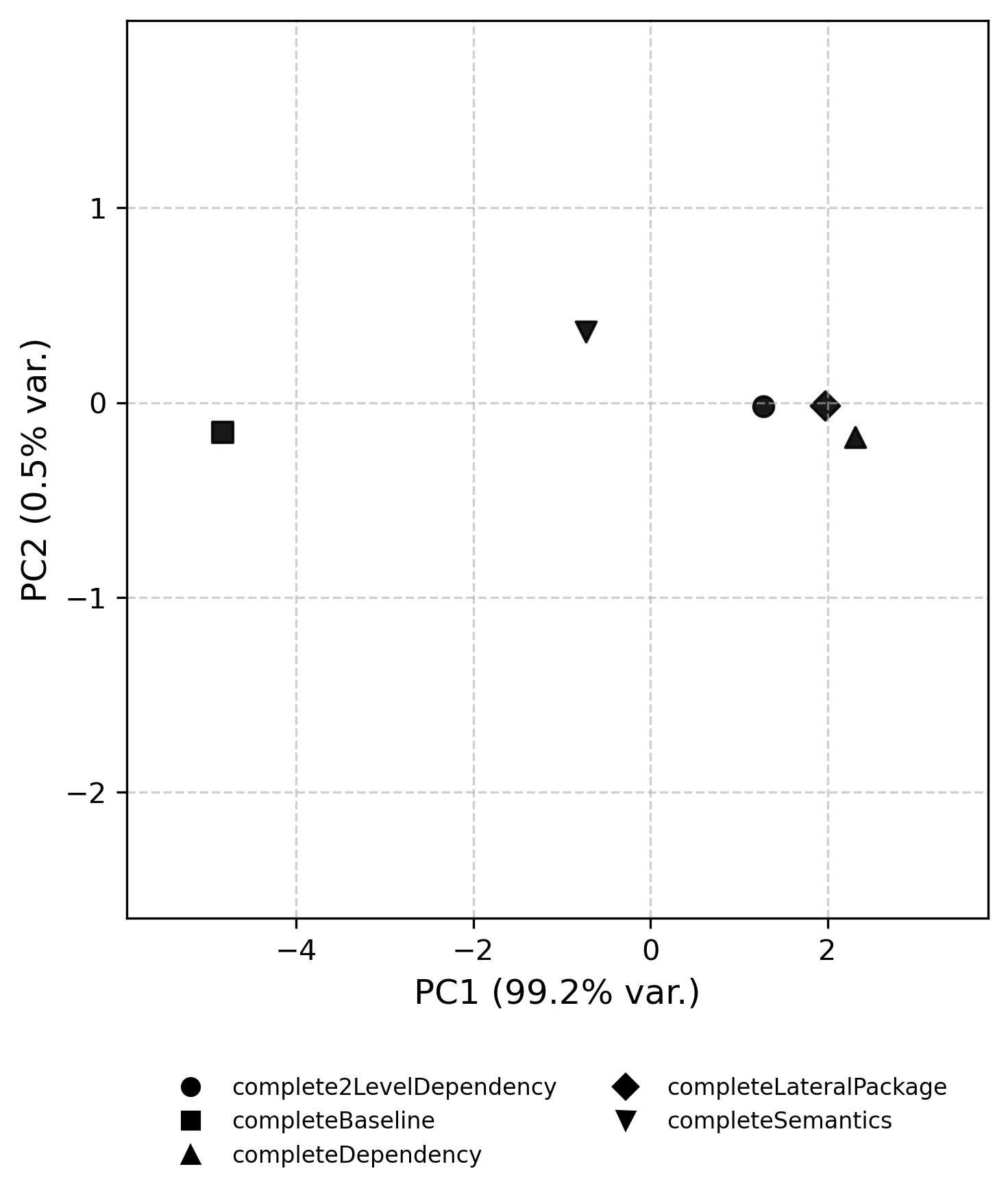}
        \caption{\textbf{PCA of Roassal \method at the \completelevel}}
        \label{fig:methodpcaroassalcomplete}
    \end{subfigure}
    \hfill
        \caption{\textbf{\completelevel VS PCA for \method in Roassal}}
 \end{figure}

In Moose, the \directdependencies strategy achieves the highest performance across all Top-K ranks (Figure~\ref{fig:methodmooseComplete}), closely followed by \lateralpackage and \transitivedependencies. This pattern reflects Moose's highly modular architecture, where core packages such as \ct{Moose-Core}, Spec are strongly interconnected through direct collaboration and extension points. The predominance of \directdependencies indicates that most relevant completions arise from immediate inter-package relationships within these core modules, while the strong performance of \lateralpackage suggests that horizontal reuse between analysis tools and model layers contributes additional context. The \semantics remains below the dependency-aware strategy, showing that type-level inference alone is less effective in this dynamic meta-modeling environment. As expected, the \baseline remains distinctly lower across all ranks, emphasizing the importance of dependency-driven reasoning for accurate completion in complex analytical frameworks such as Moose.
   
 \begin{boxB}
\textbf{Answer to RQ3: Differential Impact on Classes and Methods}

\emph{Does package awareness affect completion performance differently for \class and for \method ?}

Package awareness improves both \class and \method completion, but the effect is clearer and more stable for \class. Class-name references are directly tied to package structure: a class used in a package is often defined either in the same package or in an explicitly dependent package. Therefore, dependency-based scoping strongly reduces the candidate space and improves ranking. The completion of \method is more complex, because \method are reused across unrelated classes, inherited through class hierarchies, overridden in subclasses, and shared through polymorphic protocols. As a result, package-level information still improves completion relevance, but it does not always make one package-aware strategy clearly dominate the others. In several projects, \lateralpackage, \directdependencies, and \transitivedependencies are close for \method completion, and the best-performing strategy can depend on the project structure and prefix length. This suggests that package dependencies are a useful structural signal for \method, but that \method completion may also benefit from complementary signals such as receiver information, usage history, or type/runtime information.

\end{boxB}
   
 \subsection{\textbf{Selector Completion Discussion}}
We discuss some aspects of the selector completion. 

\paragraph{\textbf{Dependency-oriented structure}}
Across projects, package-aware heuristics improve selector completion compared to the baseline, but the separation between \lateralpackage, \directdependencies, and \transitivedependencies is much smaller than for \class completion. This is expected because selectors are not owned by a single package in the same way as class names. They are often defined, overridden, and reused across unrelated classes through inheritance and polymorphism.

\paragraph{\textbf{Semantic separation}}
The PCA (see Appendix A) shows that \semantics often occupies a region distinct from the package-aware heuristics. This suggests that semantic lookup and package-level scoping capture complementary signals rather than identical information.
% The PCA (See~\ref{appendix} and \ref{fig:appendix-iceberg-method-pca}, \ref{fig:appendix-moose-method-pca}, \ref{fig:appendix-roassal-method-pca}, \ref{fig:appendix-seaside-method-pca},    and  \ref{fig:appendix-spec-method-pca}) further highlights a strong separation of \semantics from all dependency-based heuristics. Its position along the upper-right region of PC1 indicates that semantic completion relies on an orthogonal representational basis dominated by type resolution and inheritance semantics rather than dependency topology. Despite its distinctness, \semantics remains competitive in projects with rich polymorphic structures (\eg Spec, Seaside), confirming its effectiveness in type-constrained lexical scopes.

\paragraph{\textbf{Interpretive coherence across systems}}
The selector results should therefore be interpreted more cautiously than the class-name results. Package-level scoping is beneficial, but the best package-aware strategy is project-dependent. In Moose, \directdependencies is slightly better; in Roassal, \transitivedependencies can be slightly worse; and in several cases the three package-aware strategies are practically close. This supports the conclusion that package information improves selector completion, but not that one package-aware heuristic consistently dominates all others.

%%%%%%%%%%%%
%%%%%%%%%%%%
\section{Limitations and Threats to Validity} \label{limitations}
Although our evaluation demonstrates measurable improvements in completion relevance through \directdependencies heuristics, we acknowledge several methodological limitations.

\subsection{\textbf{Methodological and Evaluation Constraints}}

\paragraph{Synthetic users} A first limitation lies in the static nature of our benchmarking protocol. The evaluation simulates completion requests by progressively truncating identifiers into prefixes ranging from 2 to 8 characters and verifying whether the correct completion appears within the top-ranked candidates. While this method aligns with standard practice in code completion research \cite{Robb08a, Robb08b, Robb10c}, it inevitably abstracts away the nuances of real-world interactive programming behavior. In an actual development session, developers frequently rely on code completion not merely to finish known names but also to explore unfamiliar APIs, correct misspellings, or recall method signatures. Consequently, completion systems are often used as a discovery and navigation mechanism rather than a mere typing aid. Our current evaluation, which isolates prefix-based retrieval performance, may therefore underestimate the broader cognitive and exploratory dimensions of completion usage.

Moreover, the benchmark is constrained by its synthetic experimental design. All evaluated completion events were generated offline, without live interaction or dynamic editing context. It excludes the fact that a developer may want to refer to a class for which its current package is not yet dependent. In such case, the dependency analysis will not be effective. From this perspective, our benchmark is biased. A possible counterbalance is that code rarely depends on another package for just a single reference. 

Future work should consider interactive user studies or telemetry-based evaluations to complement static analysis and better capture the behavioral variability of completion use in realistic settings.

\paragraph{\textbf{No direct cross-system quantitative comparison}}
We do not provide a direct quantitative comparison with completion systems evaluated in other languages or IDEs. Such a comparison would require a shared benchmark, identical completion targets, comparable project structures, and equivalent runtime assumptions. Many related systems target API recommendation, token generation, line completion, or neural repository-level generation, whereas our benchmark evaluates identifier completion for \class and \method in Pharo. We therefore compare these approaches qualitatively in the Related Work section and focus our quantitative evaluation on the controlled comparison of package-level scoping strategies inside the same completion infrastructure. This design isolates the effect of package and dependency information, while avoiding misleading conclusions from non-equivalent experimental settings.

\paragraph{Dataset composition} 
A further methodological limitation is related to the composition of the dataset. Although we selected a diverse and representative set of mature Pharo projects including Spec, Seaside, Iceberg, Roassal, and Moose, these projects share common design philosophies, coding conventions, and package structures. As such, our results may not generalize to less cohesive repositories, legacy codebases, or non-standard package organizations. In particular, the heuristics may behave differently in experimental or research projects that violate standard modular design principles, or in industrial systems that rely heavily on dynamically loaded extensions or external binary packages.

\paragraph{No development session}

Finally, the absence of development session data constitutes a further limitation. Our benchmark relies on a static snapshot of each project and therefore cannot assess how completion behavior evolves as projects grow or restructure. Prior work on change-based and usage-log-based completion~\cite{Robb08a, Robb10c, Biba22a} has shown that temporal and developer-specific patterns often influence completion accuracy and ranking preferences. Integrating change-based or version-aware benchmarking could thus provide a richer longitudinal understanding of how heuristics adapt to evolving software ecosystems.

\subsection{\textbf{Heuristic and Architectural Limitations}}

\paragraph{Different architectures} The proposed heuristics rest on the assumption that entities defined within the same package or its direct dependencies are more likely to be relevant than those located elsewhere in the system. Although this assumption holds empirically for class-name completion, it does not universally reflect actual dependency dynamics in all software architectures. Packages functioning as frameworks, utilities, or test suites often show inverse dependency directions; they are used extensively by other packages but reference few entities internally. In these cases, strict locality prioritization may degrade ranking precision by promoting rarely used \class simply because they reside in the same package.

This limitation suggests a need to empirically quantify intra-package referencing behavior across different project types. Incorporating such measurements could allow the completion engine to dynamically adjust the weighting of local versus external suggestions. For instance, in test packages or UI layers that frequently reference entities from core packages, shifting the priority toward project-level dependencies rather than package-local entities may yield better results. Conversely, in tightly cohesive modules, local prioritization remains beneficial. Adaptive, data-driven heuristics would thus generalize better across diverse development contexts.

\paragraph{Agnostic to reflective use} Another architectural limitation concerns the representation of package dependencies. Although our implementation computes the dependency graph dynamically from the current execution, it does not fully account for runtime variability introduced by reflective features or metaprogramming. Some systems establish dependencies lazily or modify them at runtime, particularly in plugin-based architectures. In such cases, a static or semi-static dependency graph may misrepresent actual usage, leading to either missing or spurious completion candidates. Addressing this issue would require integrating run time introspection or lightweight dependency profiling mechanisms capable of capturing dynamic relationships during live programming sessions.

\subsection{\textbf{Other Aspects}}

\paragraph{Long prefixes}
The evaluation currently focuses on relatively short prefixes, ranging from two to eight characters, which represent typical invocation lengths in interactive completion scenarios. However, real-world identifiers can be substantially longer and semantically compound, such as \ct{IceSBBrowserAbstractMethodCommand} or \ct{SpPresenterBuilder}. These long identifiers pose different challenges; they are often predictable only through semantic or structural cues, not through purely lexical prefix matching. Our methodology, which limits prefix length, does not consider them. 

Future work should extend the analysis to longer prefixes and compound identifiers, potentially combined with semantic decomposition or embedding-based similarity models that better capture conceptual relationships beyond shared lexical roots. 

\paragraph{Limited context}
Additionally, the current evaluation does not incorporate runtime context such as receiver types, variable bindings, or recent navigation history, which often constrain and inform relevant completions. Integrating lightweight dynamic profiling could enrich the contextual basis of completion heuristics, improving both ranking stability and developer-perceived accuracy.

\section{Responsiveness Analysis}\label{responsiveness}

While \directdependencies improves completion accuracy over the default \semantics model used in Pharo 12, the usability of the solution depends on whether the additional package analysis preserves interactive responsiveness. Code completion is invoked repeatedly while typing, so even accurate suggestions are not useful if they introduce visible latency.

The Pharo consortium integrated our implementation into Pharo 13, where it is used daily by developers. Although this production use provides practical evidence of feasibility, we complement it here with a focused responsiveness evaluation. We measure completion latency for short and longer prefixes and discuss the implementation choices used to avoid dependency-computation bottlenecks.

\paragraph{Pharo 13 default distribution}
While benchmarks may often be impacted by the code loaded in a system,  we run the benchmarks on the default Pharo 13 system. 
We report two measures: one for 2-character input and the second for 8 character inputs. It should be noted that the longer the input is, the more work may be needed because for longer inputs, the probability of being forced to use the next fetcher in the chain of fetchers is higher, involving more computation.

\begin{itemize}
\item For the \class at the \completelevel, the completion time went from \ct{1.25} to \ct{1.32} ms for the baseline to  \ct{0.64} and \ct{1} ms for the \directdependencies.
The shorter time shows that the completion found the requested names faster using the \directdependencies heuristics. 

\item For the \method at the \completelevel,  the completion time went from \ct{2} to \ct{8} ms for the baseline to  \ct{15} and \ct{25} ms for the \directdependencies. It shows that while the accuracy is better it takes more computation power. In particular method distribution cross-cut package structure since inheritance usually does it. 

\end{itemize}

This level and setup represent a default situation since Pharo 13 is about \ct{139,180} methods with \ct{62,708} unique names defined in \ct{10,665} classes. Note that these numbers are well below the 100 millisecond threshold expected by developers that is mentioned in the literature \cite{Biba22a}. Such low time to complete identifiers are due (1) to the lazy architecture of \compl that only fetches batches of identifiers one at a time, (2) the cache to limit the impact of the package dependency computation. it also shows that performing more advanced lookup strategies are worth.

\paragraph{Package dependency} Our implementation computes dynamically the dependent packages of a package. This is an expensive computation therefore, our implementation uses a simple cache. The current implementation does not invalidate the cache after each dependency modification (when a class reference is added or removed). 
From that perspective, it can provide wrong package dependencies to the completion engine, leading to wrong completion. This is one of our future works to improve these aspects of the completion.

\begin{boxB}
\textbf{Answer to RQ4: Responsiveness and Practical Viability}

\emph{Can dependency-based heuristics enhance completion relevance without compromising responsiveness and performance in large modular Pharo systems?}

The evaluation shows that dependency-based heuristics improve completion relevance while remaining within interactive latency bounds. The \directdependencies strategy introduces additional computation, especially for \method completion, but the measured times remain well below the 100 millisecond threshold commonly expected for interactive completion. For \class completion, the dependency-aware strategy can even reduce latency because relevant candidates are found earlier in the heuristic chain. For \method completion, the added cost is higher, but still compatible with interactive use.

\end{boxB}

\section{Related Work} \label{relatedwork}

Code completion has been extensively studied as a means to improve developer productivity and reduce navigation effort in modern IDEs. Over the years, research has evolved from purely syntactic \cite{Robb08a, Robb08b, Robb08c} and frequency-based techniques \cite{Bruc09a, McMi10a} to approaches that exploit semantic, statistical, and structural information about programs. In dynamically typed environments such as Pharo, these challenges are amplified by the absence of static type information and by the highly modular organization of software into interdependent packages. This section reviews prior work across several complementary dimensions: completion in dynamic languages, contextual and repository-level reasoning, use of historical and behavioral data, and recent advances in neural and hybrid models to situate our contribution within the broader landscape of intelligent code completion systems.

\subsection{\textbf{Code Completion in Dynamically-Typed Languages}}
Research on code completion has evolved from syntactic pattern matching to statistically and semantically informed models. Early systems relied on alphabetical sorting and static typing, providing low contextual relevance in dynamically typed environments. Subsequent efforts incorporated statistical frequency and contextual information to improve ranking quality, such as frequency-based ordering \cite{Bruc09a}, association rules \cite{McMi10a}, and k-nearest neighbor matching \cite{Bruc09a} over usage contexts. The “naturalness of software” paradigm \cite{Hind12a} reframed source code as a statistically predictable medium, motivating probabilistic approaches based on n-gram models and, later, neural architectures. These methods demonstrated that code exhibits repetitive patterns analogous to natural language, enabling effective token prediction. Tu \etal \cite{Tu14a} and Franks \etal \cite{Fran15a} integrated locality and cache mechanisms to capture short-term context in developer sessions, achieving measurable ranking improvements with low computational cost. 

With the rise of deep learning, transformer-based architectures brought substantial improvements to contextual prediction. Svyatkovskiy \etal \cite{Svya20a} introduced IntelliCode Compose, a transformer-based model for contextual code generation in Visual Studio, effectively demonstrating completion in dynamic and untyped environments. Building on this, Izadi \etal \cite{Izad22a} proposed CodeFill, which jointly learns from structural and naming sequences to enable multitoken and structure-aware completions. 

In dynamic environments such as Smalltalk or Pharo, static type information is limited, and code completion must instead leverage runtime or lexical context \cite{Svya19a}. Komárek \etal \cite{Koma17a} identified these constraints and proposed a modular, extensible completion engine architecture for Pharo, paving the way for heuristic-driven systems such as \compl. Romaniuk \etal \cite{Roma20a, Roma20b} extended this line by integrating unigram and bigram models into Pharo's AST-based pipeline, while Zaitsev \etal \cite{Zait20a} offered large-scale statistical analyses of Pharo code to support model training. Spasojević \etal \cite{Spas16a} further explored lightweight type-hint recovery using identifier naming, demonstrating the feasibility of enriching completion relevance even without static typing. Elkolei \etal \cite{Elko26a} introduced four complementary extensions to \compl typo-tolerant matching, implicit prefix expansion, grouped completion entries, and camel-case matching. These extensions improve matching robustness and the presentation of completion results, whereas our work changes the scope and ordering of candidates using package and dependency information. The two approaches therefore operate at different layers of the \compl architecture and can be combined.

\subsection{\textbf{Leveraging Historical and Behavioral Data}}

Another dimension in improving code completion concerns exploiting historical or behavioral signals. Robbes \etal \cite{Robb08a} pioneered change-based completion, leveraging fine-grained software history to predict future edits. Nguyen \etal \cite{Nguy16a} extended this notion through APIREC, a statistical learning model over large-scale change corpora, demonstrating that co-evolving edits reveal recurrent developer patterns. More recently, Bibaev \etal \cite{Biba22a} proposed learning-to-rank models trained on anonymous IDE usage logs, validating that user interaction data can serve as a powerful implicit supervision source for improving completion accuracy in production settings.

\subsection{\textbf{Neural Models and Hybrid Approaches}}

The advent of deep learning transformed code completion from statistical sequence modeling to structural learning. Graph-based and Transformer architectures \cite{Alla17a, Hell20a} incorporate control-flow and data-flow relations, achieving robust performance on syntactic and semantic prediction tasks. Nevertheless, such models remain computationally expensive and often overlook project-level modularity, a key characteristic of large Pharo systems. Recent hybrid approaches combine neural representations with structural reasoning. For instance, Hellendoorn \etal \cite{Hell20a} introduced graph transformer hybrids that integrate relational edges into attention mechanisms, while Li \etal \cite{Li21b} explored ensemble reranking to optimize the trade-off between suggestion relevance and developer effort. These methods demonstrate that accuracy improves when statistical learning is coupled with explicit architectural context, an insight central to the present work. Kier \etal \cite{Kier26a} recently introduced an end-to-end approach for LLM-based code completion in Pharo, combining Pharo-specific
data curation, continued pre-training, fine-tuning, and dedicated completion benchmarks. Their approach targets multi-token completion under the latency constraints of an interactive IDE. In contrast, the present work focuses on lightweight token-level candidate scoping and ranking using explicit package dependencies. These approaches are complementary: dependency information could be used to select repository context for an LLM or to rerank its suggestions.

\subsection{\textbf{Contextual and Repository-Level Completion}}
Beyond local lexical prediction, recent studies emphasize the importance of structural and contextual scoping \cite{Bruc09a}. Traditional semantic heuristics, such as those implemented in \compl, capture local binding and inheritance but treat the system namespace as flat. This omission has been increasingly recognized as a limitation in large modular systems, where developers typically operate within subsets of interdependent packages.

In broader ecosystems, repository-level completion has gained prominence, incorporating project-wide dependencies and architectural context. Systems such as RepoCoder \cite{Zhan23a}, GraphCoder \cite{Liu24a}, and RLCoder \cite{Wang25a} extend completion beyond single files by modeling repository graphs, dependency hierarchies, and global code retrieval. These approaches demonstrate that structural context, especially dependency awareness, significantly improves relevance and reduces hallucination in large-scale environments. In the specific case of Pharo, this perspective aligns with the moldable development philosophy \cite{Chis15b}, advocating adaptable tools that reflect the structure of the system being developed. Package- or repository-level reasoning thus provides a natural extension to \compl's semantic architecture, bridging the gap between language-level semantics and project-level modularity.

\subsection{\textbf{Parallels with Existing Completion Strategies}}

A direct quantitative comparison with many prior systems is difficult because these approaches target different languages, completion units, runtime assumptions, and datasets. For example, usage-log-based completion relies on IDE interaction traces \cite{Robb08a, Robb10c, Biba22a}, neural repository-level systems often target token or line completion \cite{Izad22a, Zhan23a, Liu24a, Wang25a}, and API-ranking approaches often assume static type information, type hierarchies, or library usage traces \cite{Hou10a, Spas16a}. In contrast, our evaluation focuses on identifier completion in Pharo, a live dynamically typed environment where package dependencies can be computed from the running system. Nevertheless, these differences do not prevent a conceptual comparison, since the underlying objective is shared: reducing the candidate space and improving the ranking of relevant completion proposals.

Our approach is closest to prior work that uses contextual filtering and locality to improve completion relevance. Frequency-based and Bayesian approaches exploit statistical regularities in code \cite{Bruc09a, McMi10a, Prok15a}. Best Matching Neighbors ranks candidates from similar local contexts \cite{Bruc09a}, history- and usage-log-based techniques exploit past developer behavior \cite{Robb08a, Robb10c, Biba22a}, and repository-level neural systems retrieve or model project-wide context \cite{Izad22a, Zhan23a, Liu24a, Wang25a}. Our contribution is complementary to these approaches: instead of learning a ranking model from historical traces or neural representations, we use explicit package dependencies as a lightweight structural signal. This signal is already available in modular software systems and can be integrated into a heuristic completion architecture without requiring a training corpus, user interaction logs, or a static type system.

\begin{table}[htbp]
\centering
\footnotesize
\begin{tabular}{p{0.22\linewidth}p{0.29\linewidth}p{0.37\linewidth}}
\hline
\textbf{Family of approach} & \textbf{Main contextual signal} & \textbf{Relation to our work} \\
\hline
Frequency-based, statistical, and Bayesian completion \cite{Bruc09a, McMi10a, Hind12a, Prok15a} & Global or local occurrence frequencies and statistical regularities in code & Shares the goal of ranking likely candidates, but uses statistical regularities rather than package structure. \\
\hline
Best Matching Neighbors and locality-based completion \cite{Bruc09a, Tu14a, Fran15a, Robb08a} & Similar local usage contexts, recent usage, and lexical proximity & Shares the intuition that nearby context is more predictive than the global namespace; our notion of proximity is package/dependency based. \\
\hline
History- and usage-log-based completion \cite{Robb08a, Robb10c, Nguy16a, Biba22a} & Developer behavior, recent edits, change history, or IDE interaction traces & Complementary to our approach; such signals could further rank candidates after dependency-based scoping. \\
\hline
Type- and API-aware completion \cite{Hou10a, Spas16a} & Static types, type hierarchies, API structures, or identifier-based type hints & Similar in spirit because it filters irrelevant candidates, but our approach targets a dynamically typed setting without relying on static type information. \\
\hline
Neural and hybrid completion models \cite{Svya20a, Izad22a, Hell20a, Li21b} & Learned code representations, structural relations, and reranking models & Complementary to our approach; package dependencies could be used as an additional structural signal before or during neural reranking. \\
\hline
Repository-level completion \cite{Zhan23a, Liu24a, Wang25a} & Project-wide retrieval, repository graphs, dependency hierarchies, or global code context & Shares the repository-level perspective; our approach provides a lightweight structural alternative based on explicit package dependencies. \\
\hline
This work & Current package, lateral packages, direct dependencies, and two-level dependencies & Provides a systematic evaluation of package-level scoping strategies for \class and \method completion in Pharo. \\
\hline
\end{tabular}
\end{table}

\subsection{\textbf{Positioning of the Present Work}}

The present study contributes to this line of research by evaluating package- and repository-level scoping as a lightweight source of context for code completion. We do not claim to introduce a new general-purpose learning algorithm. Rather, our scientific contribution is empirical and system-oriented: we formulate several package-level scoping strategies, integrate them into the \compl architecture, and systematically evaluate their impact on completion ranking and responsiveness.

Compared with previous Pharo completion models, which focused mainly on semantic scoping or statistical ranking \cite{Koma17a, Roma20a, Roma20b, Zait20a}, our approach explicitly uses dependency graphs to constrain the candidate space to structurally relevant packages. Compared with repository-level neural completion \cite{Zhan23a, Liu24a, Wang25a}, our approach is deliberately lightweight: it does not require model training or repository-scale retrieval at completion time. Compared with history- or usage-log-based completion \cite{Robb08a, Robb10c, Biba22a}, it does not require previous developer interaction traces. The results show that, on the Pharo projects evaluated in this study, \directdependencies provides the strongest overall ranking improvement, especially for \class completion, while remaining within interactive latency bounds. This positions dependency-based scoping as a practical bridge between language-level semantic completion and repository-level structural awareness.

%%%%%%%%%%%%
%%%%%%%%%%%%
\section{Conclusion and future works} \label{conclusion}

This work enhances \compl, Pharo's semantic code completion engine, by introducing package-aware and dependency-driven heuristics. Traditional \compl treated the environment as a flat namespace, which limited completion precision in large modular projects. The new approach introduces locality and structural awareness, prioritizing entities from the current package and its dependencies before falling back to global scope. Our evaluation on five Pharo projects shows that \directdependencies achieves the best overall Mean Reciprocal Rank (MRR) and top-K accuracy, especially for \class completion. The responsiveness analysis shows that dependency-aware completion may introduce additional computation, particularly for \method completion, but the measured times remain within interactive latency bounds. This demonstrates that dependency context improves completion relevance while preserving practical responsiveness. However, naive prefix-based ranking can perform poorly in terms of accuracy in packages with extensive cross-references (\eg test packages), suggesting the need for refined dependency weighting.

Future research will extend this approach toward advanced dependency scoping and hybrid statistical and structural models. By combining lightweight frequency analysis with dependency graphs, \compl can better adapt to varying usage patterns and code architectures. The next steps will also explore leveraging historical and behavioral data as well as integrating GUI-level interaction insights, ensuring the engine evolves toward a fully adaptive, repository-level completion system for large-scale Pharo development. Overall, the results show that package dependencies provide a practical middle ground between purely lexical completion and heavier learned repository-level models. The approach does not replace statistical, neural, type-based, or history-based ranking strategies; rather, it provides a lightweight structural scoping layer that can reduce the candidate space before such ranking strategies are applied. This is particularly relevant for live, dynamically typed environments such as Pharo, where static type information is limited but package and dependency information is available at runtime.

\paragraph{\textbf{Acknowledgments}}
We thank Inria and the LLM4Code challenge for the funding of the first author. 

\newpage
\begingroup
\small   
\bibliographystyle{abbrv}
\bibliography{rmod,others}
\endgroup

%\setcounter{tocdepth}{5}
%\tableofcontents

\newpage
\appendix \label{appendix}
\section{Principale Components Analysis}

%iceberg and iceberg-tests
\begin{figure}[H]
    \centering
    % ---- Row 1 ----
    \begin{subfigure}[b]{0.45\textwidth}
        \includegraphics[width=\linewidth]{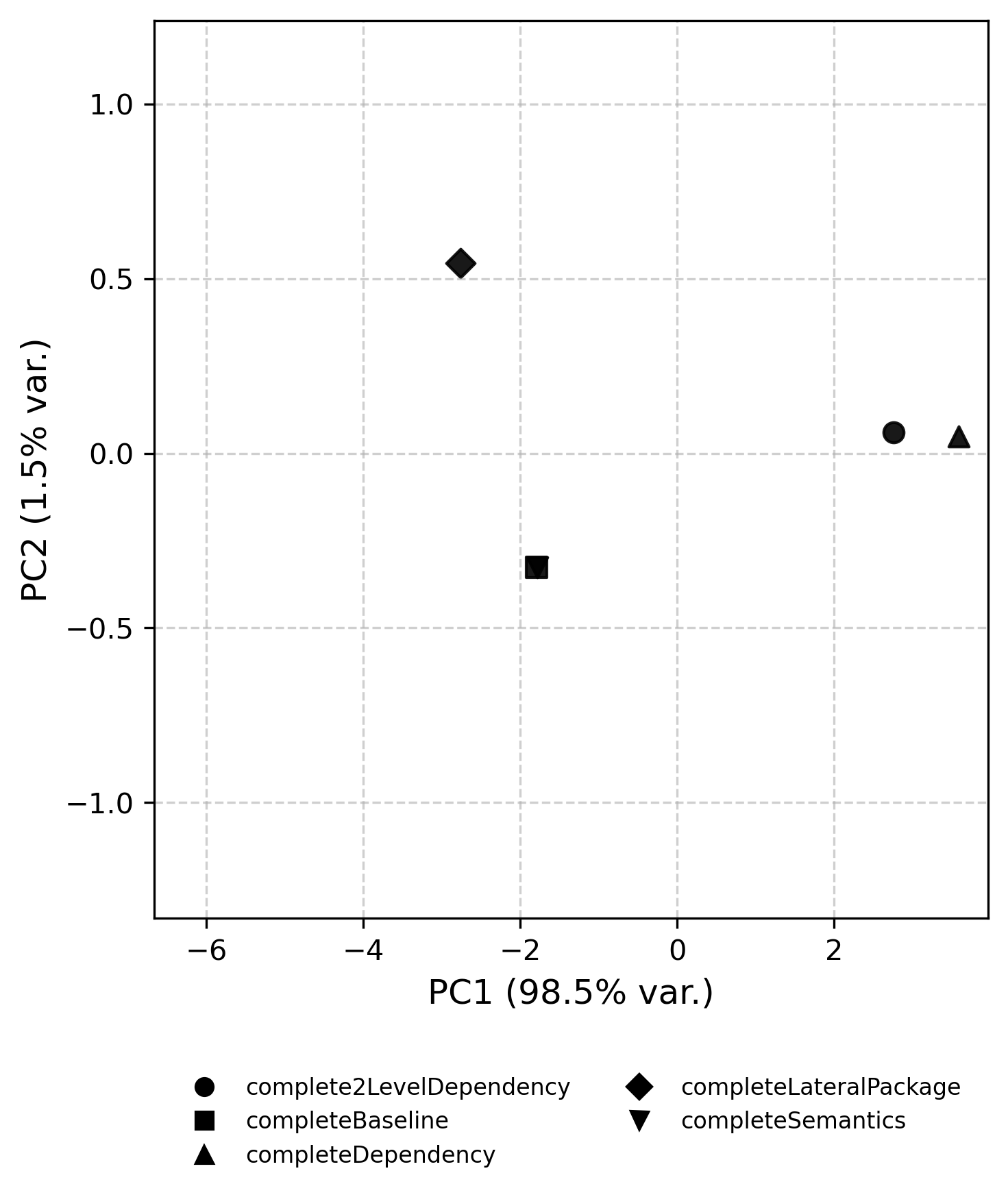}
        \caption{\textbf{\completelevel for \class in Iceberg}}
        \label{fig:appendix-iceberg-class-pca}
    \end{subfigure}
    \hfill
    \begin{subfigure}[b]{0.45\textwidth}
        \includegraphics[width=\linewidth]{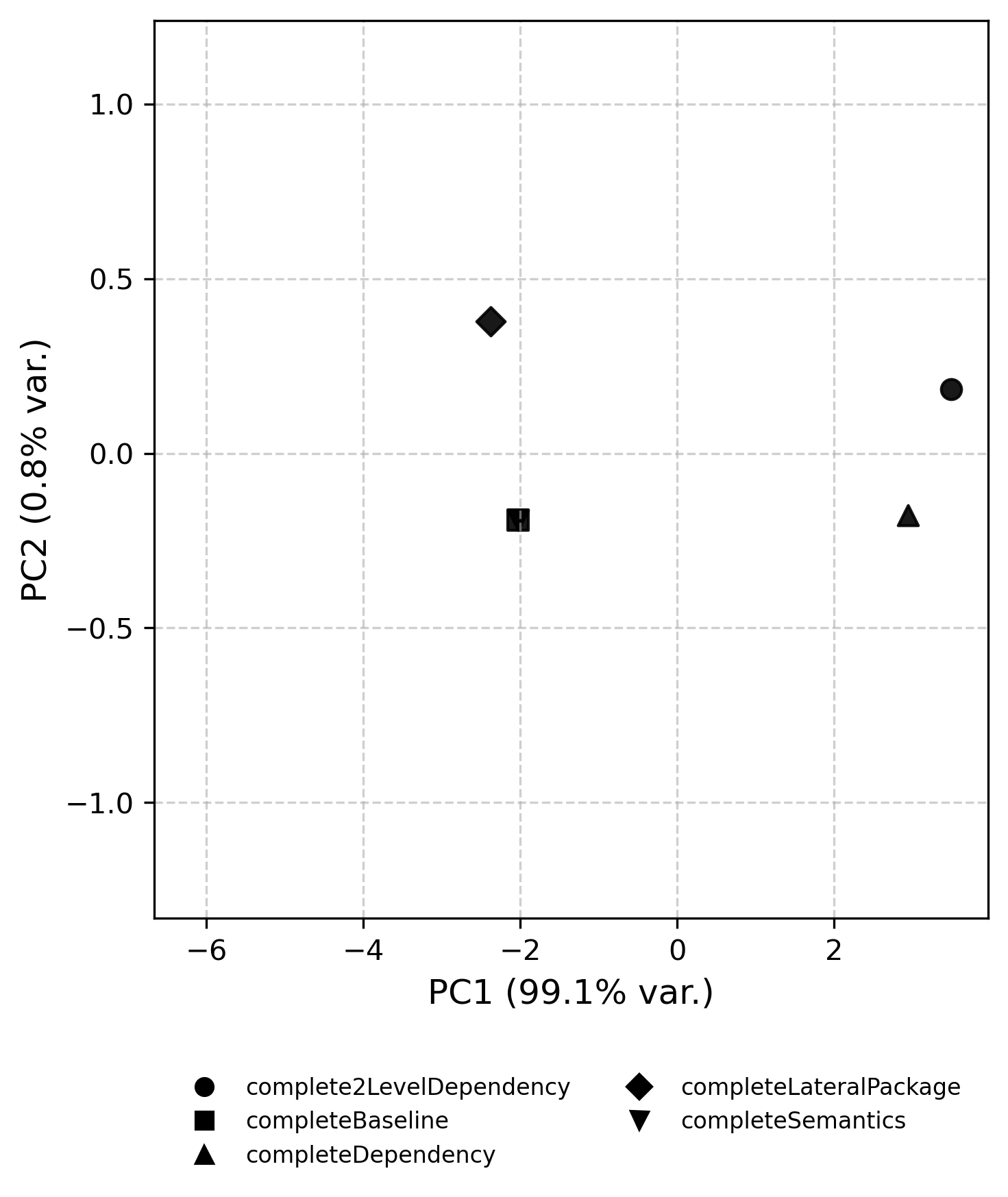}
        \caption{\textbf{\completelevel for \class in Iceberg-tests}}
        \label{fig:appendix-icebergtests-class-pca}
    \end{subfigure}

    \vspace{0.25cm}

    % ---- Row 2 ----
    \begin{subfigure}[b]{0.45\textwidth}
        \includegraphics[width=\linewidth]{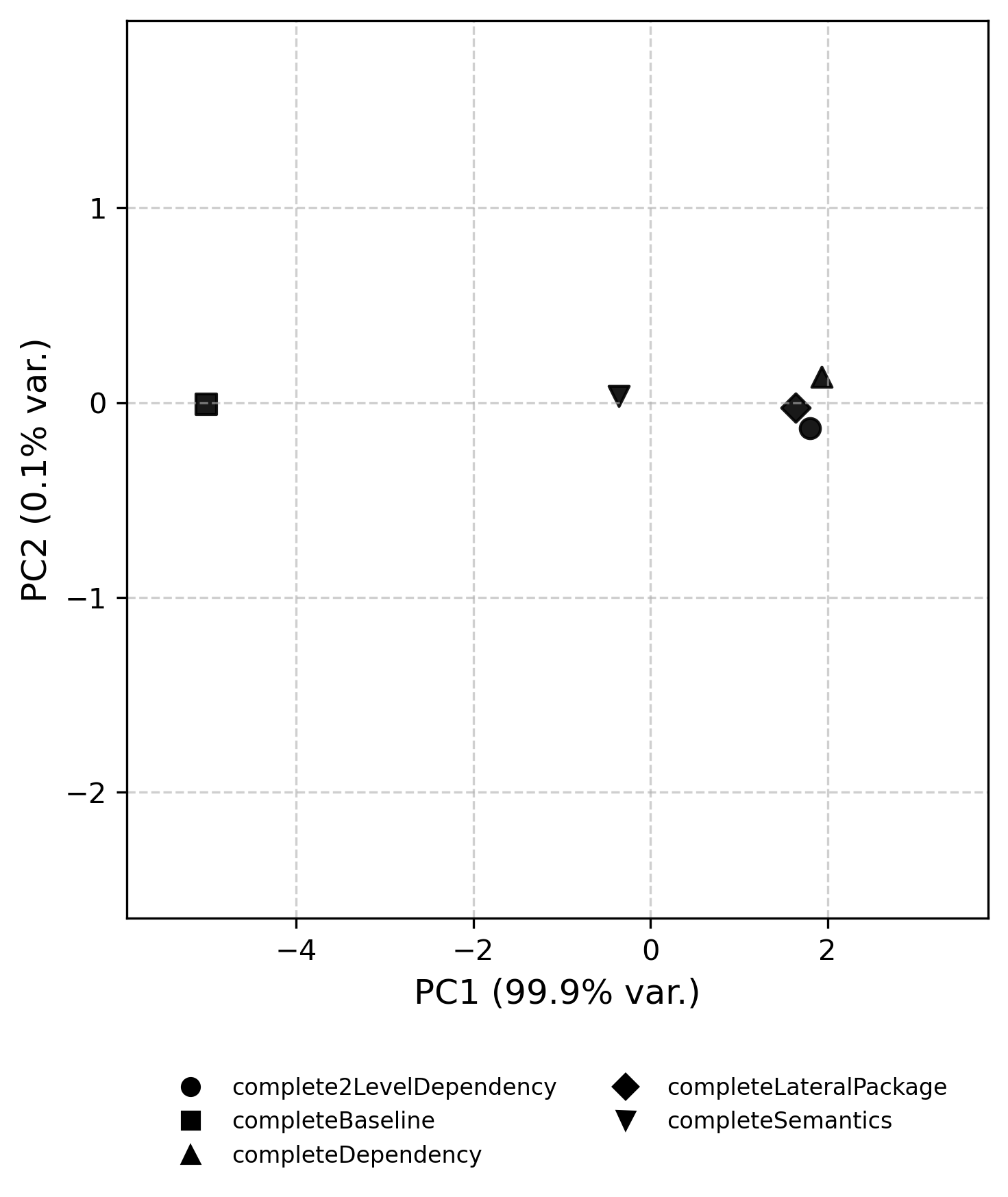}
        \caption{\textbf{\completelevel for \method in Iceberg}}
        \label{fig:appendix-iceberg-method-pca}
    \end{subfigure}
    \hfill
    \begin{subfigure}[b]{0.45\textwidth}
        \includegraphics[width=\linewidth]{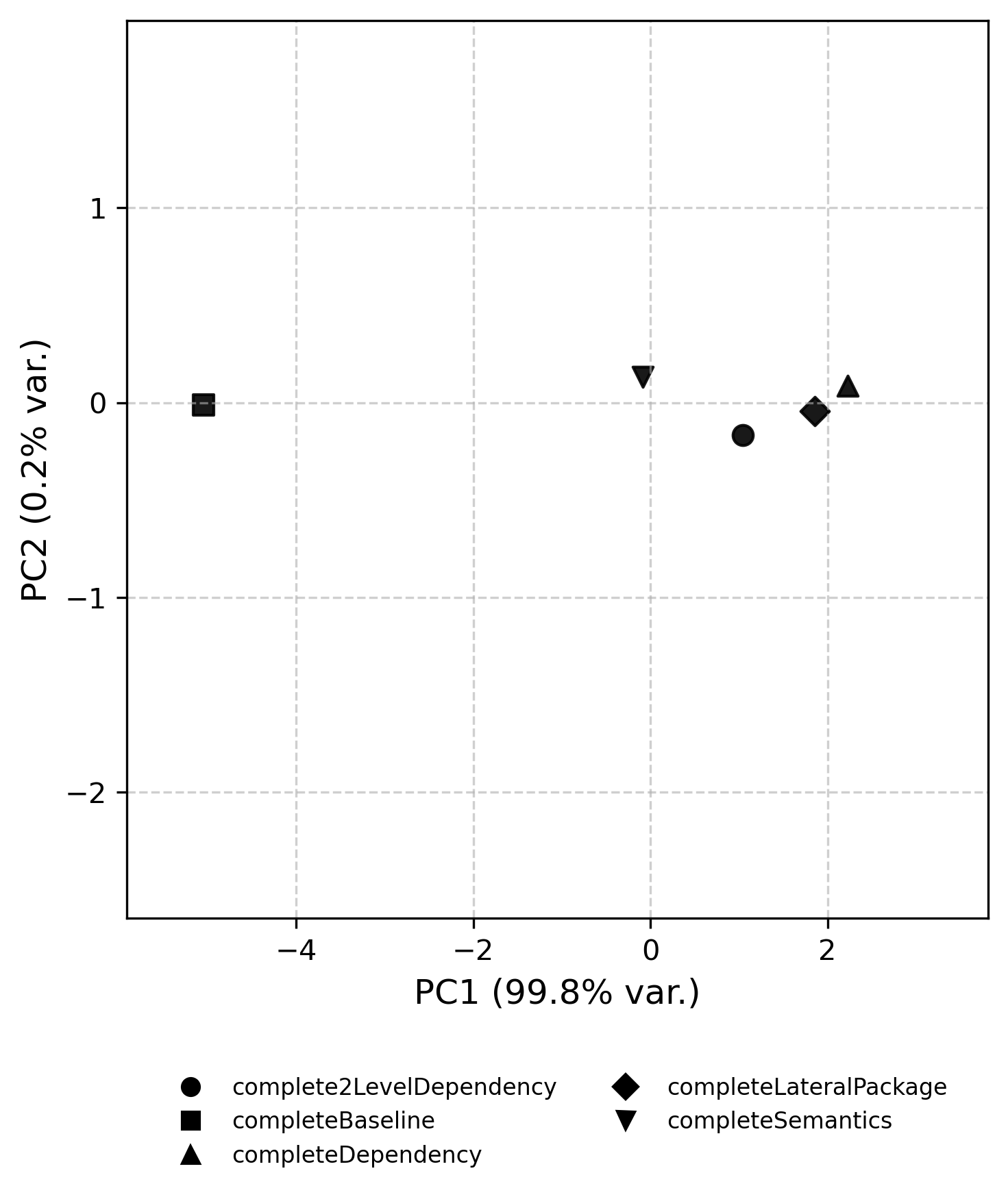}
        \caption{\textbf{\completelevel for \method in Iceberg-tests}}
        \label{fig:appendix-icebergtests-method-pca}
    \end{subfigure}

    \caption{\textbf{Comparison of \completelevel for  \class \& \method Iceberg \& Iceberg-tests}}
    \label{fig:appendix-iceberg-pca}
\end{figure}

%moose and mosse-tests
\begin{figure}[H]
    \centering
    % ---- Row 1 ----
    \begin{subfigure}[b]{0.45\textwidth}
        \includegraphics[width=\linewidth]{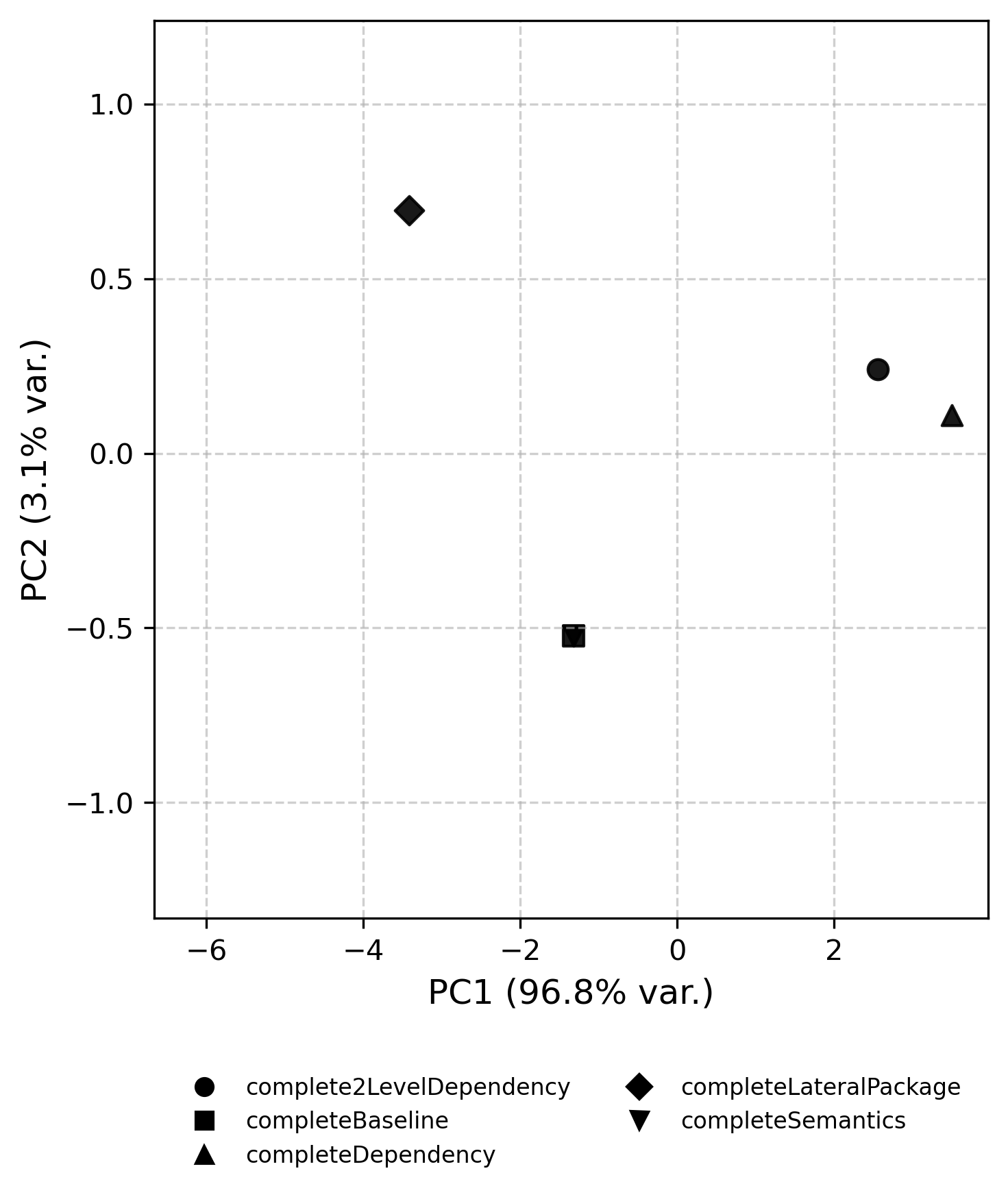}
        \caption{\textbf{\completelevel for \class in Moose}}
        \label{fig:appendix-moose-class-pca}
    \end{subfigure}
    \hfill
    \begin{subfigure}[b]{0.45\textwidth}
        \includegraphics[width=\linewidth]{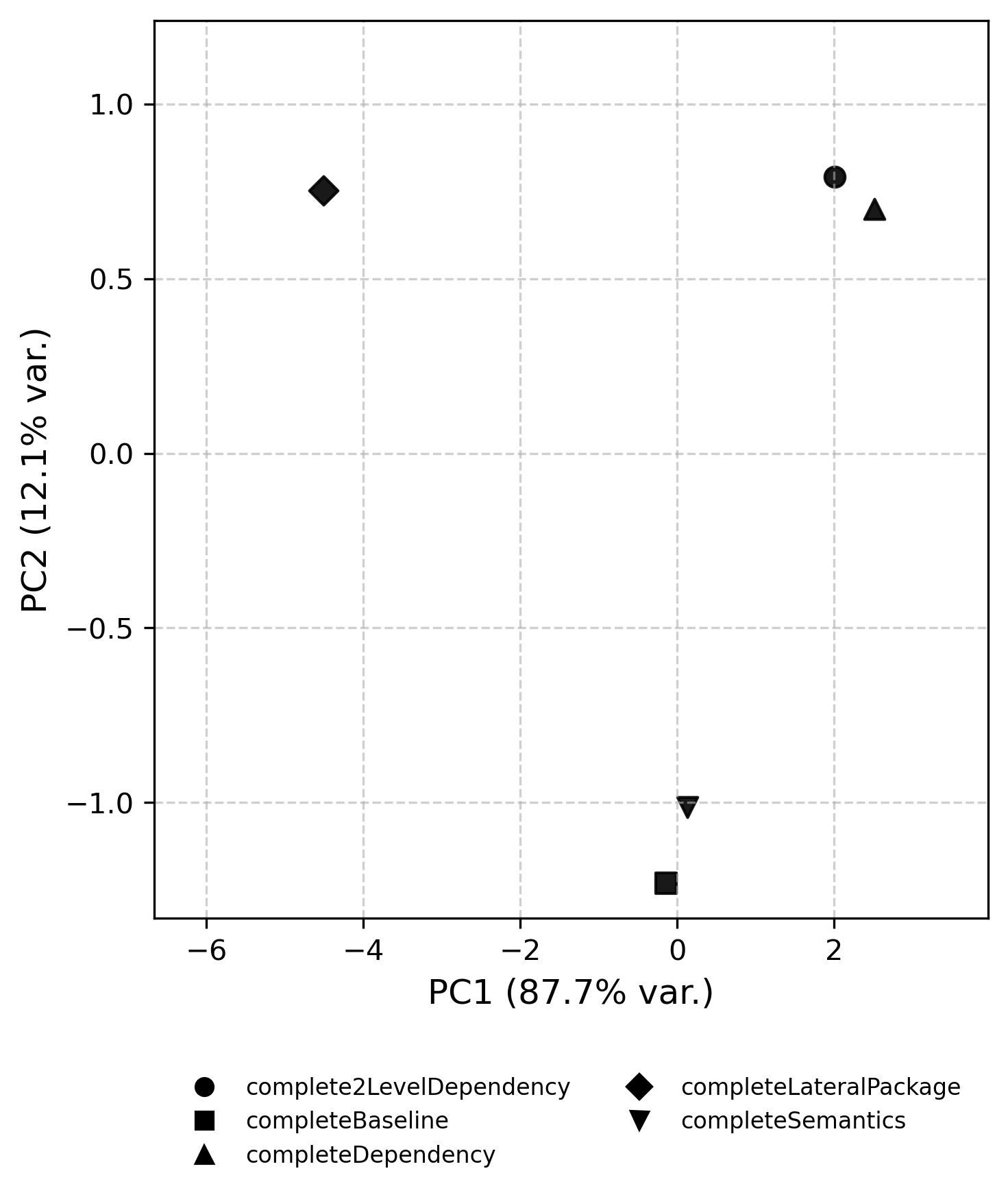}
        \caption{\textbf{\completelevel for \class in Moose-tests}}
        \label{fig:appendix-moose-tests-class-pca}
    \end{subfigure}

    \vspace{0.25cm}

    % ---- Row 2 ----
    \begin{subfigure}[b]{0.45\textwidth}
        \includegraphics[width=\linewidth]{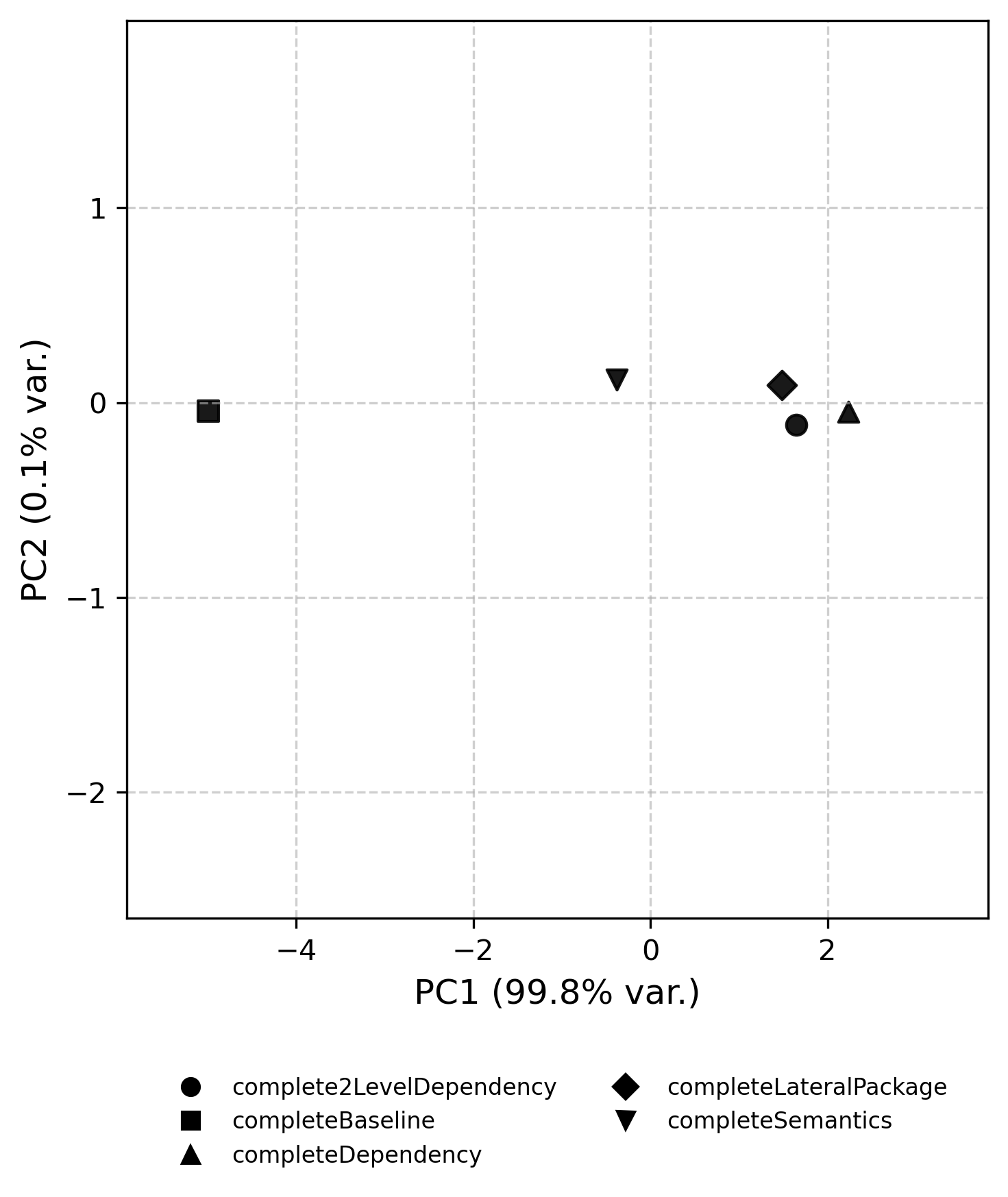}
        \caption{\textbf{\completelevel for \method in moose}}
        \label{fig:appendix-moose-method-pca}
    \end{subfigure}
    \hfill
    \begin{subfigure}[b]{0.45\textwidth}
        \includegraphics[width=\linewidth]{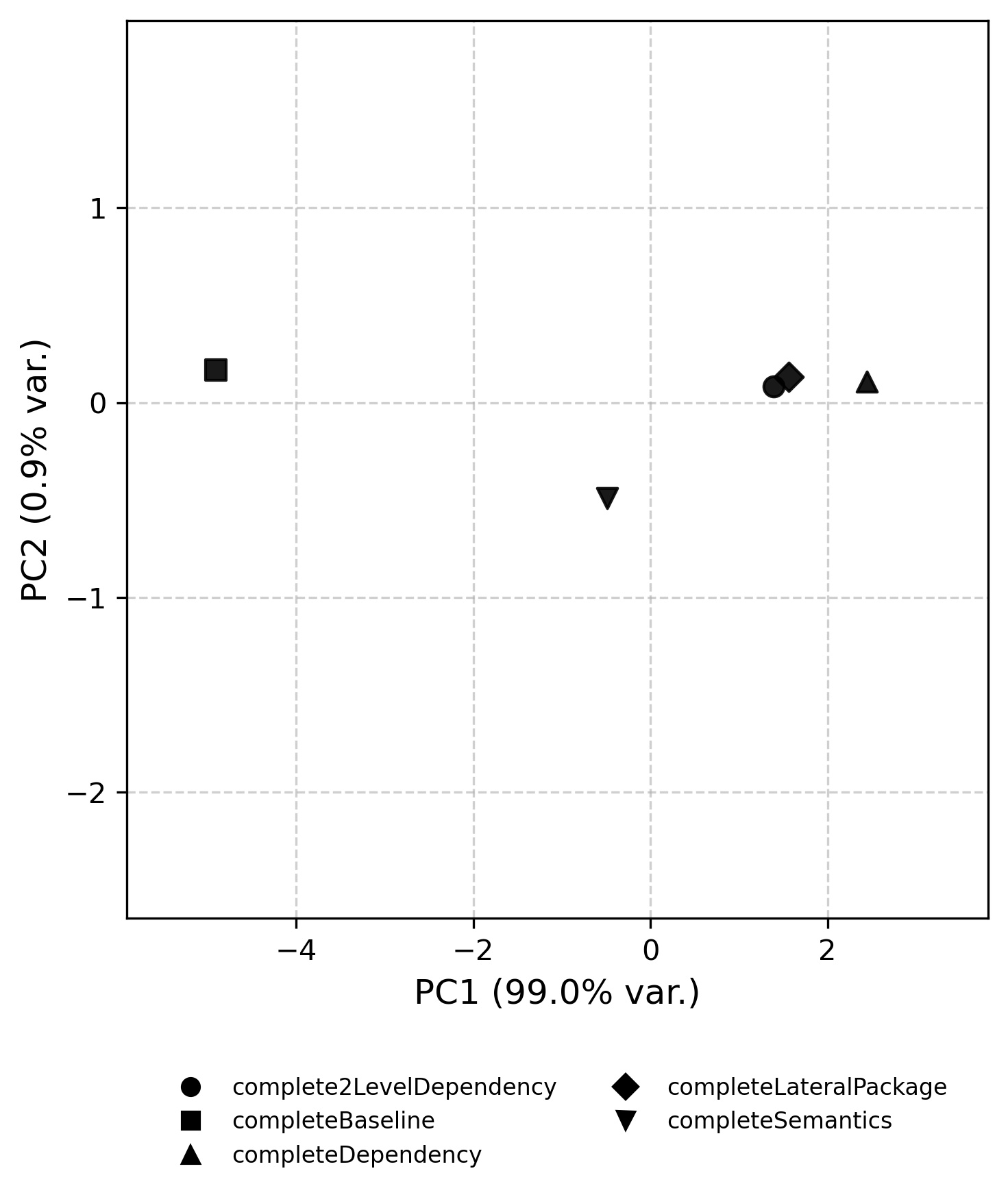}
        \caption{\textbf{\completelevel for \method in Moose-tests}}
        \label{fig:appendix-moosetests-method-pca}
    \end{subfigure}

    \caption{\textbf{Comparison of \completelevel for  \class \& \method Moose \& Moose-tests}}
    \label{fig:appendix-moose-pca}
\end{figure}

%roassal and roassal-tests
\begin{figure}[H]
    \centering
    % ---- Row 1 ----
    \begin{subfigure}[b]{0.45\textwidth}
        \includegraphics[width=\linewidth]{variables-roassal-pca-complete.png}
        \caption{\textbf{\completelevel for \class in Roassal}}
        \label{fig:appendix-roassal-class-pca}
    \end{subfigure}
    \hfill
    \begin{subfigure}[b]{0.45\textwidth}
        \includegraphics[width=\linewidth]{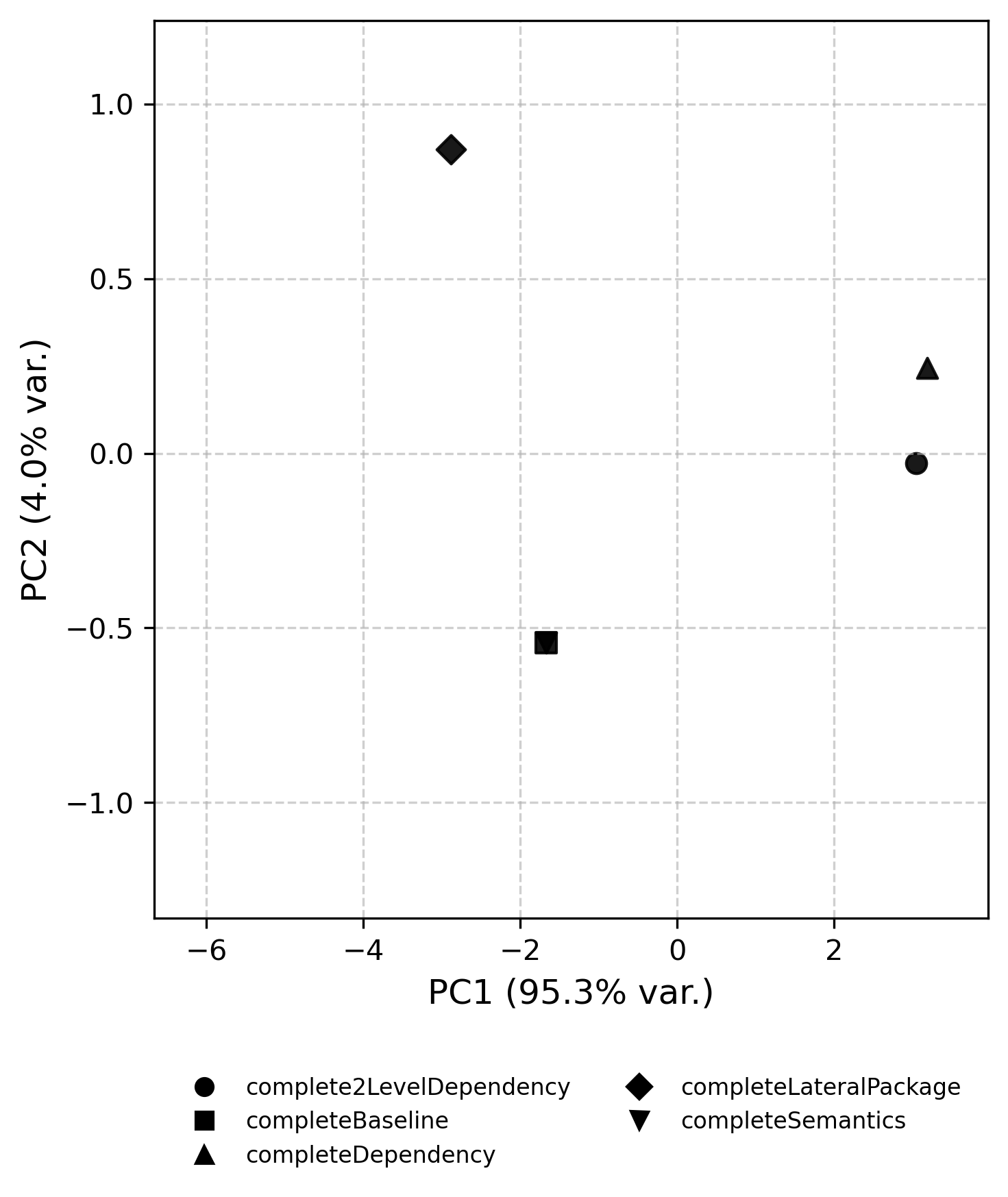}
        \caption{\textbf{\completelevel for \class in Roassal-tests}}
        \label{fig:appendix-roassaltests-class-pca}
    \end{subfigure}

    \vspace{0.25cm}

    % ---- Row 2 ----
    \begin{subfigure}[b]{0.45\textwidth}
        \includegraphics[width=\linewidth]{method-roassal-pca-complete.png}
        \caption{\textbf{\completelevel for \method in Roassal}}
        \label{fig:appendix-roassal-method-pca}
    \end{subfigure}
    \hfill
    \begin{subfigure}[b]{0.45\textwidth}
        \includegraphics[width=\linewidth]{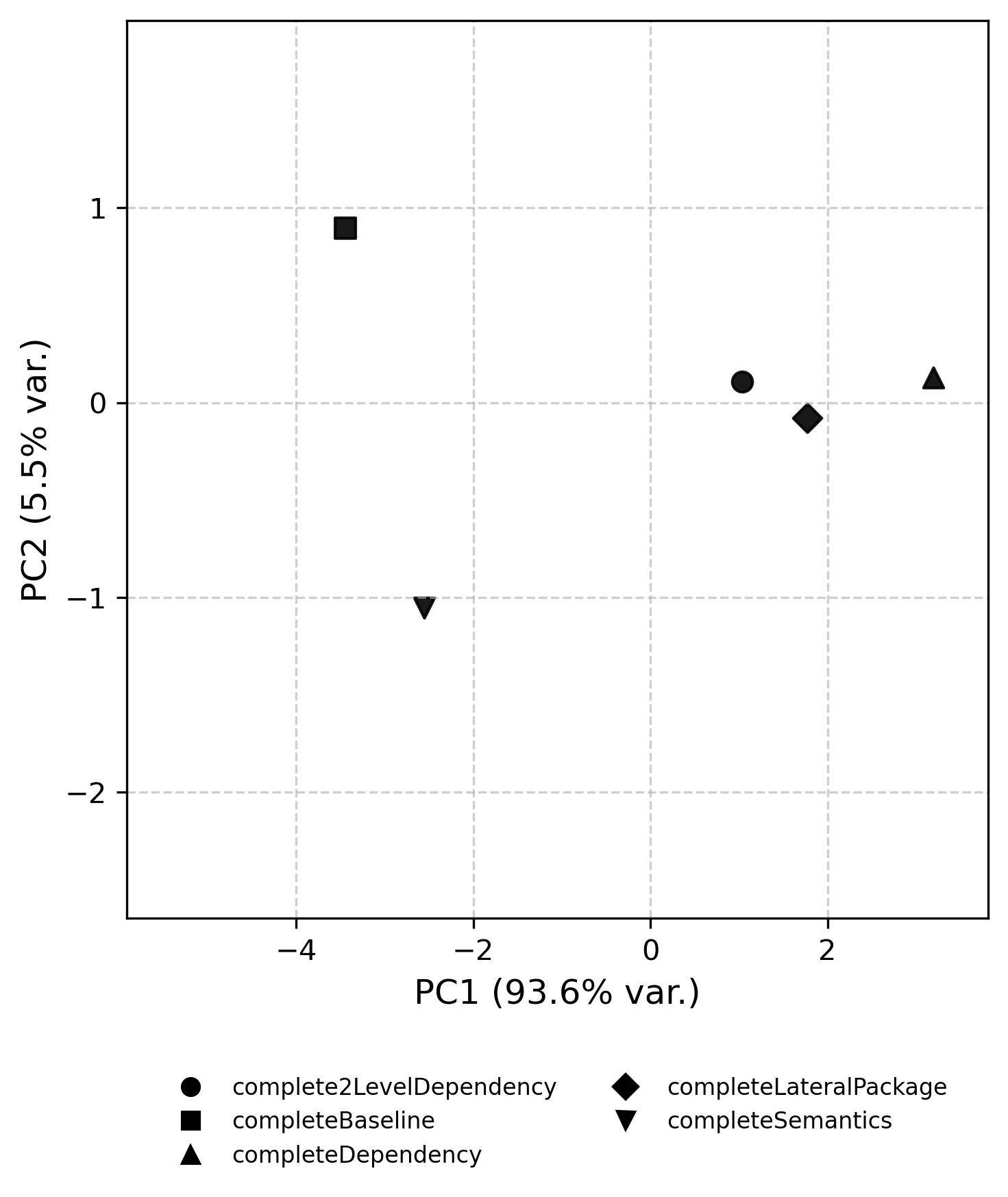}
        \caption{\textbf{\completelevel for \method in Roassal-tests}}
        \label{fig:appendix-roassaltests-method-pca}
    \end{subfigure}

    \caption{\textbf{Comparison of \completelevel for  \class \& \method Roassal \& Roassal-tests}}
    \label{fig:appendix-roassal-pca}
\end{figure}

%seaside and seaside-tests
\begin{figure}[H]
    \centering
    % ---- Row 1 ----
    \begin{subfigure}[b]{0.45\textwidth}
        \includegraphics[width=\linewidth]{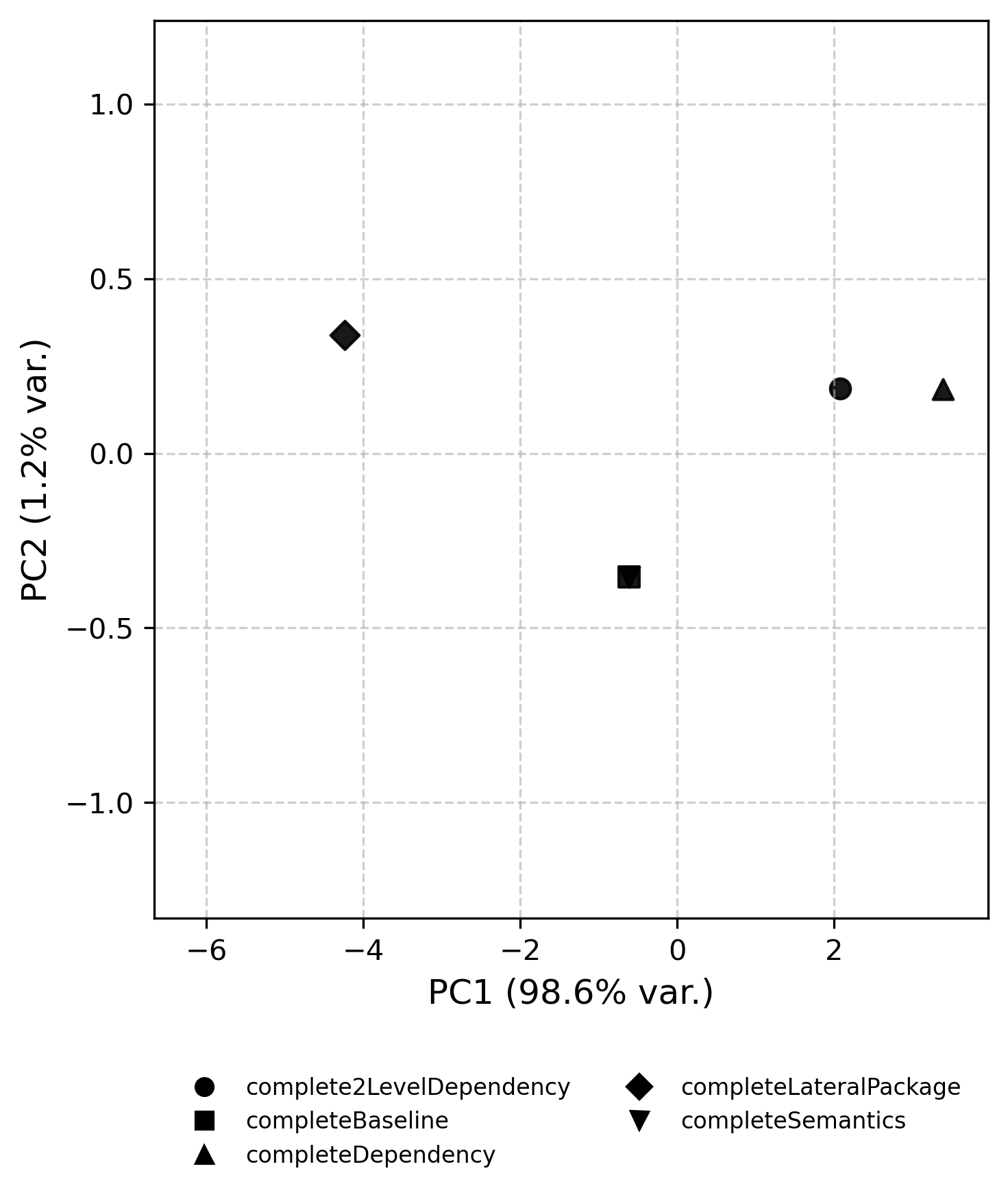}
        \caption{\textbf{\completelevel for \class in Seaside}}
        \label{fig:appendix-seaside-class-pca}
    \end{subfigure}
    \hfill
    \begin{subfigure}[b]{0.45\textwidth}
        \includegraphics[width=\linewidth]{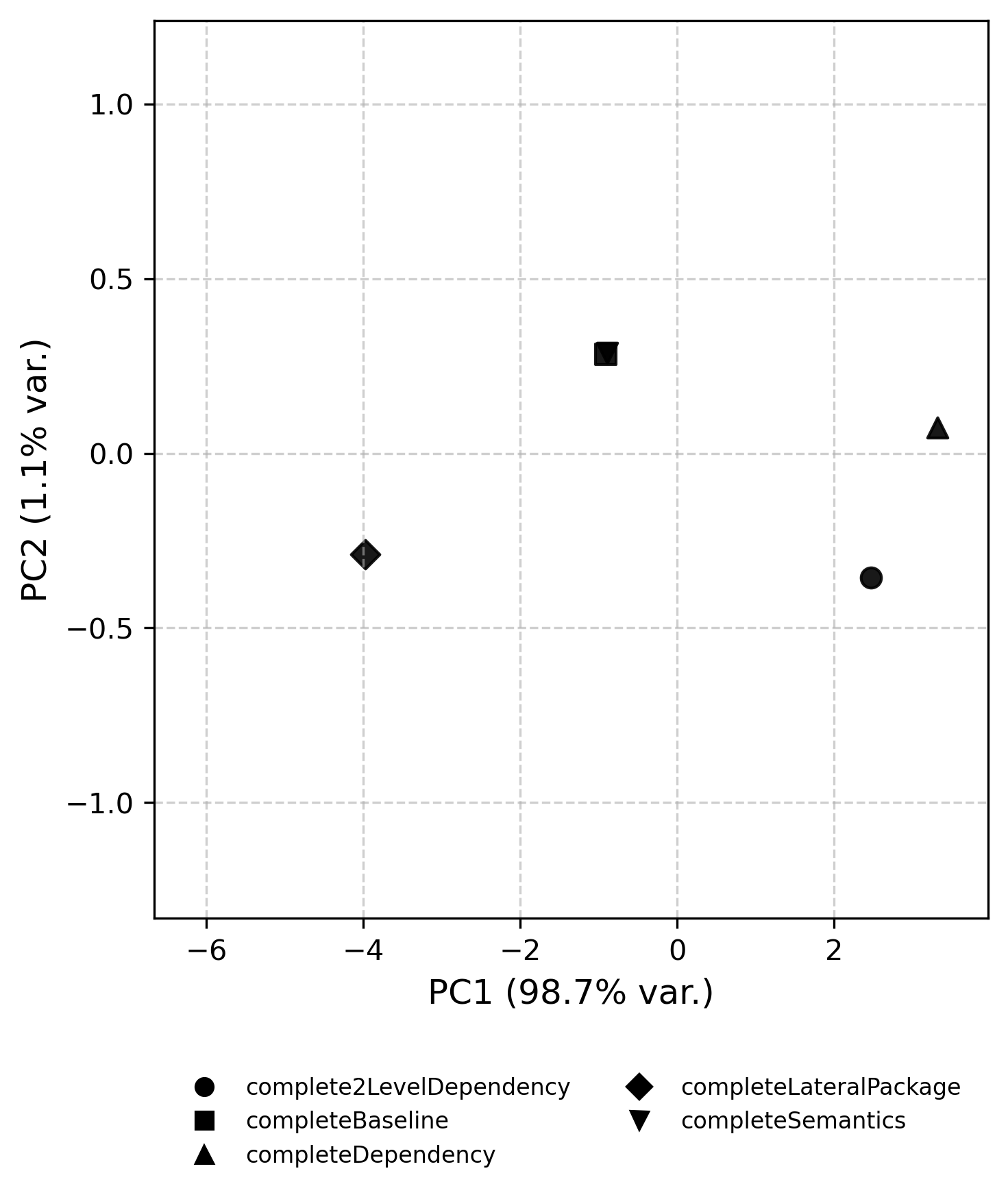}
        \caption{\textbf{\completelevel for \class in Seaside-tests}}
        \label{fig:appendix-seasidetests-class-pca}
    \end{subfigure}

    \vspace{0.25cm}

    % ---- Row 2 ----
    \begin{subfigure}[b]{0.45\textwidth}
        \includegraphics[width=\linewidth]{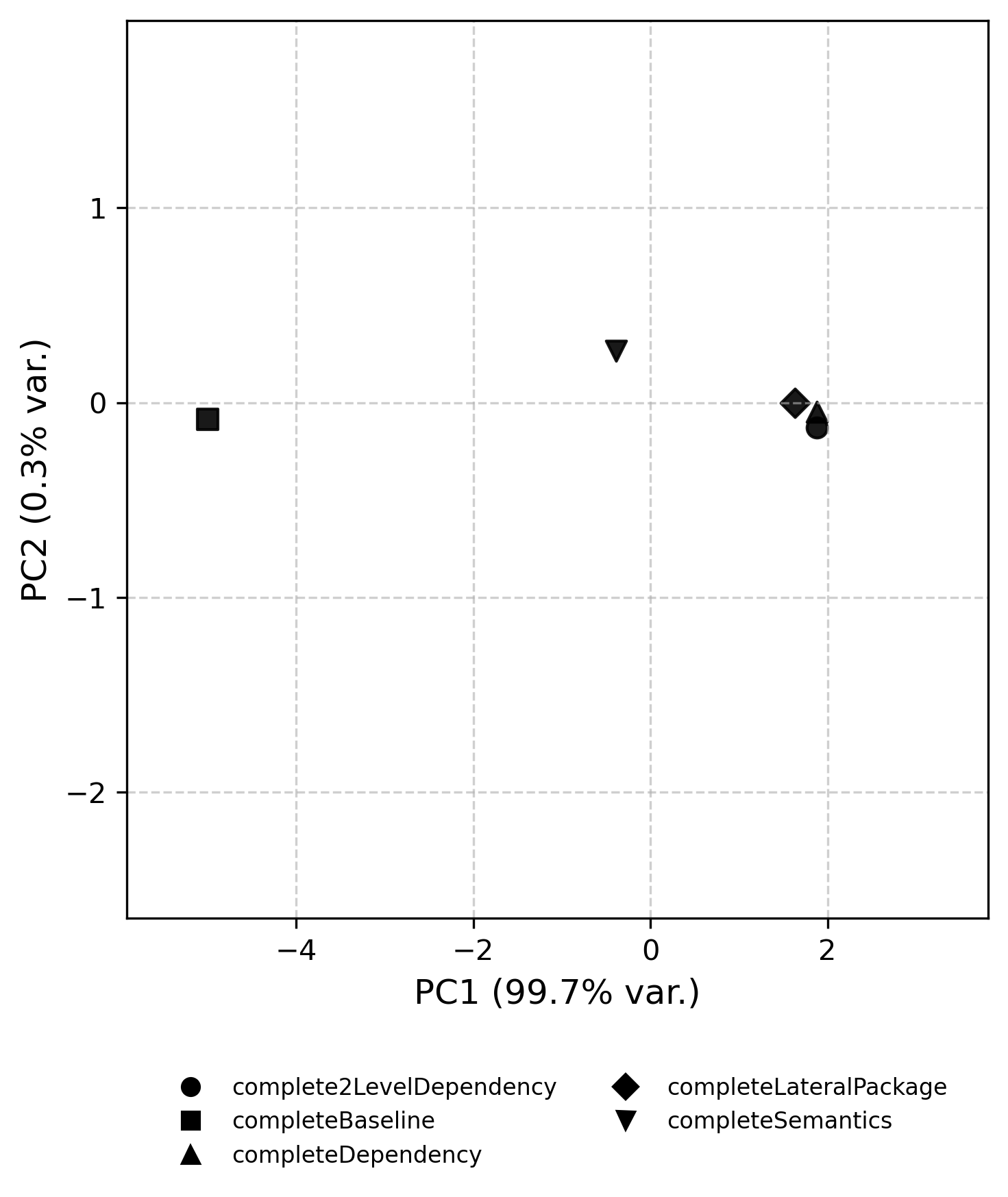}
        \caption{\textbf{\completelevel for \method in Seaside}}
        \label{fig:appendix-seaside-method-pca}
    \end{subfigure}
    \hfill
    \begin{subfigure}[b]{0.45\textwidth}
        \includegraphics[width=\linewidth]{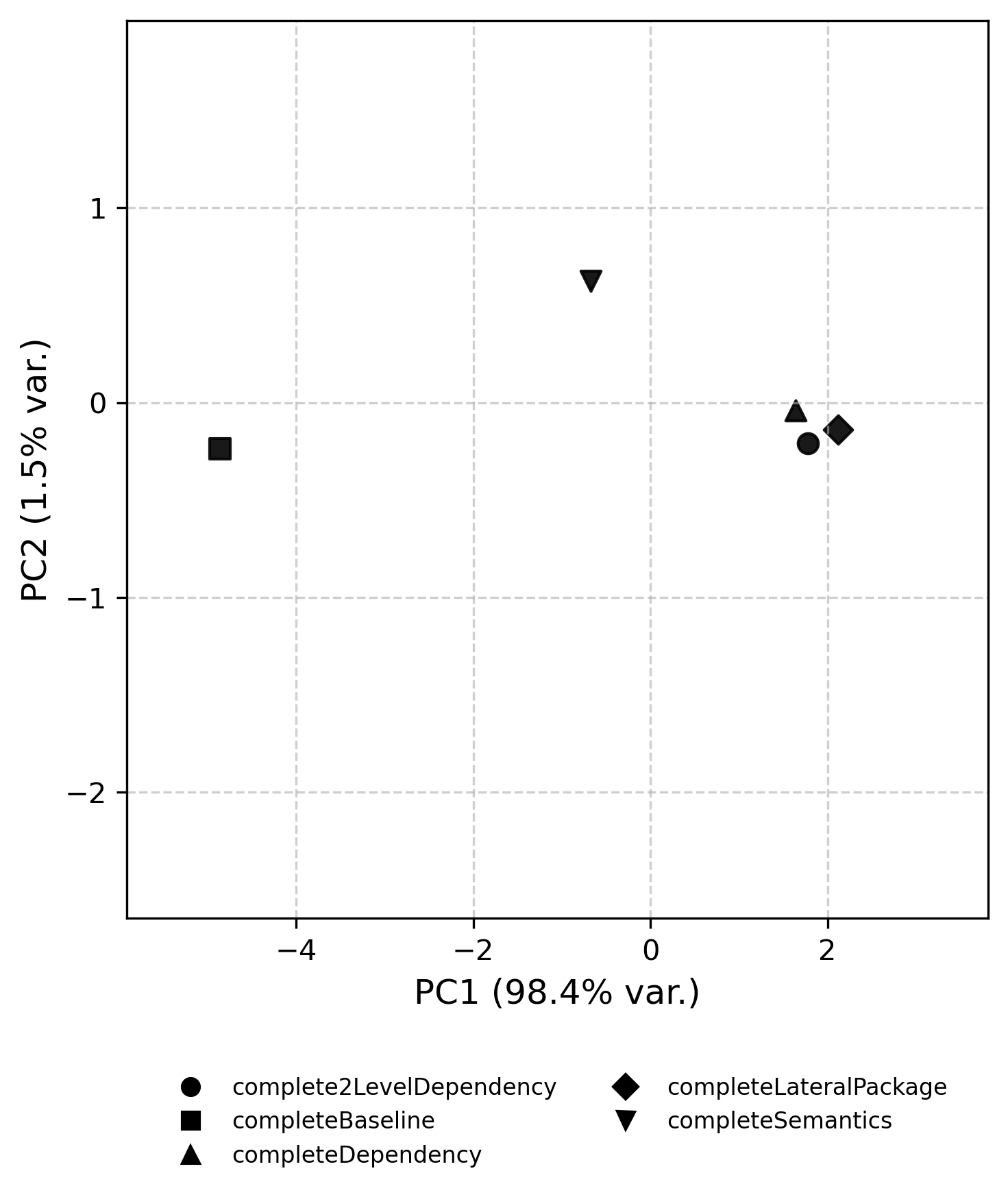}
        \caption{\textbf{\completelevel for \method in Seaside-tests}}
        \label{fig:appendix-seasidetests-method-pca}
    \end{subfigure}

    \caption{\textbf{Comparison of \completelevel for  \class \& \method Seaside \& Seaside-tests}}
    \label{fig:appendix-seaside-pca}
\end{figure}

%spec and spec-tests
\begin{figure}[H]
    \centering
    % ---- Row 1 ----
    \begin{subfigure}[b]{0.45\textwidth}
        \includegraphics[width=\linewidth]{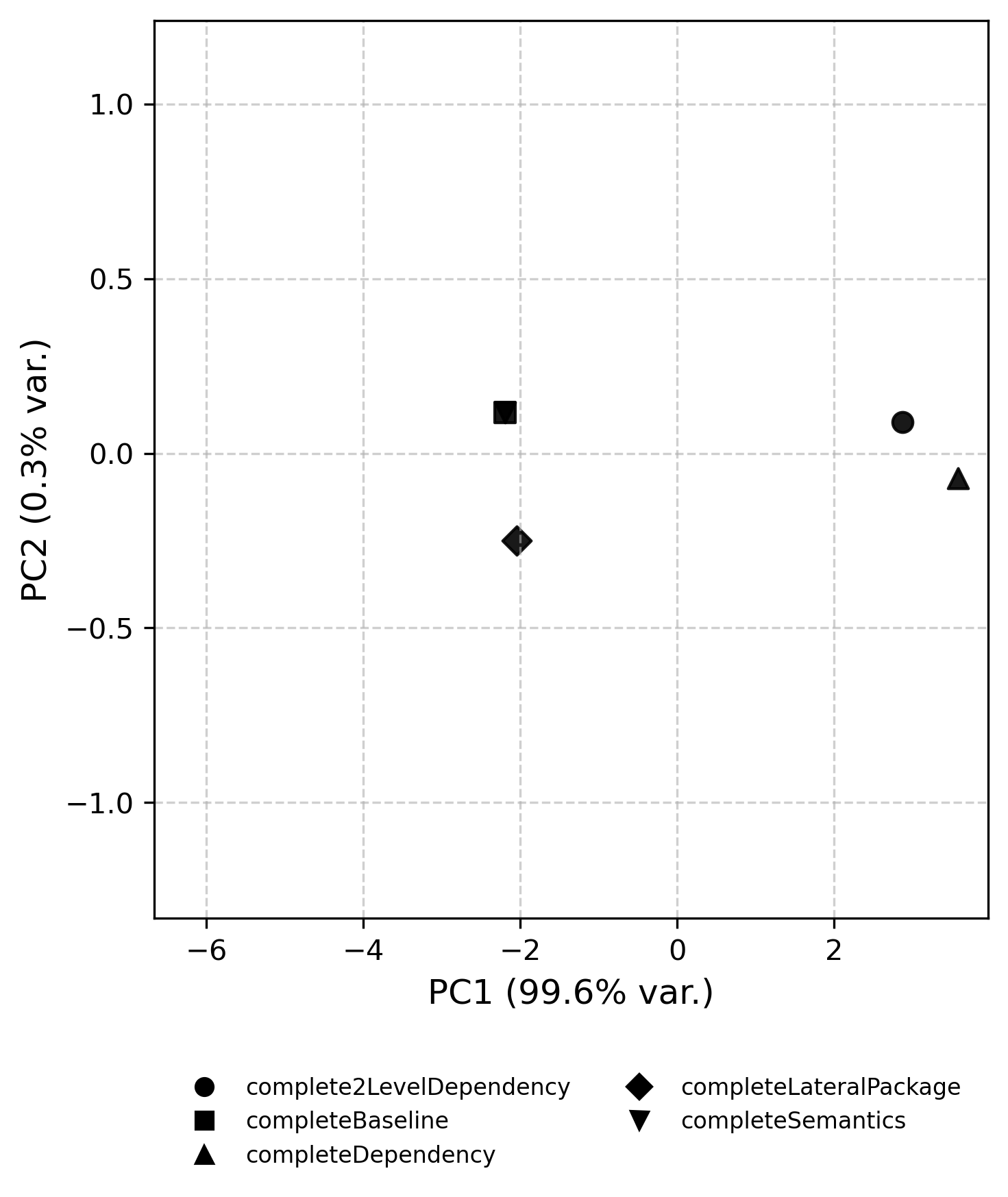}
        \caption{\textbf{\completelevel for \class in Spec}}
        \label{fig:appendix-spec-class-pca}
    \end{subfigure}
    \hfill
    \begin{subfigure}[b]{0.45\textwidth}
        \includegraphics[width=\linewidth]{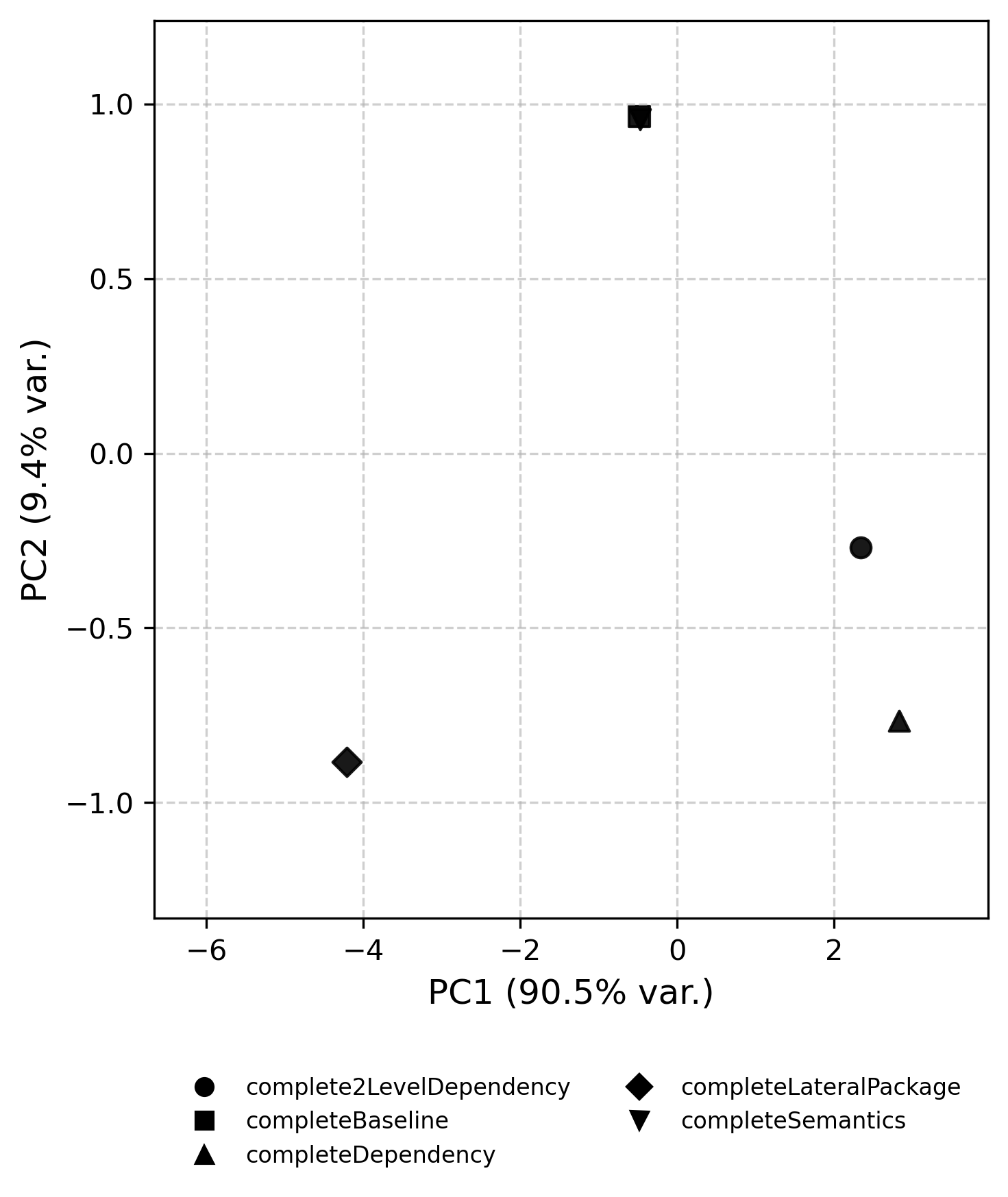}
        \caption{\textbf{\completelevel for \class in Spec-tests}}
        \label{fig:appendix-spectests-class-pca}
    \end{subfigure}

    \vspace{0.25cm}

    % ---- Row 2 ----
    \begin{subfigure}[b]{0.45\textwidth}
        \includegraphics[width=\linewidth]{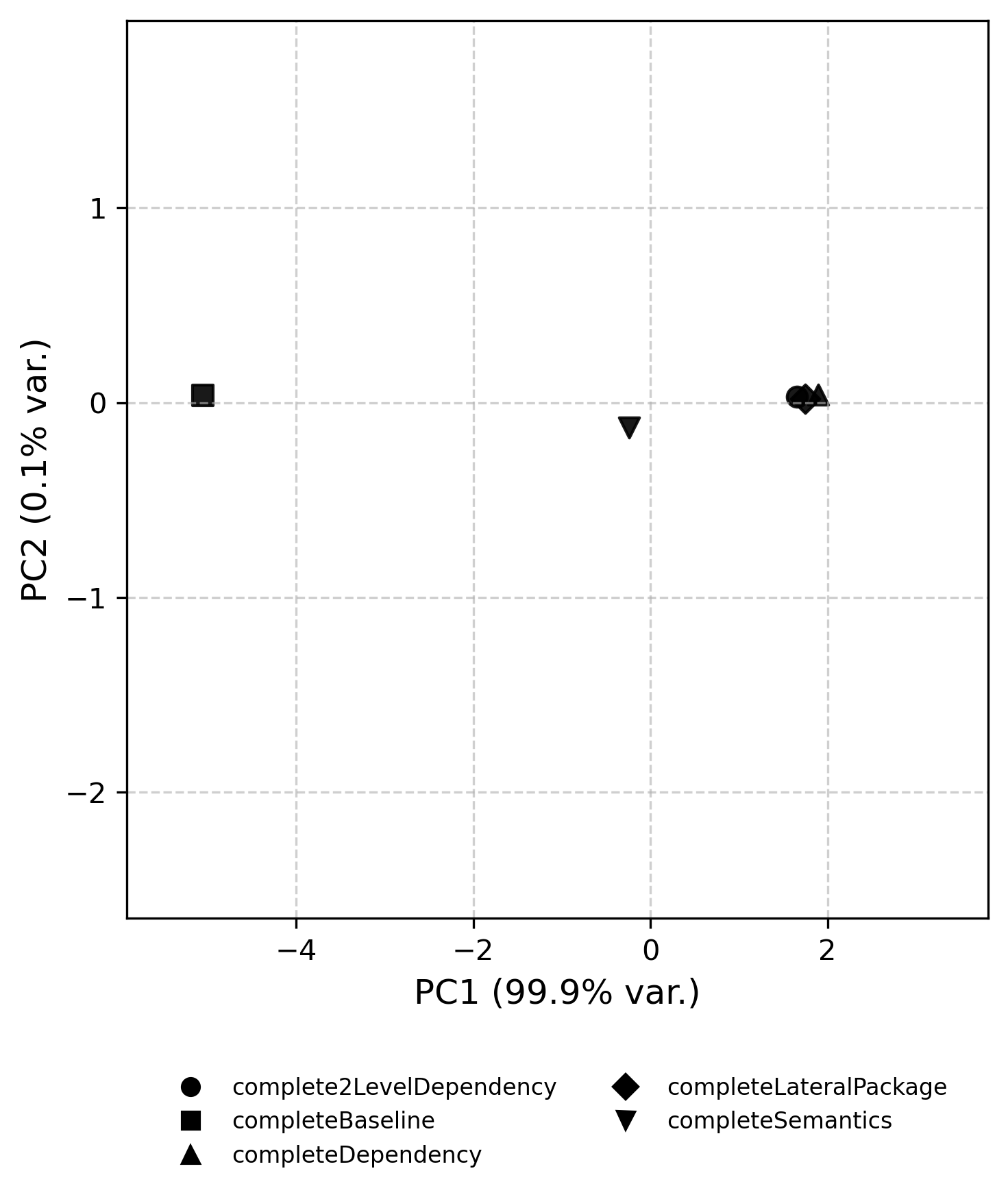}
        \caption{\textbf{\completelevel for \method in Spec}}
        \label{fig:appendix-spec-method-pca}
    \end{subfigure}
    \hfill
    \begin{subfigure}[b]{0.45\textwidth}
        \includegraphics[width=\linewidth]{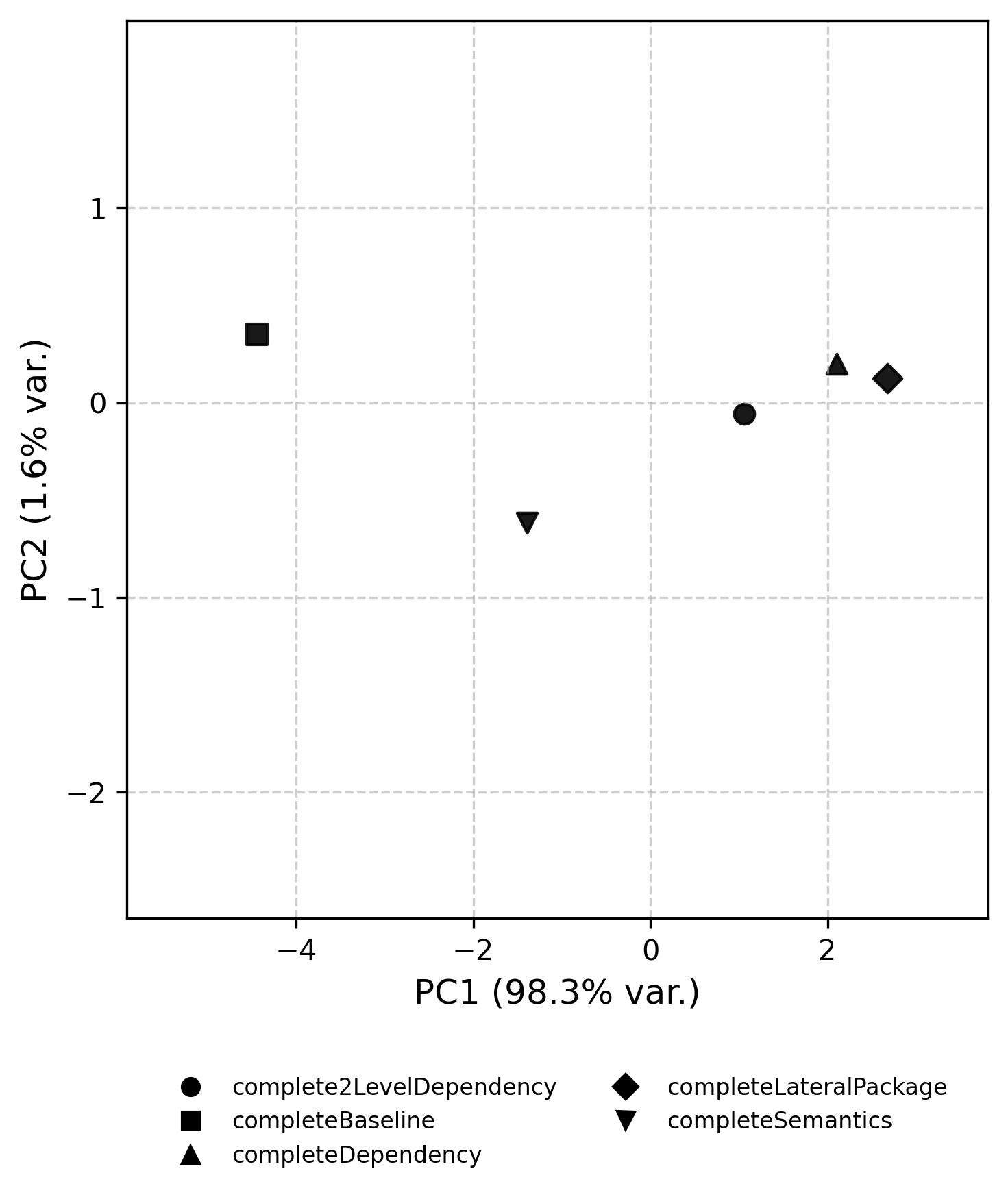}
        \caption{\textbf{\completelevel for \method in Spec-tests}}
        \label{fig:appendix-spectests-method-pca}
    \end{subfigure}

    \caption{\textbf{Comparison of \completelevel for  \class \& \method Spec \& Spec-tests}}
    \label{fig:appendix-spec-pca}
\end{figure}

\end{document}